\documentclass[11pt,preprint,showpacs]{revtex4}
\usepackage{graphicx}
\usepackage{dcolumn}
\usepackage{bm}
\usepackage{amsmath}
\usepackage{amssymb}
\usepackage{amsfonts}

\begin{document}

\title{Numerical investigations of the three-dimensional proton-proton screened Coulomb t-matrix}

\author{R.~Skibi\'nski}
\affiliation{M. Smoluchowski Institute of Physics, Jagiellonian
University,
                    PL-30059 Krak\'ow, Poland}
\author{J.~Golak}
\affiliation{M. Smoluchowski Institute of Physics, Jagiellonian
University,
                    PL-30059 Krak\'ow, Poland}
\author{H.~Wita{\l}a}
\affiliation{M. Smoluchowski Institute of Physics, Jagiellonian
University,
                    PL-30059 Krak\'ow, Poland}

\date{\today}
\begin{abstract}
We demonstrate behaviour of the momentum space screened 
Coulomb t-matrix, obtained by a numerical solution of the three-dimensional 
Lippmann-Schwinger equation. Examples are given for 
different types of screening.  
They prove that it is possible to obtain numerically
a reliable three-dimensional screened Coulomb t-matrix, 
what is important in view of its application in  
few-body calculations.
\end{abstract}


\maketitle \setcounter{page}{1}

\section{Introduction}

The few nucleon systems can be nowadays studied using 
realistic nuclear forces in all their complexity.
However, addition of the Coulomb force acting between protons poses serious obstacle
and is still a challenging task.
Up to now the long range Coulomb potential could be exactly implemented
only into calculations of $^3$He and $^4$He bound states and
of scattering states up to three nucleons.

In the proton-proton (pp) elastic scattering Coulomb force can be included exactly and 
observables can 
be calculated using e.g. the Vincent-Phatak method ~\cite{Vincent74}.
For three-nucleon (3N) reactions it is possible to include the Coulomb force
when performing calculations of the elastic proton-deuteron (pd) scattering both 
in a coordinate~\cite{kievski}
as well as in a momentum space representation~\cite{deltuva,witala}.
The pd breakup observables can be predicted only in the momentum space using  
two approaches~\cite{deltuva2,witala2}. Both rely on the screened Coulomb 
potential and renormalization, 
what permits to apply standard methods valid for short-range interactions. 

In the formalism of Refs.~\cite{witala} and~\cite{witala2} the Coulomb 
interaction enters Faddeev equations via the screened
Coulomb t-matrix. This t-matrix 
appears in a partial wave decomposed
form as well as in its direct three-dimensional form for which 
no partial wave decomposition is performed.
This allows to include the full Coulomb interaction, not
restricted to the finite number of the lowest partial waves.
Since that three-dimensional screened Coulomb t-matrix is a basic
component of that formulation, therefore one has to put a special
attention to its numerical realization.
 
Precise numerical calculations of the three dimensional screened Coulomb t-matrix
are possible for any type of screening. They are numerically quite demanding but in principle
use the concepts developed in~\cite{elster1998}
for the Malfliet-Tjion potential.
This potential consists of two terms which have the same functional form as 
an exponentially screened Coulomb potential with the screening parameter $n=1$ 
(for details see below). 
Such particular value of the screening parameter $n$ allows one
to perform many steps analytically when obtaining momentum space 
partial wave decomposed as well as direct three-dimensional t-matrix elements.
The analogous procedure can be adapted to the higher values of $n$, as done
in~\cite{skib09}, where observables for pp scattering were calculated.
Also there on-, half- and off-the-energy-shell elements of the t-matrix are compared 
with the analytical expressions in the screening limit from Refs.~\cite{kok81,henrybook}.

In this paper we present details of three-dimensional screened Coulomb  
t-matrix calculations. The behaviour of the numerically obtained 
t-matrix and its dependence on 
the screening parameters for different types of screening 
will be presented.
In Sect.\ref{sec1} we describe the numerical procedure 
to get the three dimensional screened Coulomb t-matrix.
In Sect.\ref{sec3} the behaviour of resulting t-matrices 
obtained with different forms of screening will be presented and 
t-matrices 
will be compared
to the screened Coulomb potential.
In Sect.\ref{sec4} the convergence of the partial wave
decomposition of the screened Coulomb t-matrix
will be shown by comparing it with the 
generated three-dimensional screened Coulomb t-matrix.   
Finally, we summarize in Sect.\ref{summary}.

\section{Numerical calculation of the screened Coulomb t-matrix.}
\label{sec1}

The three-dimensional momentum space matrix elements of the screened Coulomb potential 
$< \vec p~' | V_c^R | \vec p > \equiv V_c^R (p',p,x\equiv\ \cos(\theta) = \hat p \cdot \hat{p'})$ and the
3-dimensional screened Coulomb t-matrix elements
$< \vec p~' | t_c^R(E) | \vec p > \equiv
t_c^R(p',p, 
\hat p \cdot {\hat p}';E)$ at energy $E=\frac{p_0^2}{m}$ are related by
~\cite{elster1998,skib09}
\begin{equation}
t_c^R(p',p,x;E) = \frac {1} {2\pi} v_c^R(p',p,x,1) +
\int_0^{\infty} dp"p"^2 \int_{-1}^1 dx" v_c^R(p',p",x,x")
\frac{1} {E+i\epsilon - \frac {p"^2} {m_p} } t_c^R(p",p,x";E) \label{eq.1}
\end{equation}
with
\begin{equation}
v_c^R(p',p,x',x) \equiv \int_0^{2\pi} d\phi
V_c^R(p',p,x'x+\sqrt{1-x'^2} \sqrt{1-x^2} cos\phi) ~.
\label{eq.1a}
\end{equation}

We solve Eq.~(\ref{eq.1}) by generating its
Neumann series  and applying  Pad\`e summation~\cite{skib09} for three types of screening:

1. the sharp cut-off screening 
\begin{equation}
V_c^R(r) = \Theta(R-r) \frac{e^ 2}{r}\;
\label{eq.3sharp}
\end{equation}
with the unit step function $\Theta$. This screening leads to matrix elements
\begin{equation}
v_c^R(p',p,x',x)=\frac{e^2}{2\pi^2q^2} \int_0^{2\pi} d\phi (1-\cos(qR))
\end{equation}
where $q \equiv \sqrt{p^2+p'^2-2pp'(x'x+\sqrt{1-x'^2}\sqrt{1-x^2}cos\phi)}$.

2. the localized screening~\cite{deltuva09}, where the transition from the pure Coulomb potential to zero values takes place
smoothly in a finite interval $[R,3R]$
\begin{equation}
V_c^R(r) = \frac{e^ 2}{r} ( \Theta(R-r) + \frac12 \Theta(r-R)\Theta(3R-r)(1+\sin(\frac{\pi r}{2R})))\;.
\label{localized}
\end{equation}
The corresponding $v_c^R(p',p,x',x)$ is 
\begin{equation}
v_c^R(p',p,x',x)=\frac{e^2}{4\pi^2q^2} \int_0^{2\pi} d\phi \frac{2\pi^2-8q^2R^2-\pi^2(\cos(qR)+\cos(3qR))}
{\pi^2-4q^2R^2}\;.
\end{equation}

3. the exponential screening, dependent on two parameters: the power $n$ and the
screening radius $R$: 
\begin{equation}
V_c^R(r) = \frac{e^ 2}{r} e^{-{(\frac {r} {R})}^n} ~.
\label{eq.2}
\end{equation} 
In this case 
\begin{equation}
v_c^R(p',p,x',x) = \frac{e^2} {2\pi^2} \int_0^{2\pi} d\phi
\int_0^{\infty} dr \frac
{\sin(qr)}{q}e^{-(\frac {r} {R})^n}
\equiv
\frac{e^2} {2\pi^2} \int_0^{2\pi} d\phi \frac{I_{n,R}(q)}{q}~, \label{eq.4}
\end{equation}
where the function $I_{n,R}(q)=\int_0^{\infty} dr \sin(qr) e^{-(\frac {r} {R})^n }$ 

For the value of $n=1$ the integration over $r$ can be performed analytically resulting in 
\begin{equation}
v_c^R(p',p,x',x) = \frac{e^2}{\pi} \frac{1}{\sqrt{(p'^2+p^2-2pp'x+\frac{1}{R^2})^2-4p'^2p^2(1-x')(1-x)}}\;.
\end{equation}

\begin{figure}
\includegraphics[scale=0.9]{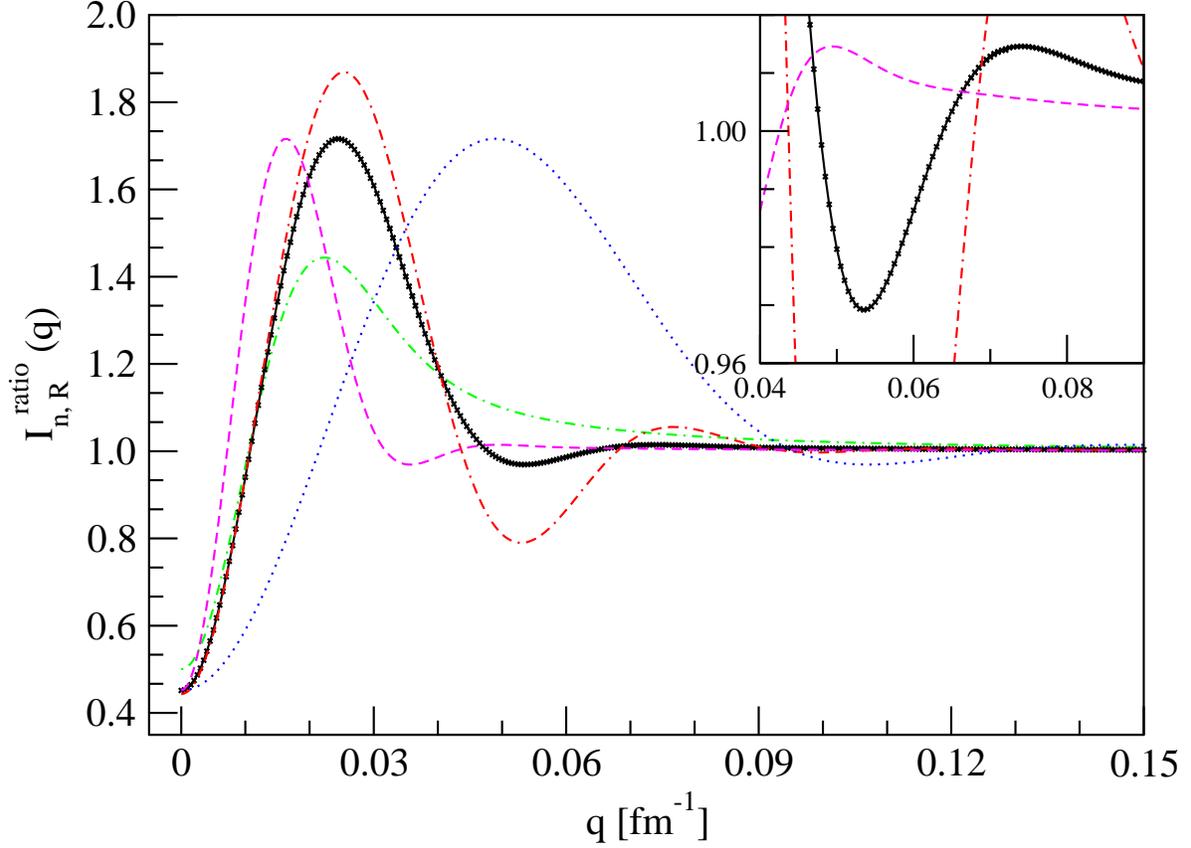}
\caption{(color online) The $I_{n,R}^{ratio}(q)$ for different combinations 
of screening parameters and the quality of polynomial fit for $I_{3,120}^{ratio}(q)$.
The solid (black) line represents $I_{3,120}^{ratio}(q)$ calculated by 
direct integration over $r$ in (\ref{eq.4}) as a function of momenta q.
The black x-es show the values of $I_{3,120}^{ratio}(q)$ obtained 
from its polynomial representation.
Other curves: dash-dotted (green), double-dashed-dotted (red), dotted (blue) and
dashed (magenta) represents 
$I_{2,120}^{ratio}, I_{4,120}^{ratio}, I_{3,60}^{ratio}$
and $I_{3,180}^{ratio}$, respectively. For $0.15 < q < 100$ fm the $I_{n,R}^{ratio}(q)$
is practically equal 1.
}
\label{fig1}
\end{figure}

For $n > 1$ a two-dimensional numerical integration is required to obtain $v_c^R(p',p,x',x)$.
Due to strong oscillations of the integrand in (\ref{eq.4})
the big number of integration r-points is needed to achieve sufficient precision. This 
significantly increases the computer time needed 
for the t-matrix calculation, which has to be done 
on a big grid of $p,p',x$ and $x'$ points. 
Typically, we solve the Lippmann-Schwinger equation~(\ref{eq.1}) using
120 p-points and 190 x-points for the sharply cut off potential
and 95 p-points with 130 x-points for other screenings, what 
requires over $1.5\times10^8$ calculations of the $v_c^R(p',p,x',x)$ function.
The integration over $\phi$ in~(\ref{eq.4}) can be performed with 
relatively small number of $\phi$-points  
and thus the whole numerical difficulty is shifted to 
calculation of $I_{n,R}(q)$. In order to speed it up we use the following method:
in the first step we prepare the $I_{n,R}(q)$ on a grid of 300 q-points in the range of 0-100 fm$^{-1}$. 
In order to calculate the integral over $r$ we use the Filon's integration formula~\cite{abram} 
which is dedicated to 
integrals of the product of the sine (or cosine) with some 
nonoscillatory function $f(x)$.
The upper limit of integration $r_{max}$ is chosen sufficiently large so that 
the integrand approaches zero ($e^{-(\frac {r_{max}} {R})^n }=10^{-20}$).
Since the resulting function $I_{n,R}(q)$ undergoes changes of
10 orders of magnitude in a rather small region of $q$,
it is very difficult to handle it properly in further 
interpolations and integrations.
A way out is to 
perform interpolations 
for the ratio 
$I_{n,R}^{ratio}(q)= I_{n,R}(q)/I_{1,R}(q)$ with
analytically known $I_{1,R}=\frac{q}{q^2+R^{-2}}$.   
Variation of that ratio $I_{n,R}^{ratio}(q)$ is much more restricted, as shown in Fig.\ref{fig1},
and
we use its polynomial representation to get $I_{n,R}^{ratio}(q)$ at any value of $q$.
For each value of $n$ and $R$ we
divide the interpolation region into some optimal number of intervals, optimizing their
length as well as degree of the polynomial. Typically we have 6 intervals while the
degree of the polynomial varies between 6 to 12. This allows us to
describe the oscillating function $I_{n,R}^{ratio}(q)$ with a sufficiently high precision,
what is exemplified in Fig.~\ref{fig1} for n=3 and R=120 fm.
In addition to the solid line for $I_{3,120}^{ratio}(q)$, the x-es in Fig.\ref{fig1} show 
the $I_{3,120}^{ratio}(q)$ values
obtained from the polynomial formula.
The agreement is perfect.
The examples of the polynomial representation of $I_{n,R}^{ratio}(q)$ for two sets of 
screening parameters are given in the Appendix.
The important and useful feature of $I_{n,R}^{ratio}(q)$, and thus also of the fit parameters,
is its independence from the scattering energy. Due to that the interpolation can be done once for 
a given set of $n$ and $R$ values.

Once the $I_{n,R}^{ratio}(q)$ is calculated also $I_{n,R}(q)$ is known and the final integration 
over $\phi$ leads to $v_c^R(p',p,x',x)$. In our calculations we use typically 
a set of 500 gaussian points for the $\phi$-integration for all types of screening.

The properties of the resulting screened t-matrices  
are presented in the next Section.
 
\section{The screened Coulomb t-matrix properties}
\label{sec3}

In this section the three-dimensional screened Coulomb t-matrix $t(p',p,x=\cos(\theta))$ will be shown 
as a function of momenta $p,p'$ at given scattering angle $\theta$.
We choose as examples of backward, intermediate and forward angles the following values of
$\theta$: $134^{\circ}, 45^{\circ}, 10^{\circ}$ and $5^{\circ}$. 

\begin{figure}
\includegraphics[scale=0.4,clip=true,angle=-90]{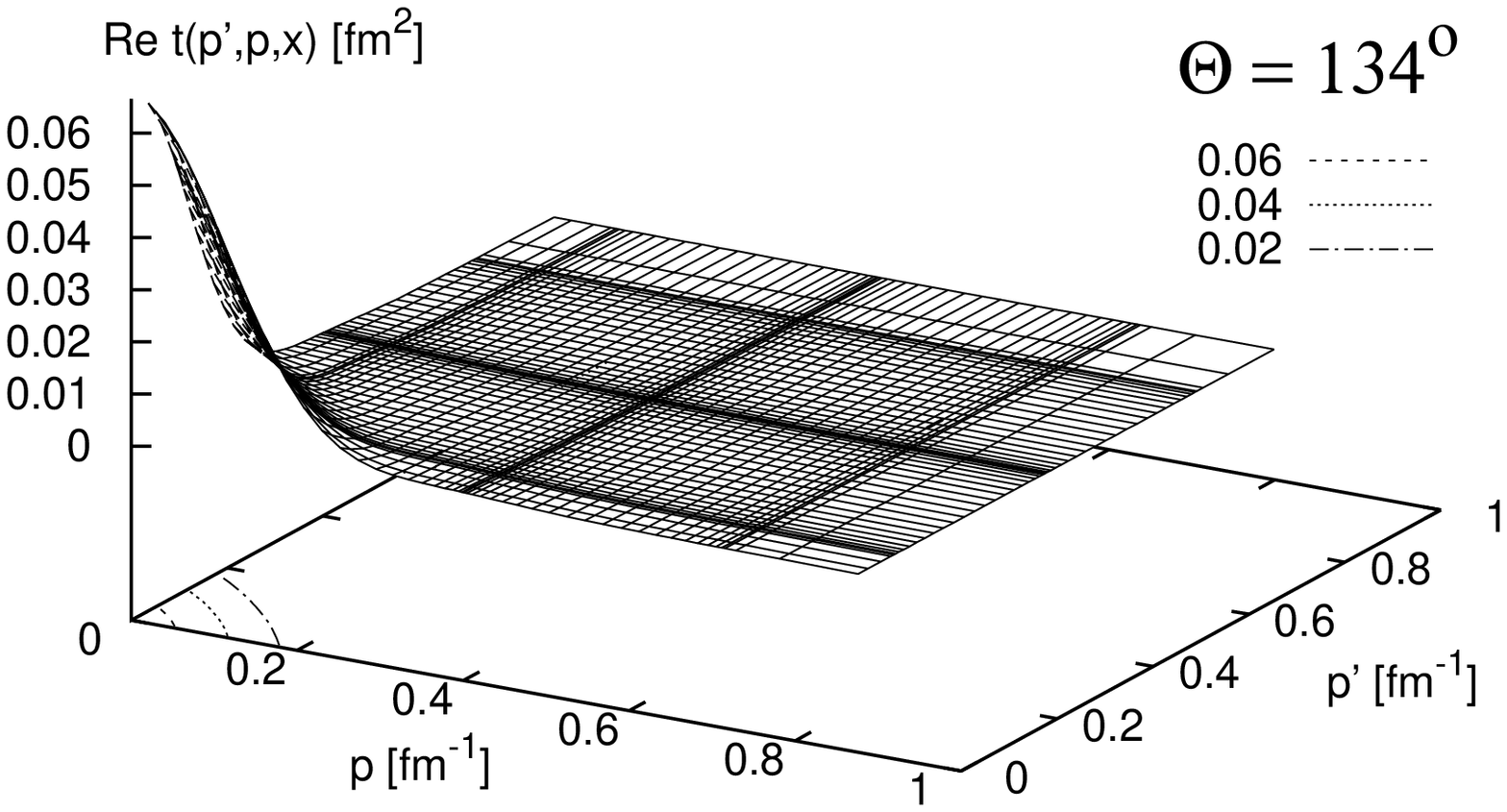}
\includegraphics[scale=0.4,clip=true,angle=-90]{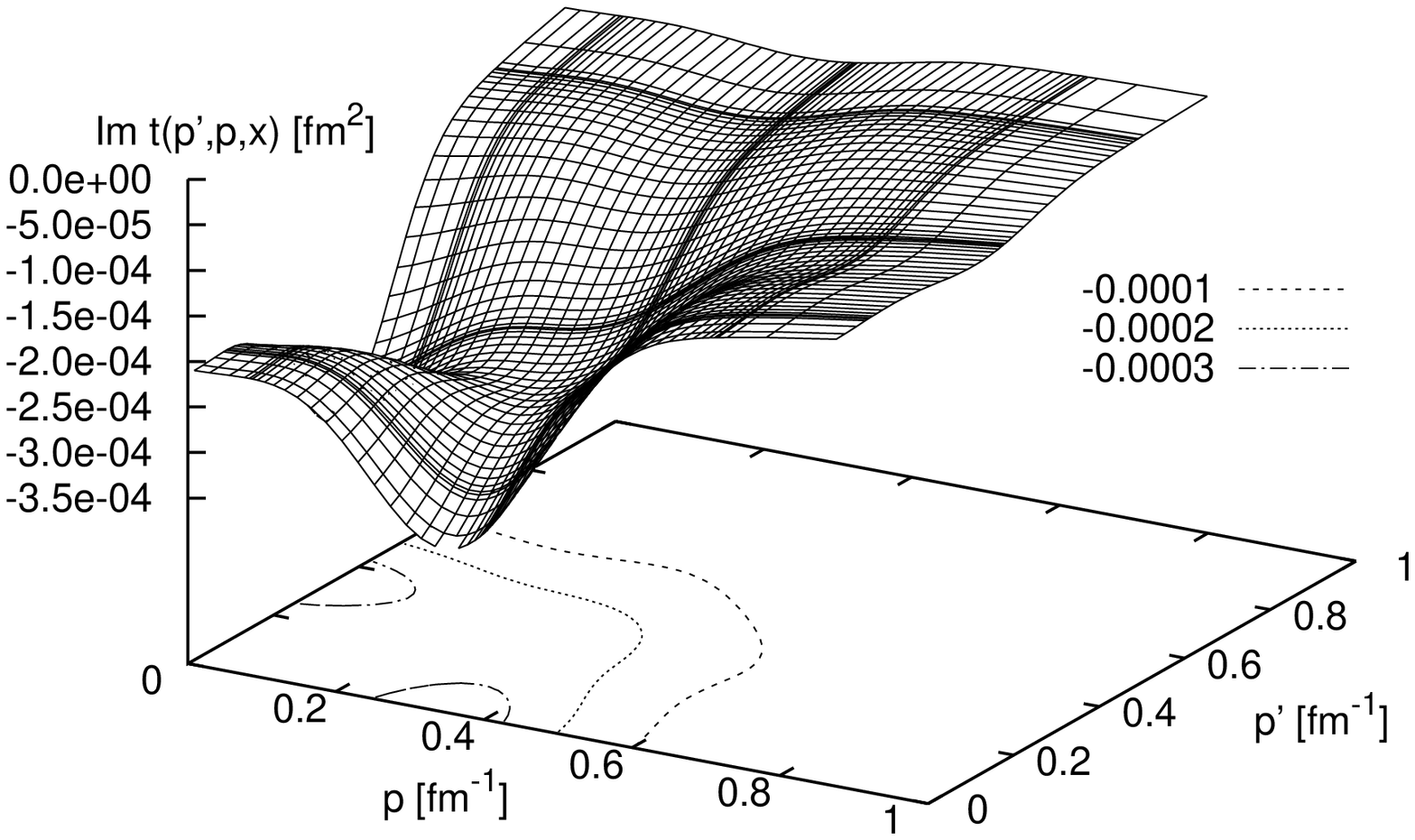}
\includegraphics[scale=0.4,clip=true,angle=-90]{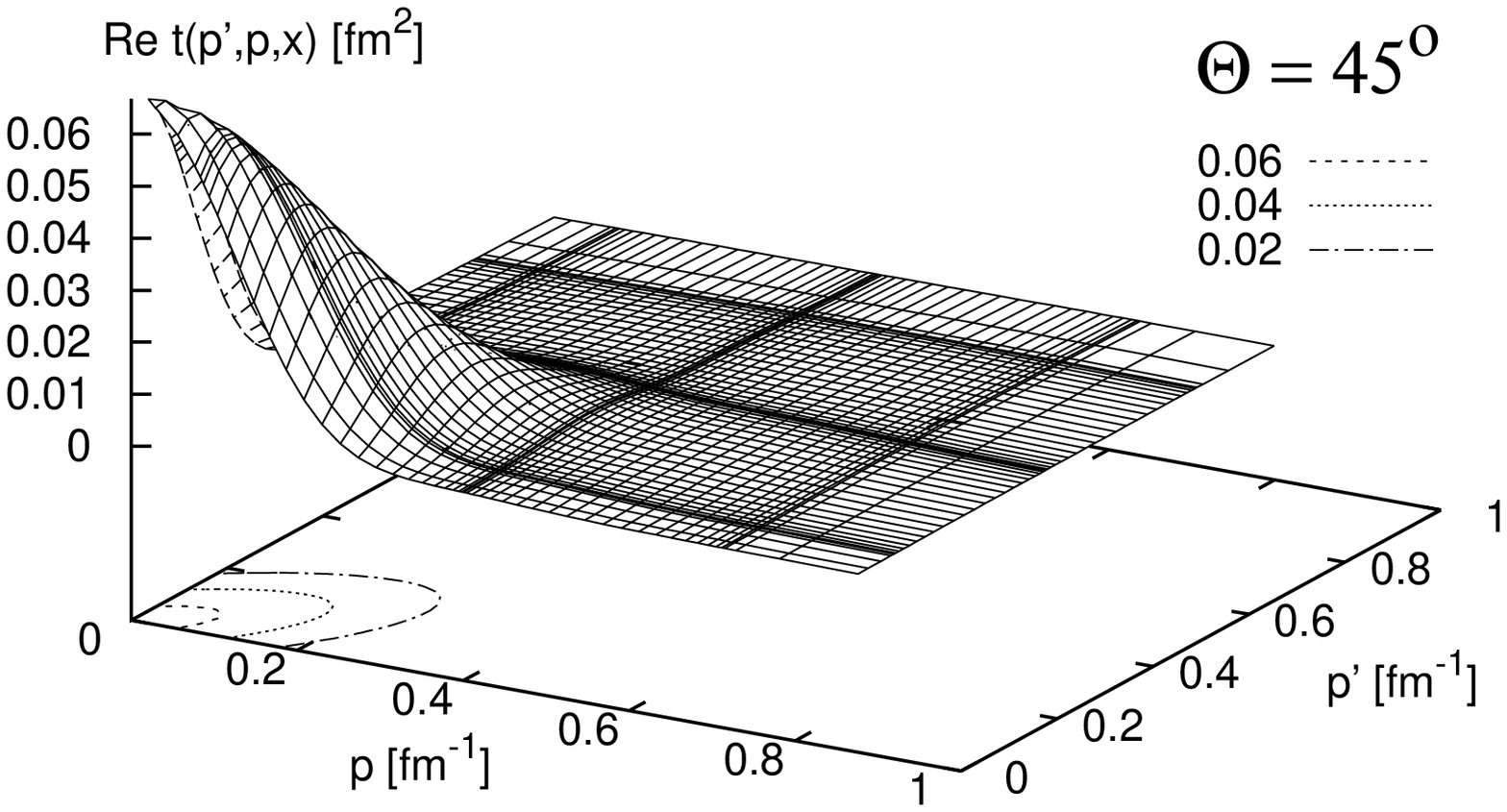}
\includegraphics[scale=0.4,clip=true,angle=-90]{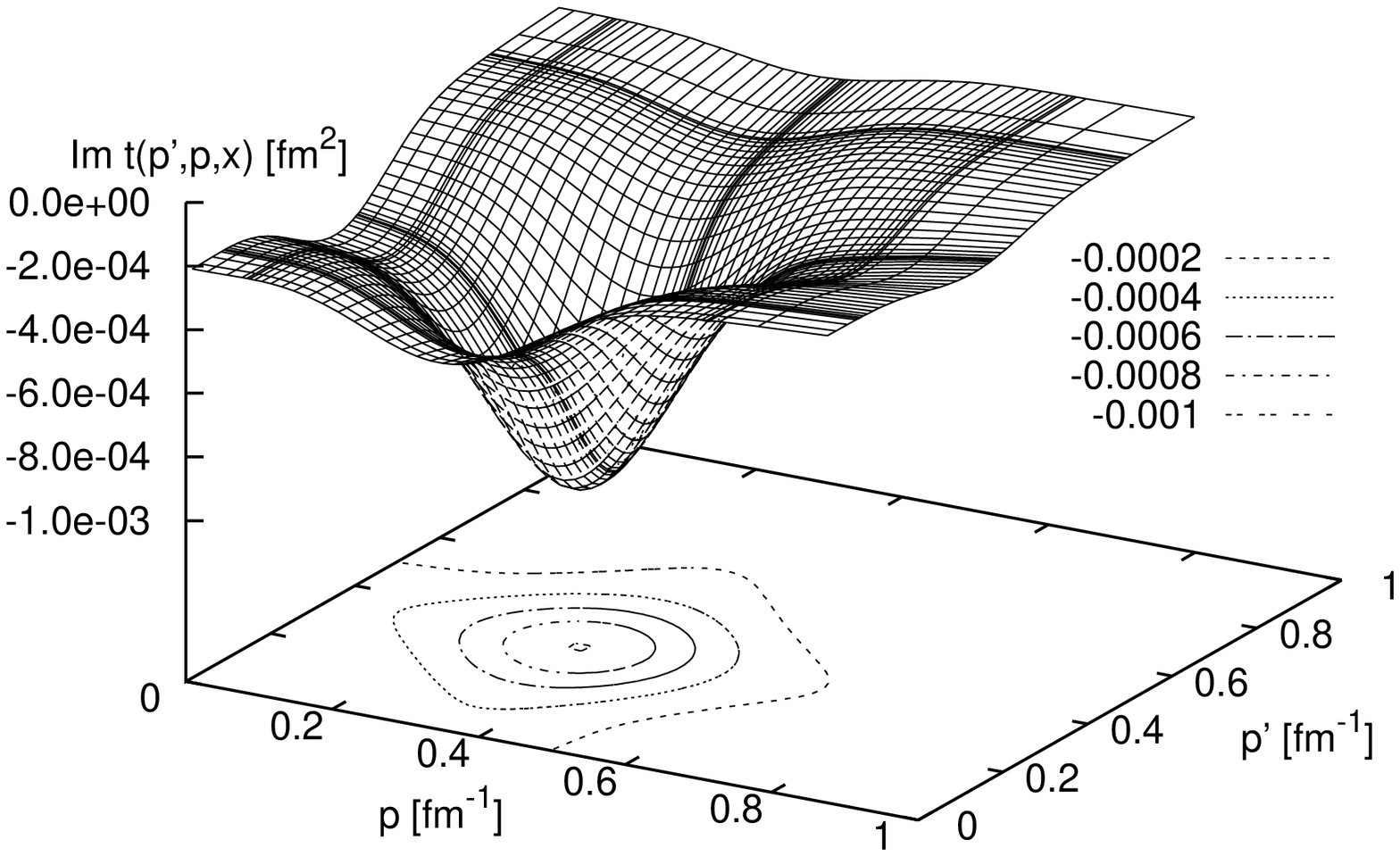}
\includegraphics[scale=0.4,clip=true,angle=-90]{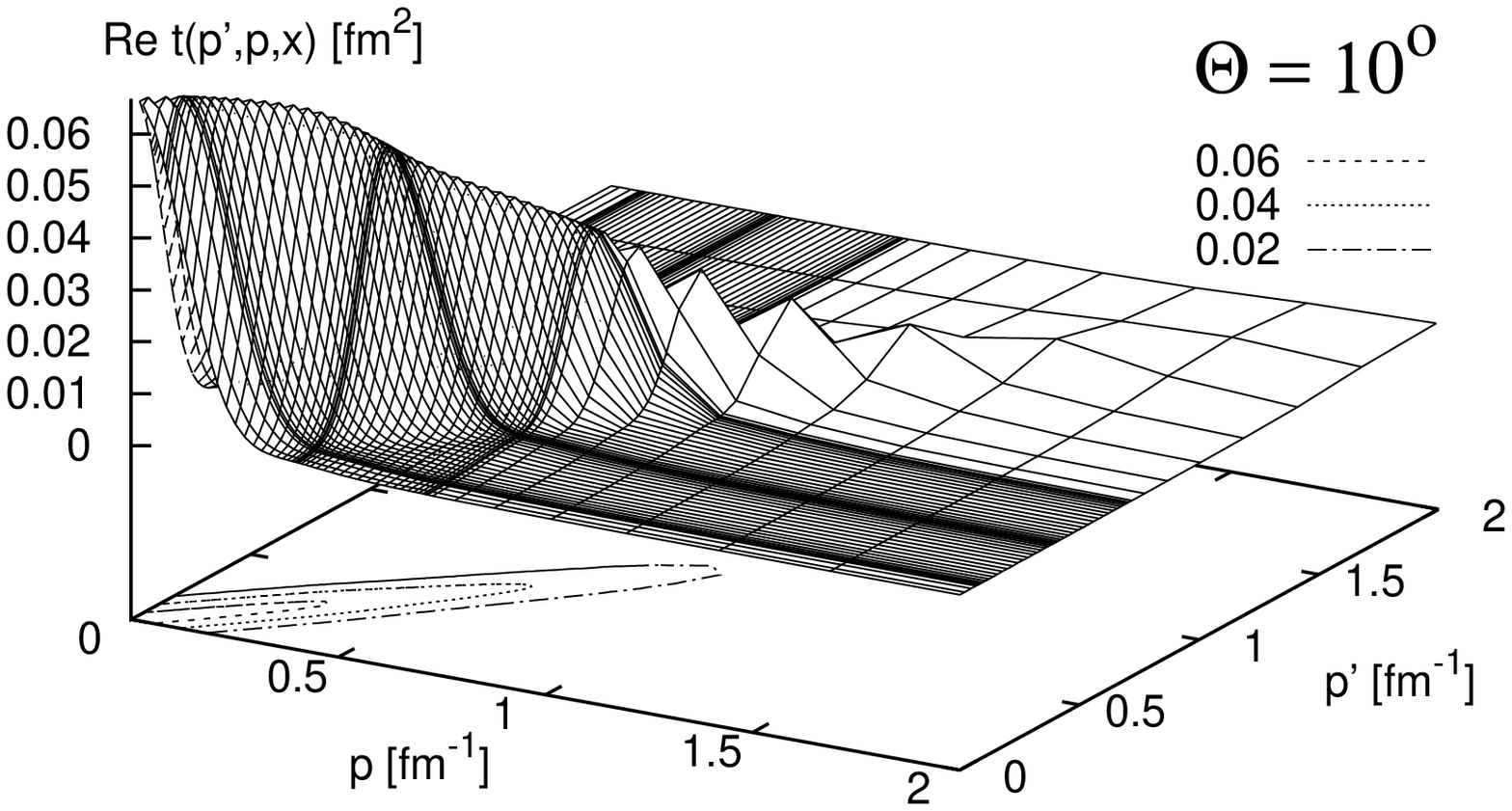}
\includegraphics[scale=0.4,clip=true,angle=-90]{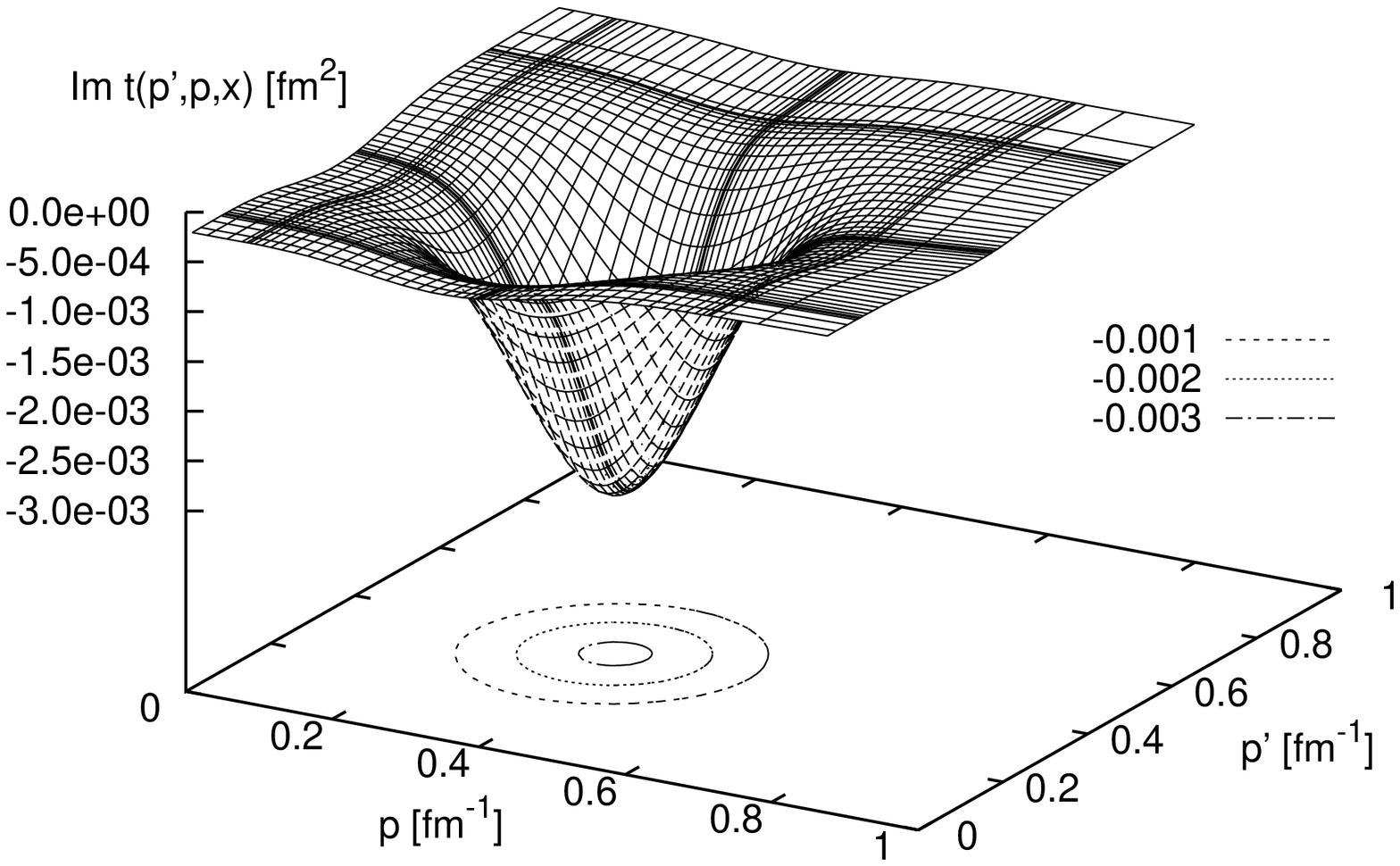}
\includegraphics[scale=0.4,clip=true,angle=-90]{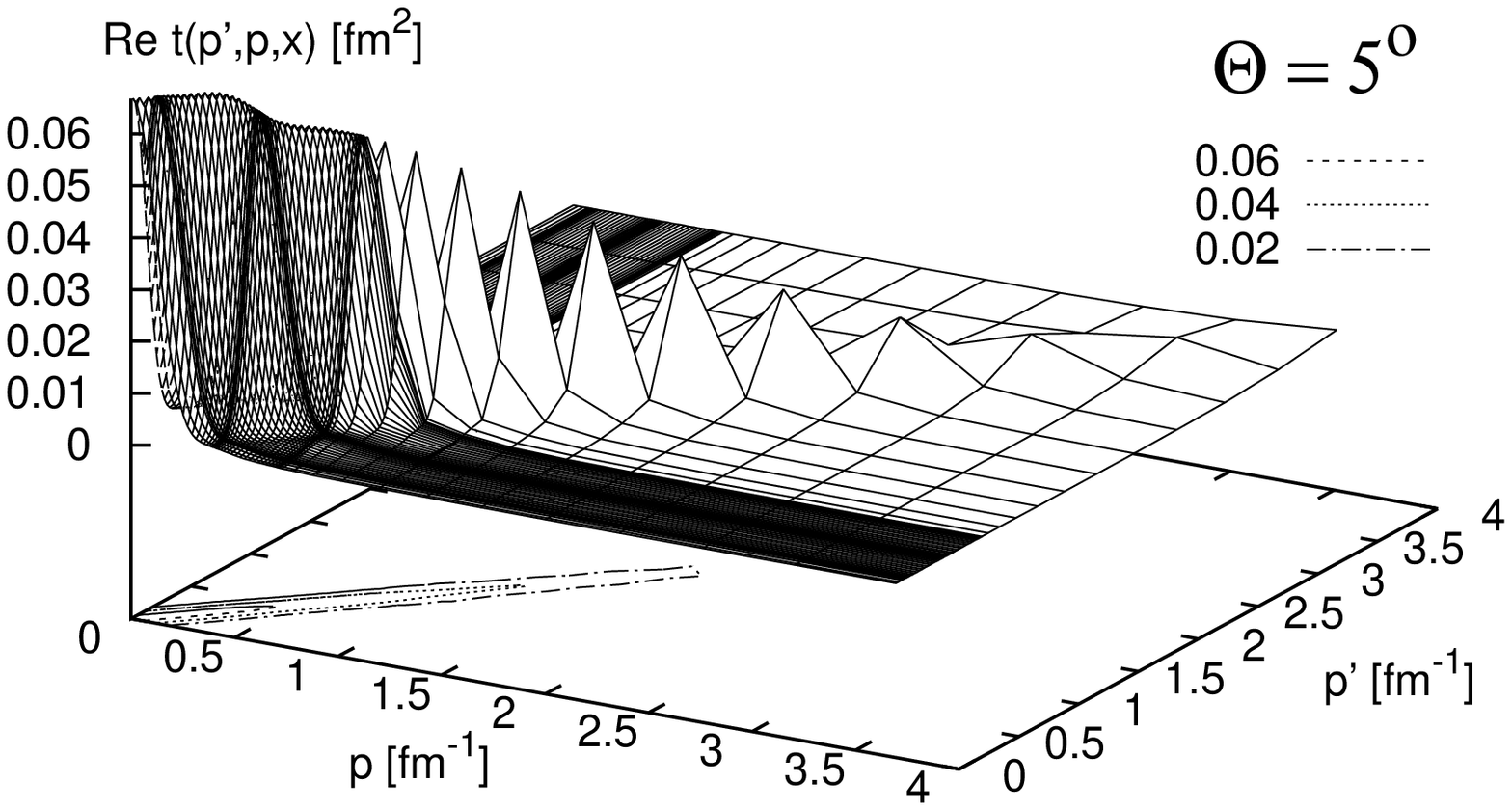}
\includegraphics[scale=0.4,clip=true,angle=-90]{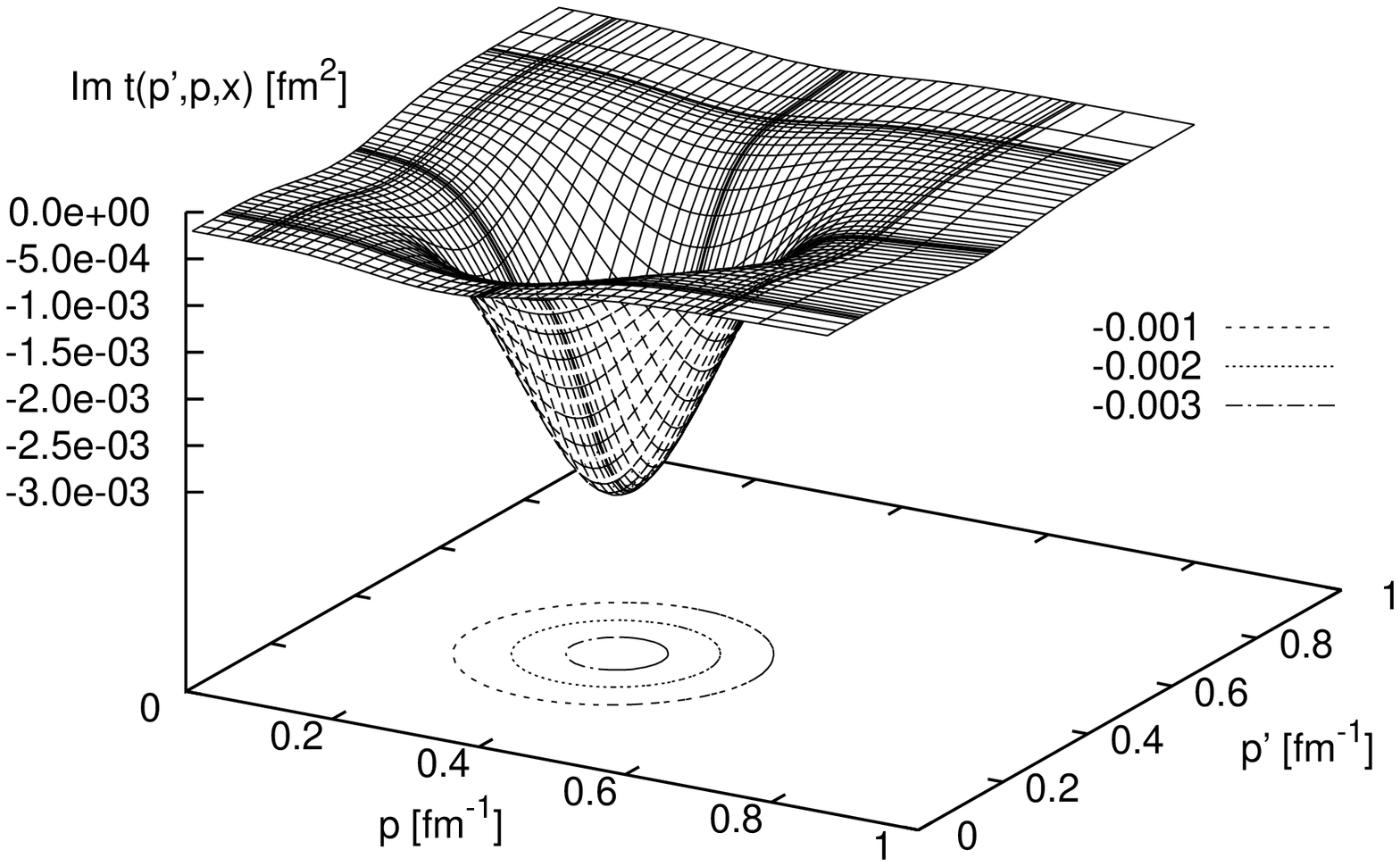}
\caption{The real (left) and imaginary (right) parts of the screened Coulomb 
t-matrix $t(p,p',x=\cos(\theta))$ at E=13 MeV for different  
scattering angles $\theta=$ 
$134^{\circ}$(1-st row), $\theta=45^{\circ}$(2-nd row), 
$\theta=10^{\circ}$(3-rd row) and $\theta=5^{\circ}$(4-th row).
The exponential screening with R=20 fm and n=4 has been applied.}
\label{fig5}
\end{figure}

In Fig.~\ref{fig5} the real and imaginary parts of the exponentially screened Coulomb t-matrix with 
n=4 and R=20 fm at
E=13 MeV are shown.
The real part of $t$ has a high and steep maximum at small momenta at $\theta=134^{\circ}$,
which evolves to a ridge lying along diagonal $p'=p$ for smaller angles. 
The spiky structure seen for the smallest angle comes only from the graphical 
representation on the finite grid of $p$ and $p'$-points. 
The increasing range of the ridge shows that action of the screened Coulomb force
becomes more and more important at bigger momenta 
when moving to smaller scattering angles.
The imaginary part 
has a minimum at the on-shell point $p=p'=p_0$~$(\approx 0.396$ fm$^{-1}$
for E=13 MeV$)$. Its absolute value is about one order of magnitude smaller than 
the maximum of the real part. The minimum of the imaginary part becomes deeper and narrower with decreasing angle.

\begin{figure}
\includegraphics[scale=0.4,clip=true,angle=-90]{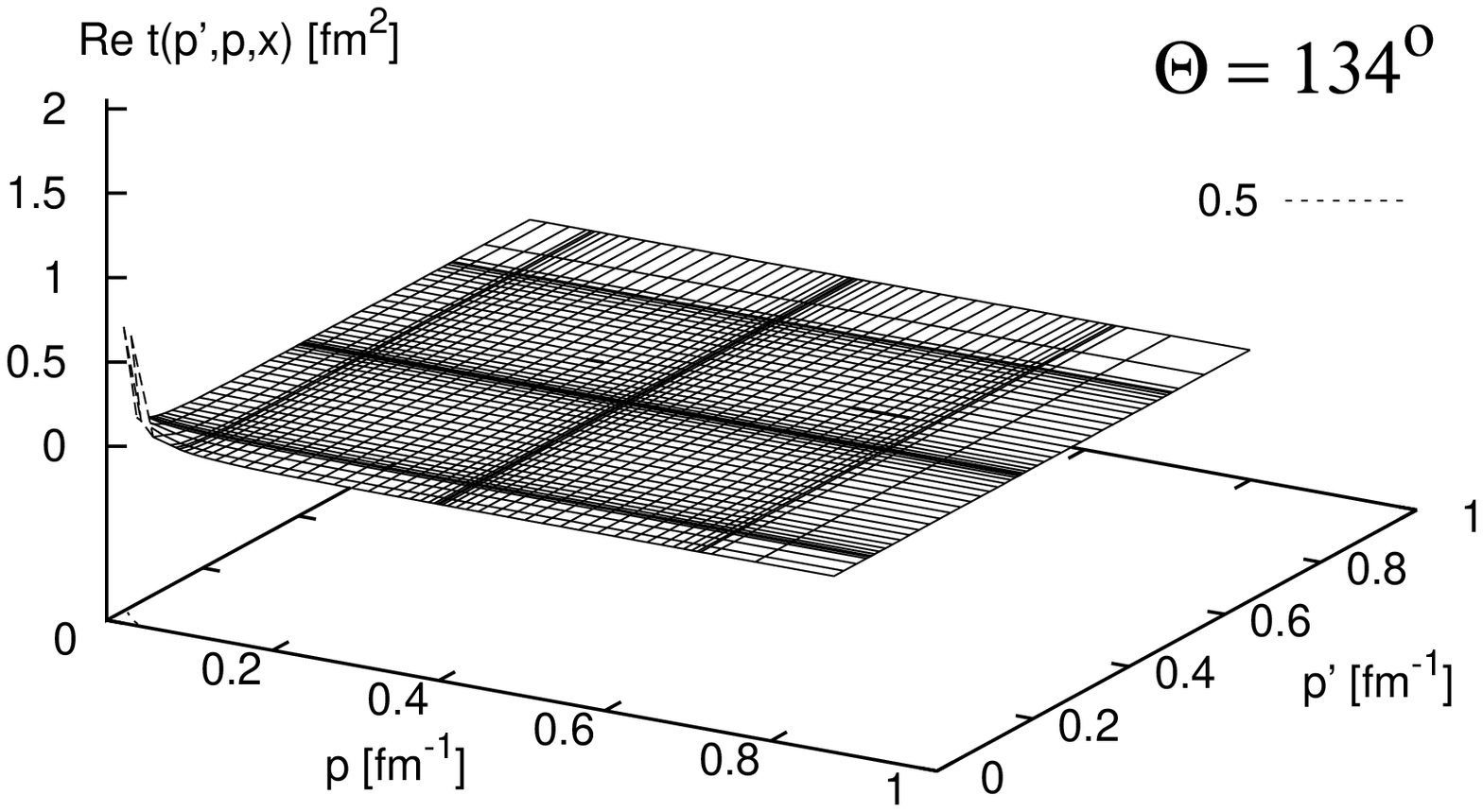}
\includegraphics[scale=0.4,clip=true,angle=-90]{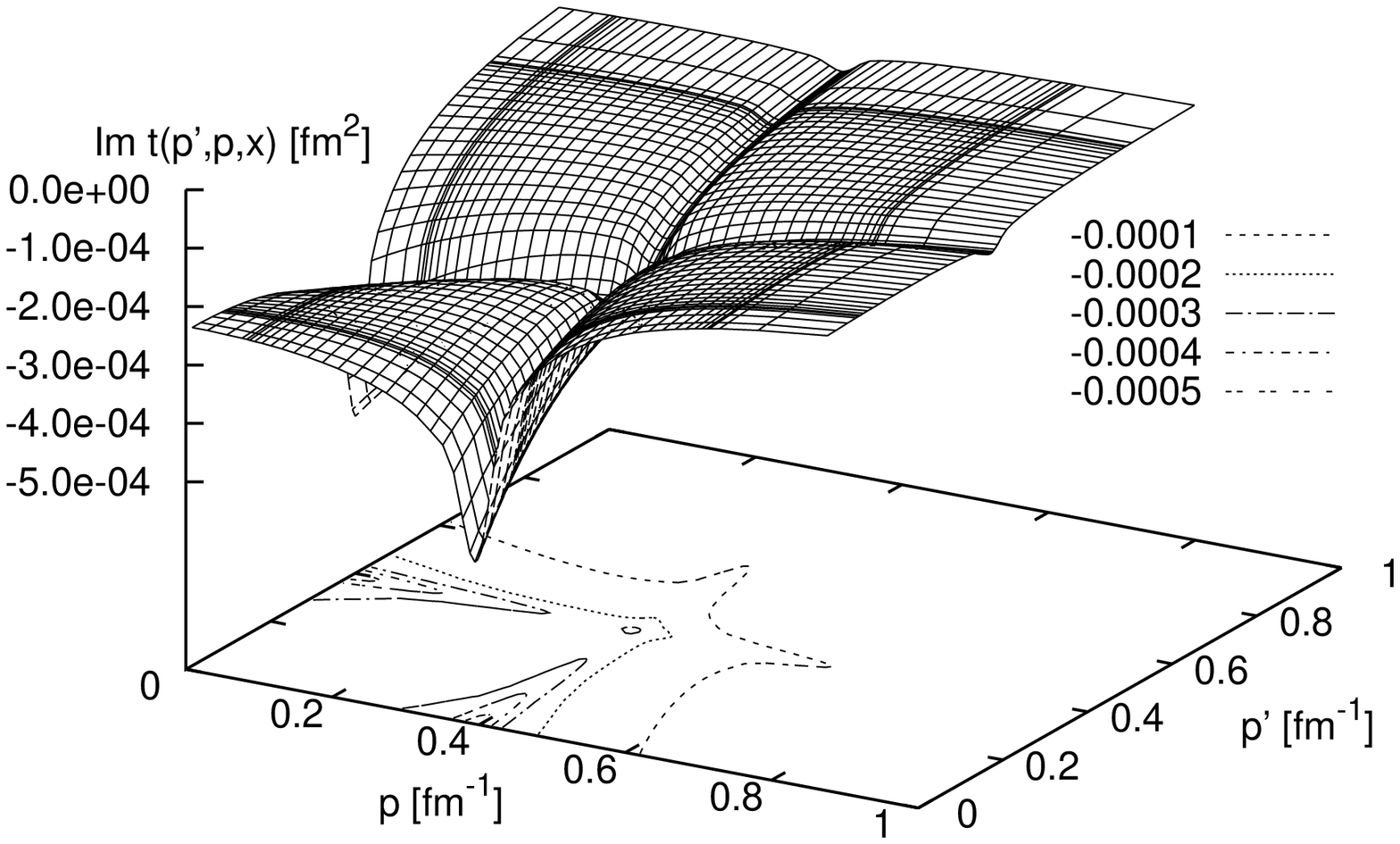}
\includegraphics[scale=0.4,clip=true,angle=-90]{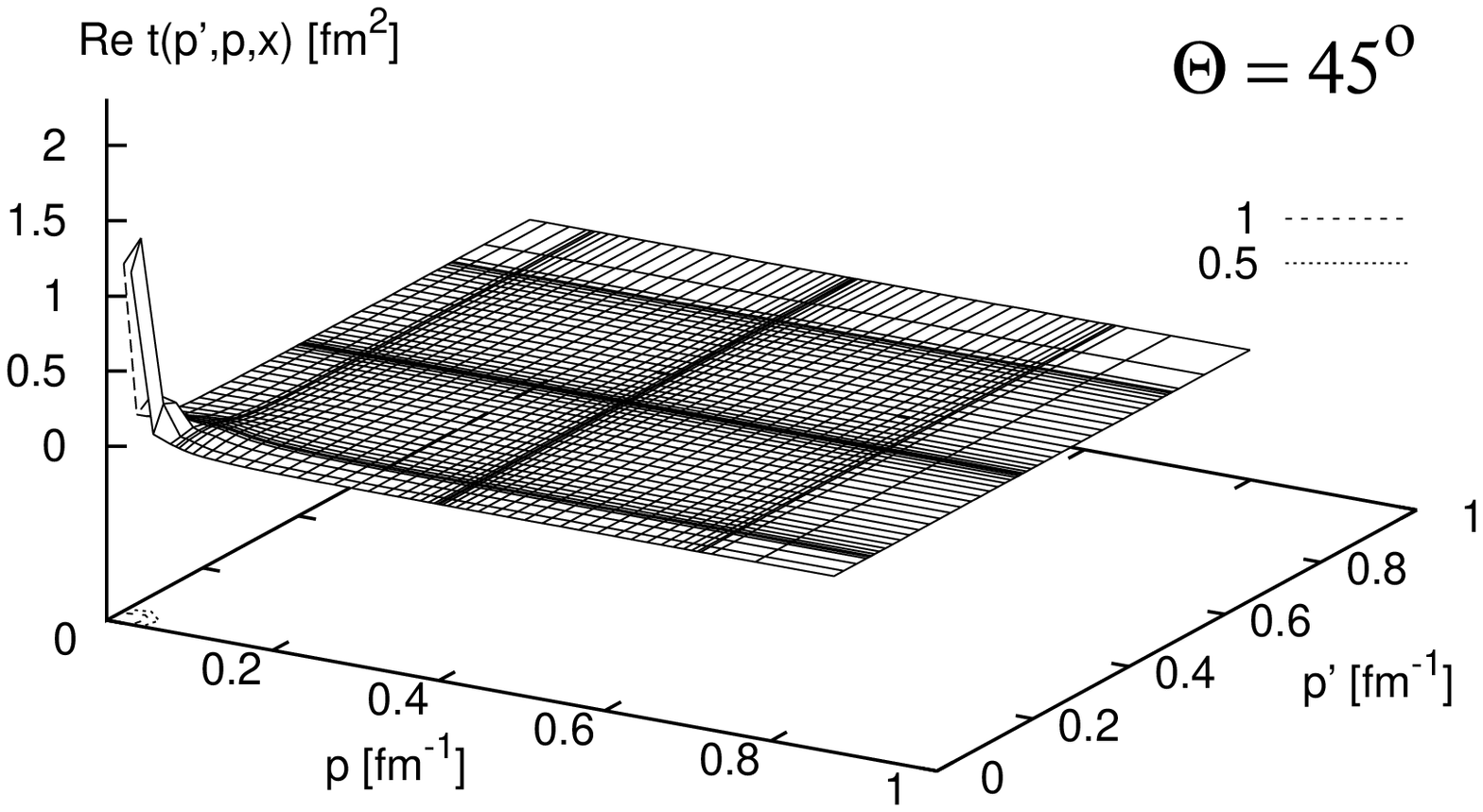}
\includegraphics[scale=0.4,clip=true,angle=-90]{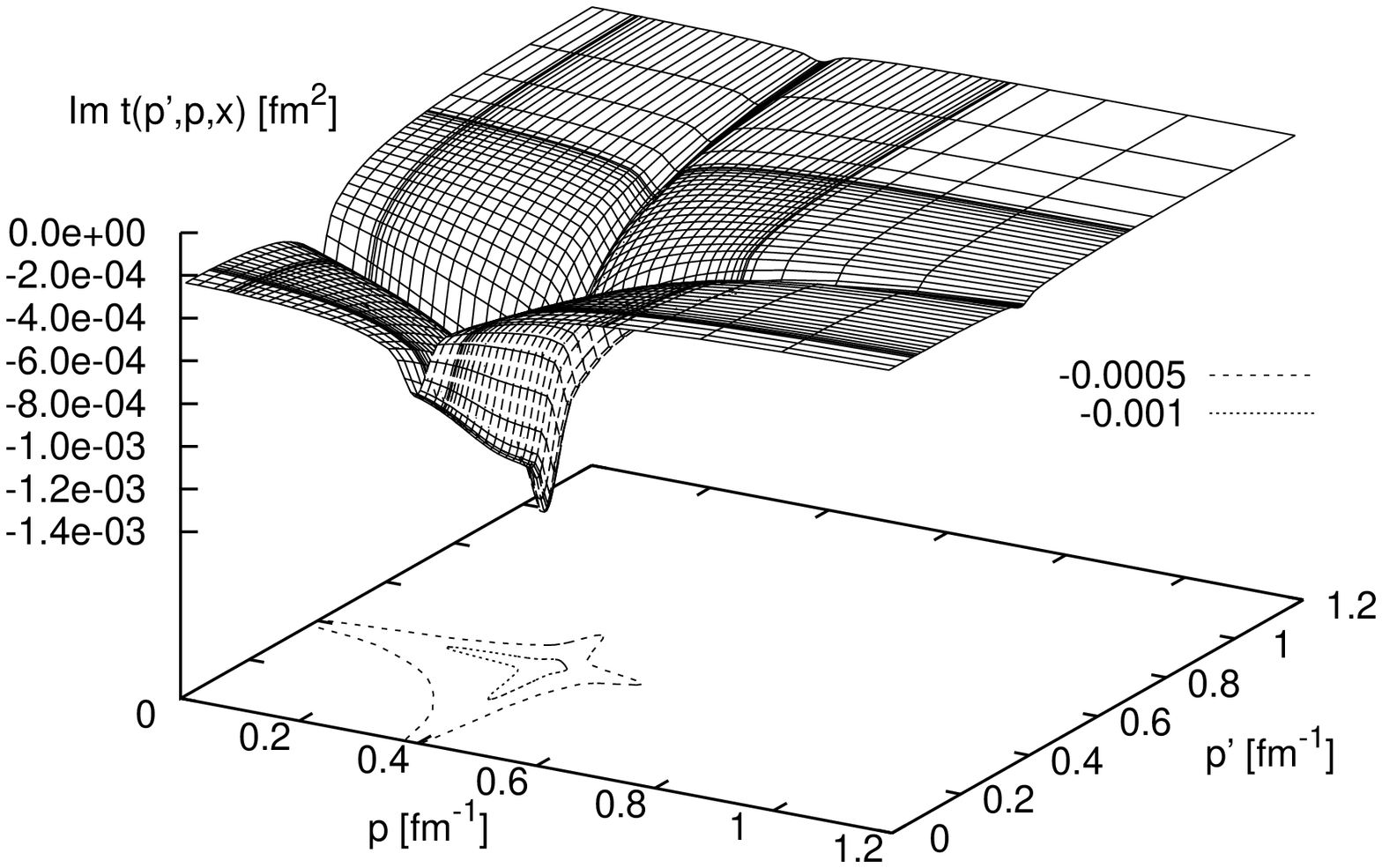}
\includegraphics[scale=0.4,clip=true,angle=-90]{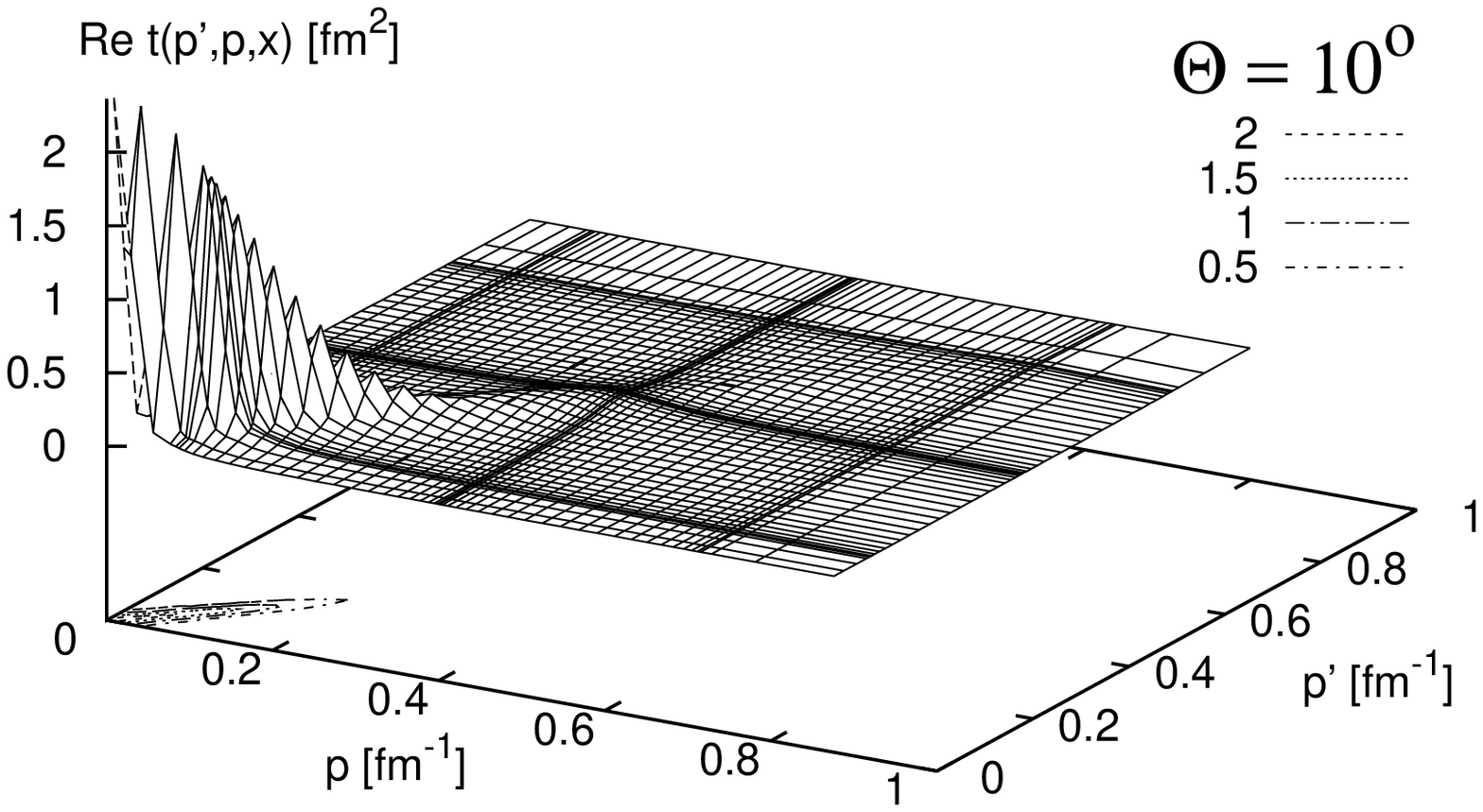}
\includegraphics[scale=0.4,clip=true,angle=-90]{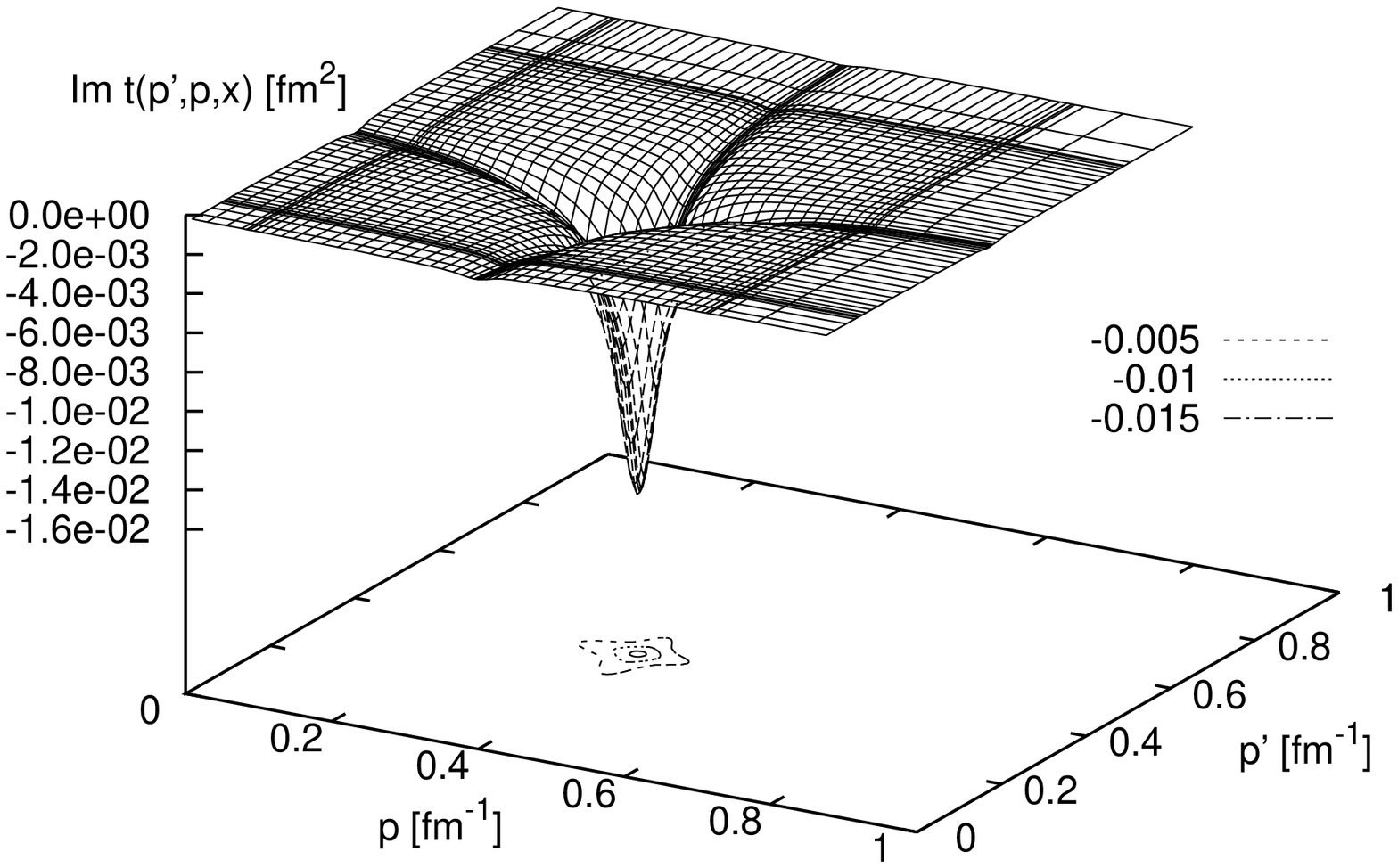}
\includegraphics[scale=0.4,clip=true,angle=-90]{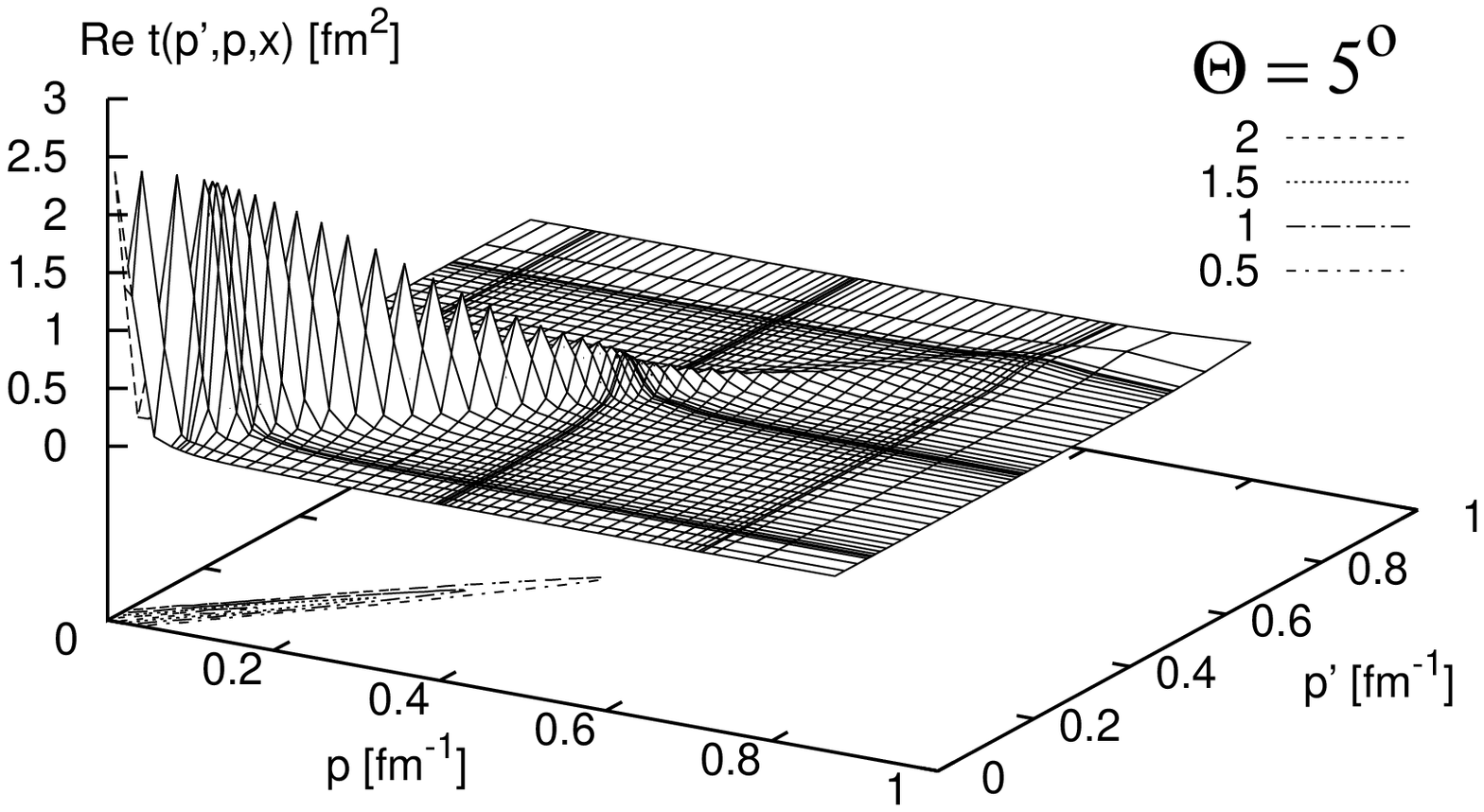}
\includegraphics[scale=0.4,clip=true,angle=-90]{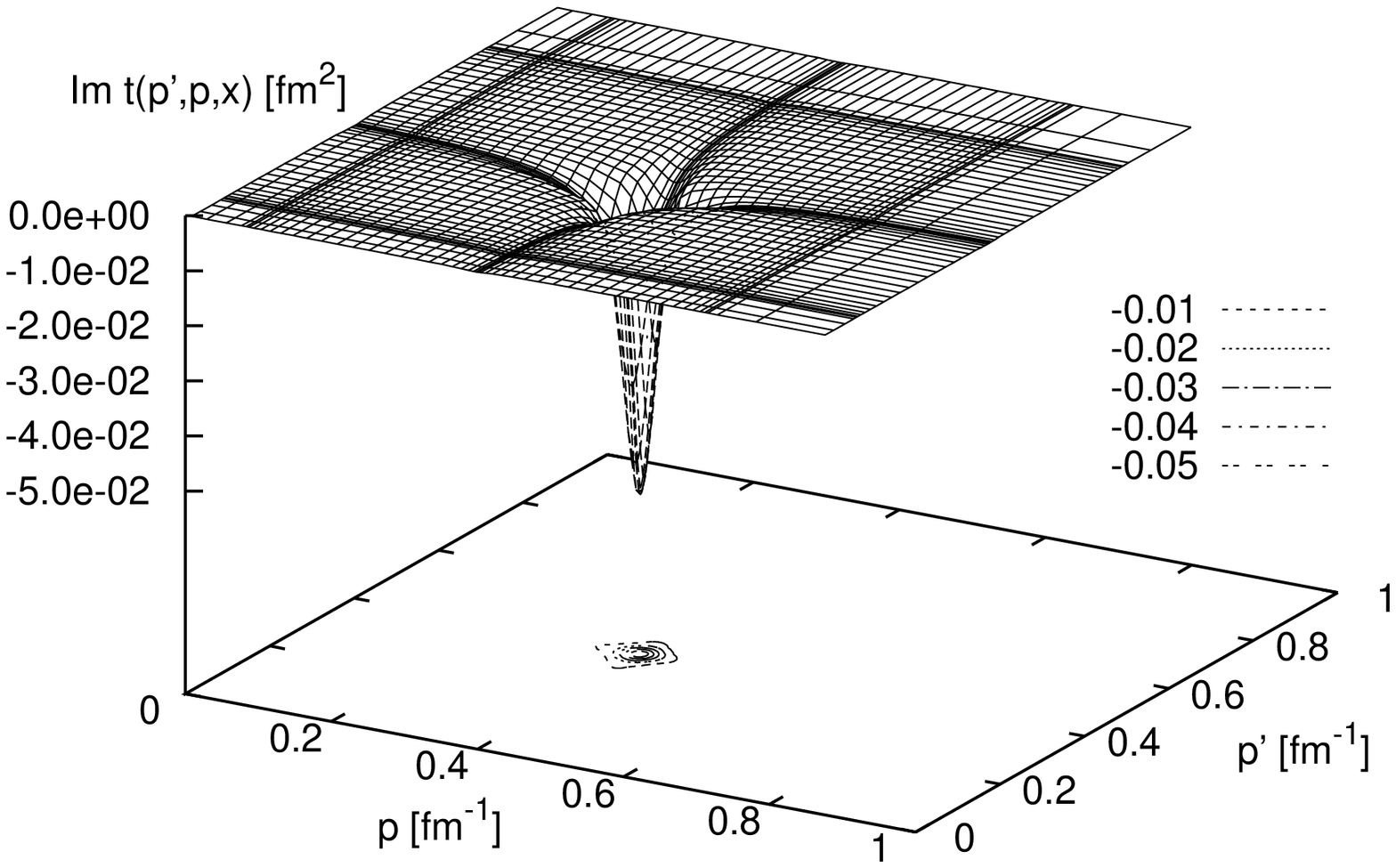}
\caption{The same as in Fig.\ref{fig5} but for R=120 fm.}
\label{fig6}
\end{figure}

The similar behaviour exists also for other values of the
screening radius $R$. This is exemplified in Fig.~\ref{fig6} for $R=120$ fm.
Taking higher values of $R$ leads 
to a more restricted range of momenta, at which the real part of $t$ takes large values.
However, the maximum of $Re(t)$ is much higher than for $R=20$ fm.
Also the range of momenta, where $Im(t)$ has the deep minimum is 
much smaller than for $R=20$ fm. 

\begin{figure}
\includegraphics[scale=0.4,clip=true,angle=-90]{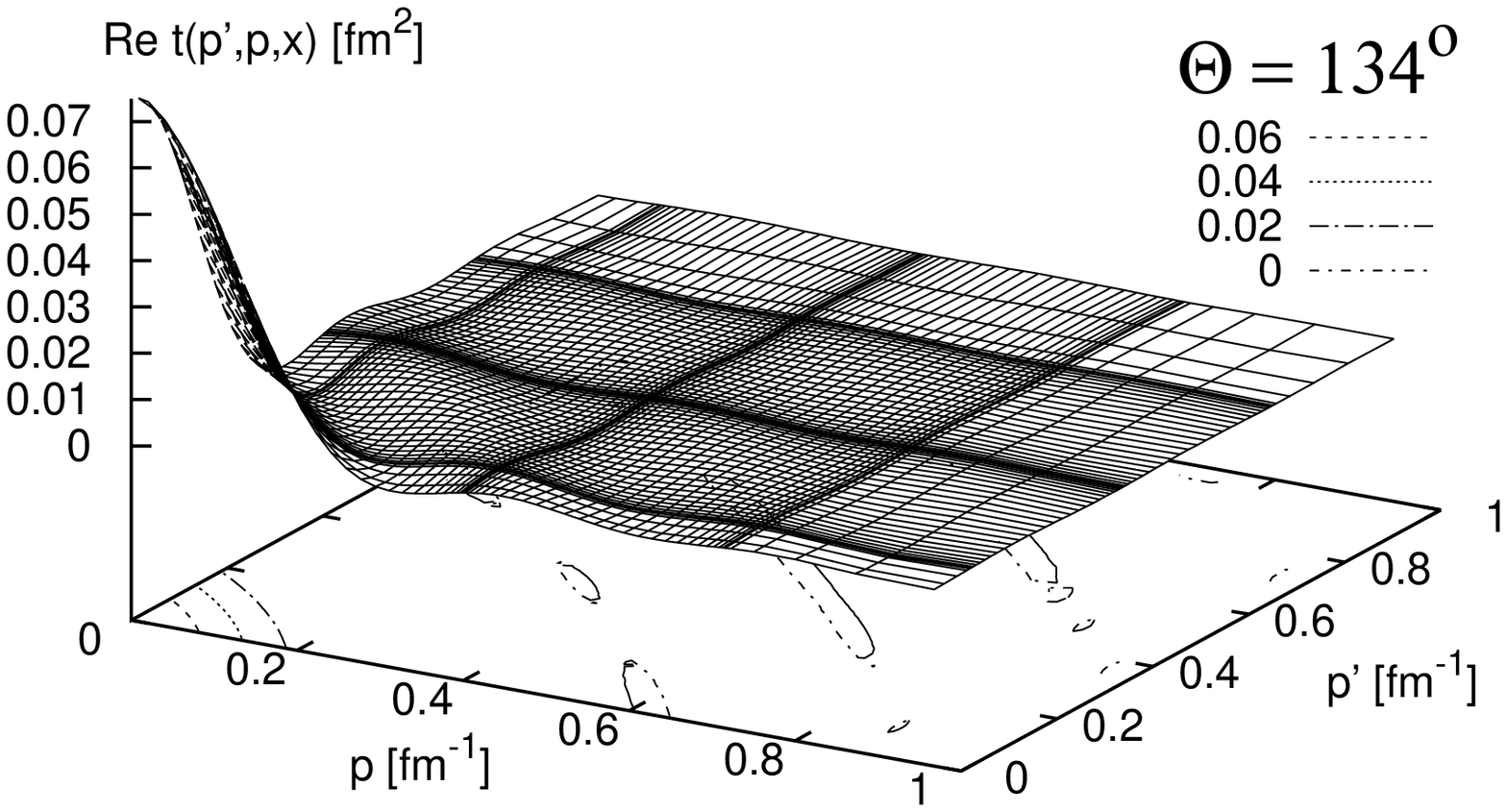}
\includegraphics[scale=0.4,clip=true,angle=-90]{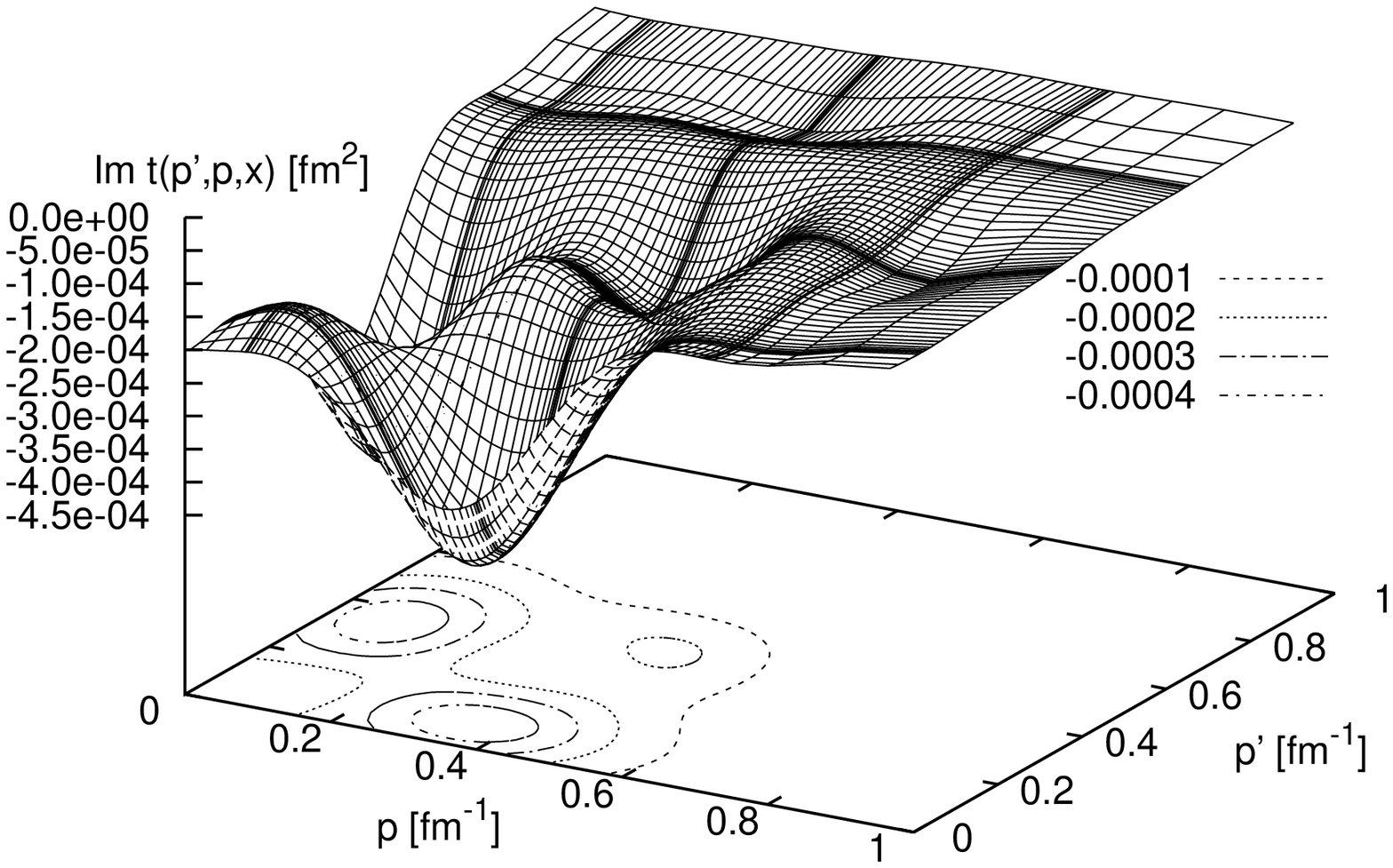}
\includegraphics[scale=0.4,clip=true,angle=-90]{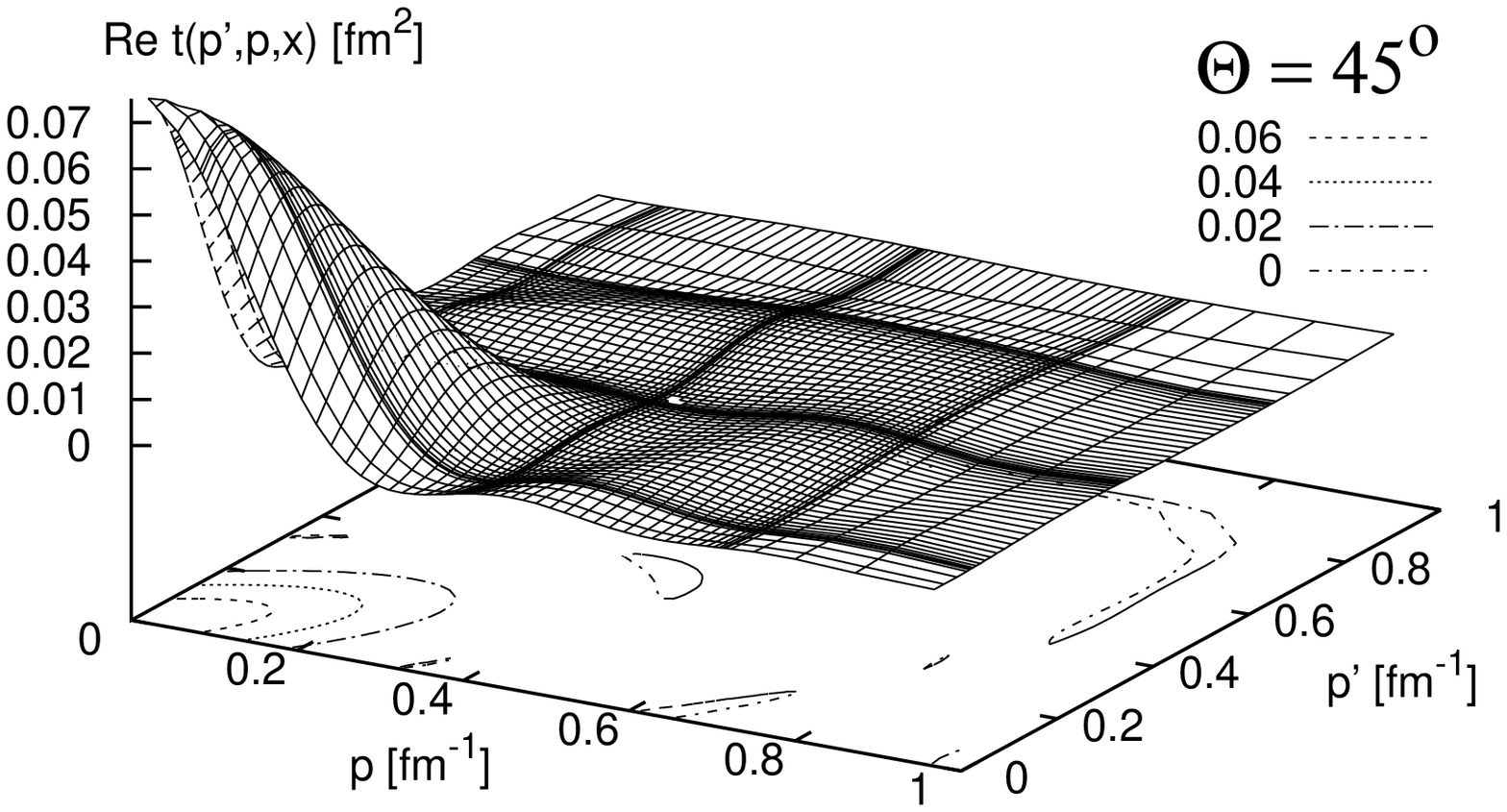}
\includegraphics[scale=0.4,clip=true,angle=-90]{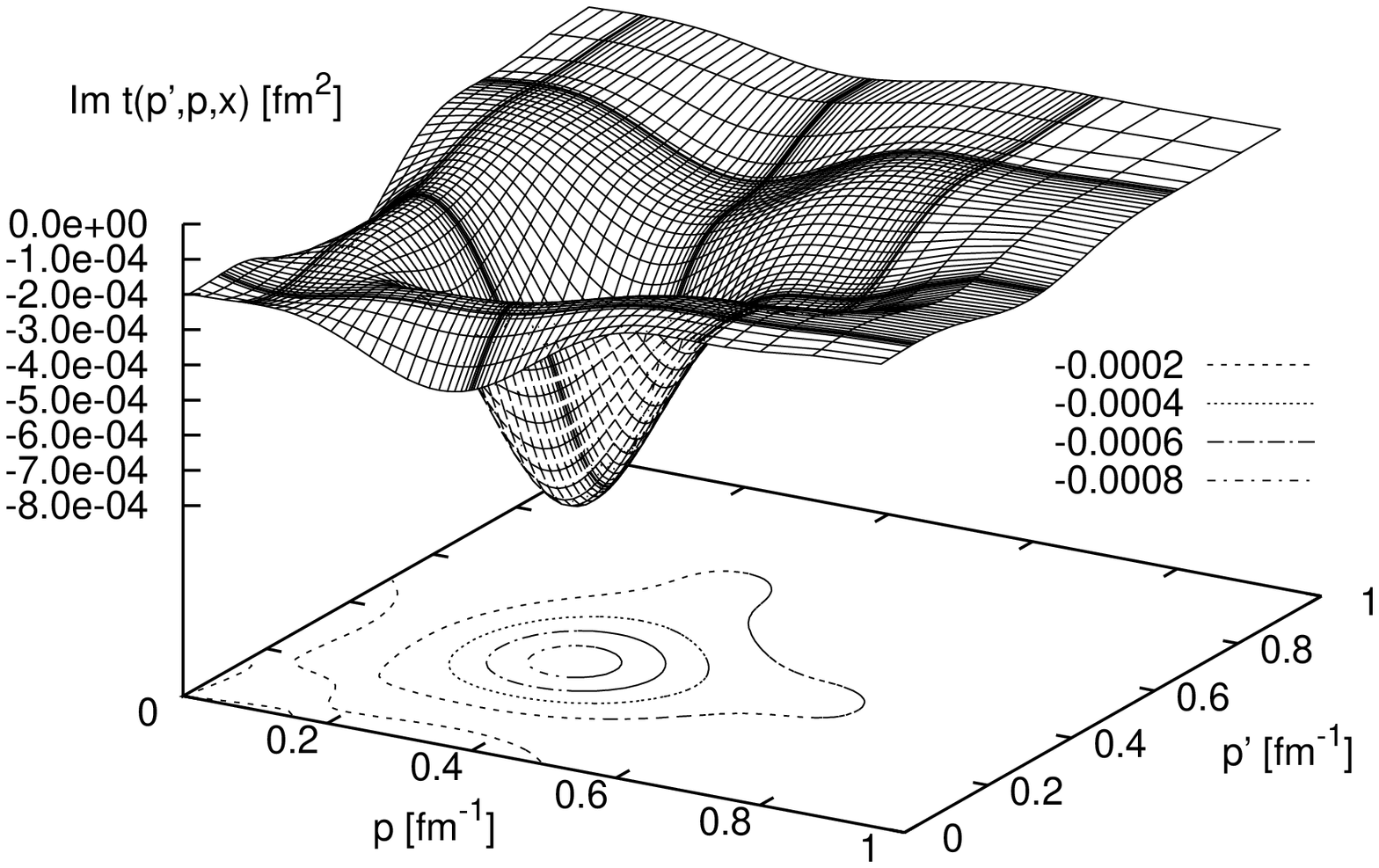}
\includegraphics[scale=0.4,clip=true,angle=-90]{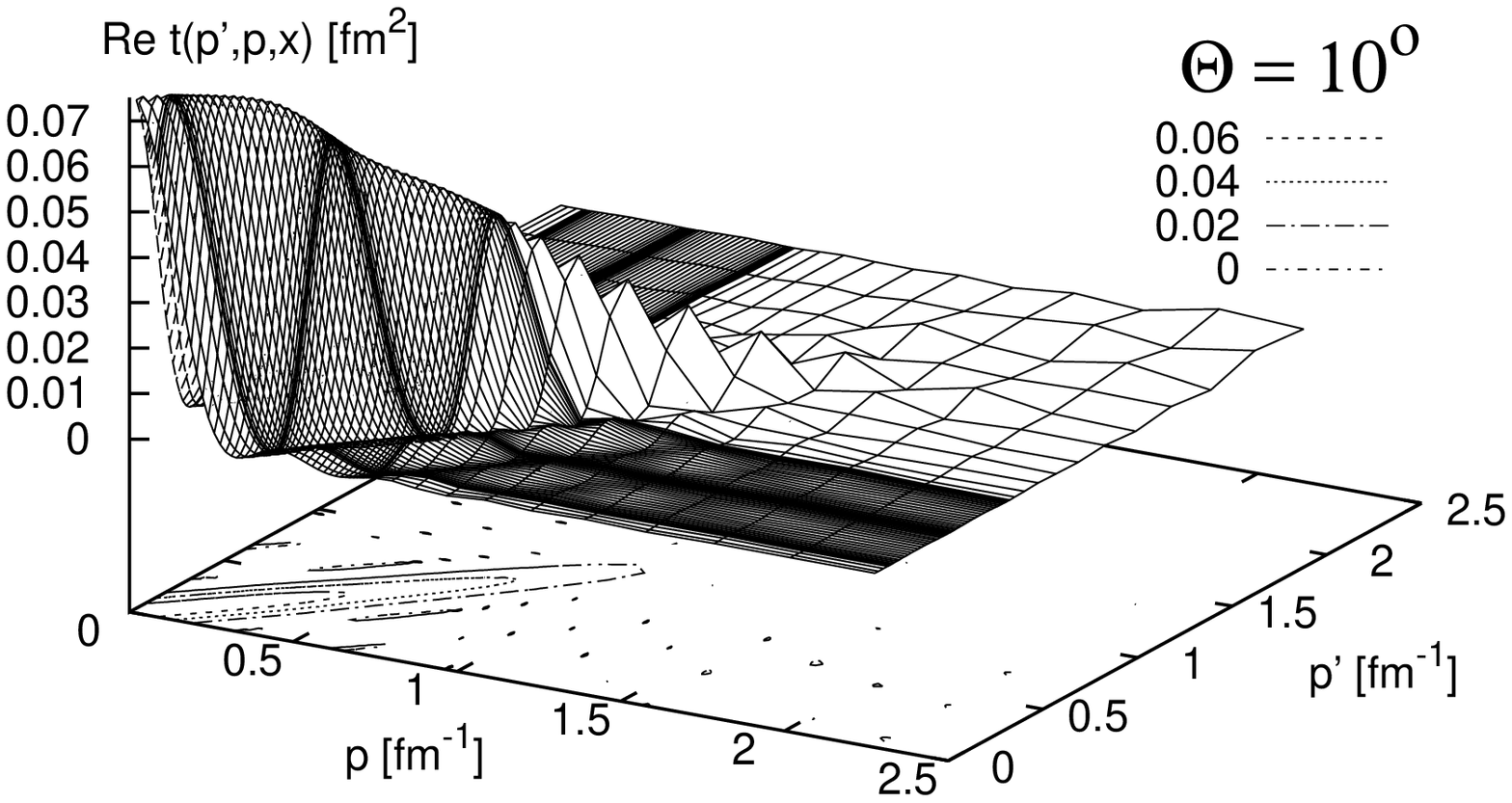}
\includegraphics[scale=0.4,clip=true,angle=-90]{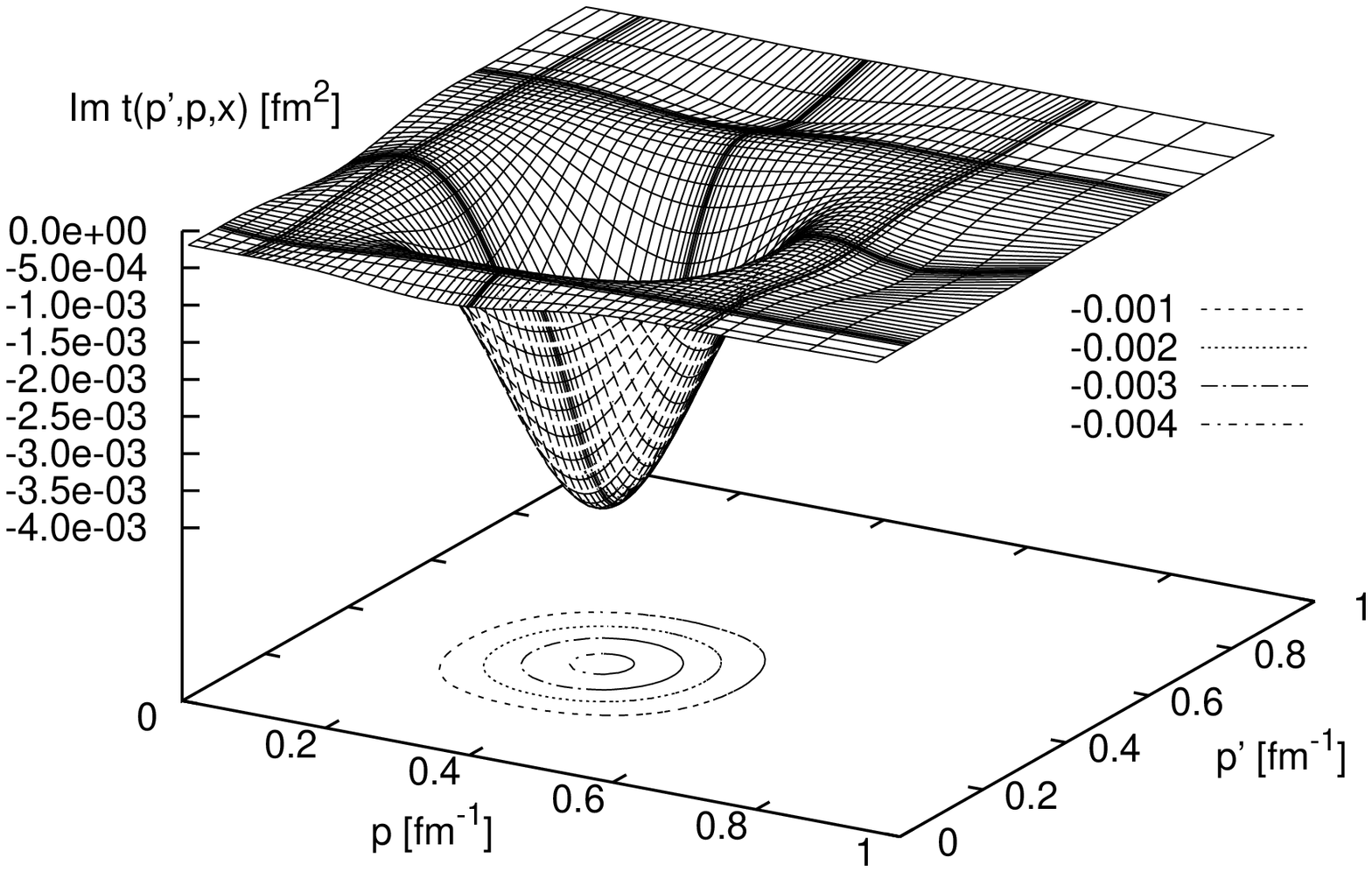}
\includegraphics[scale=0.4,clip=true,angle=-90]{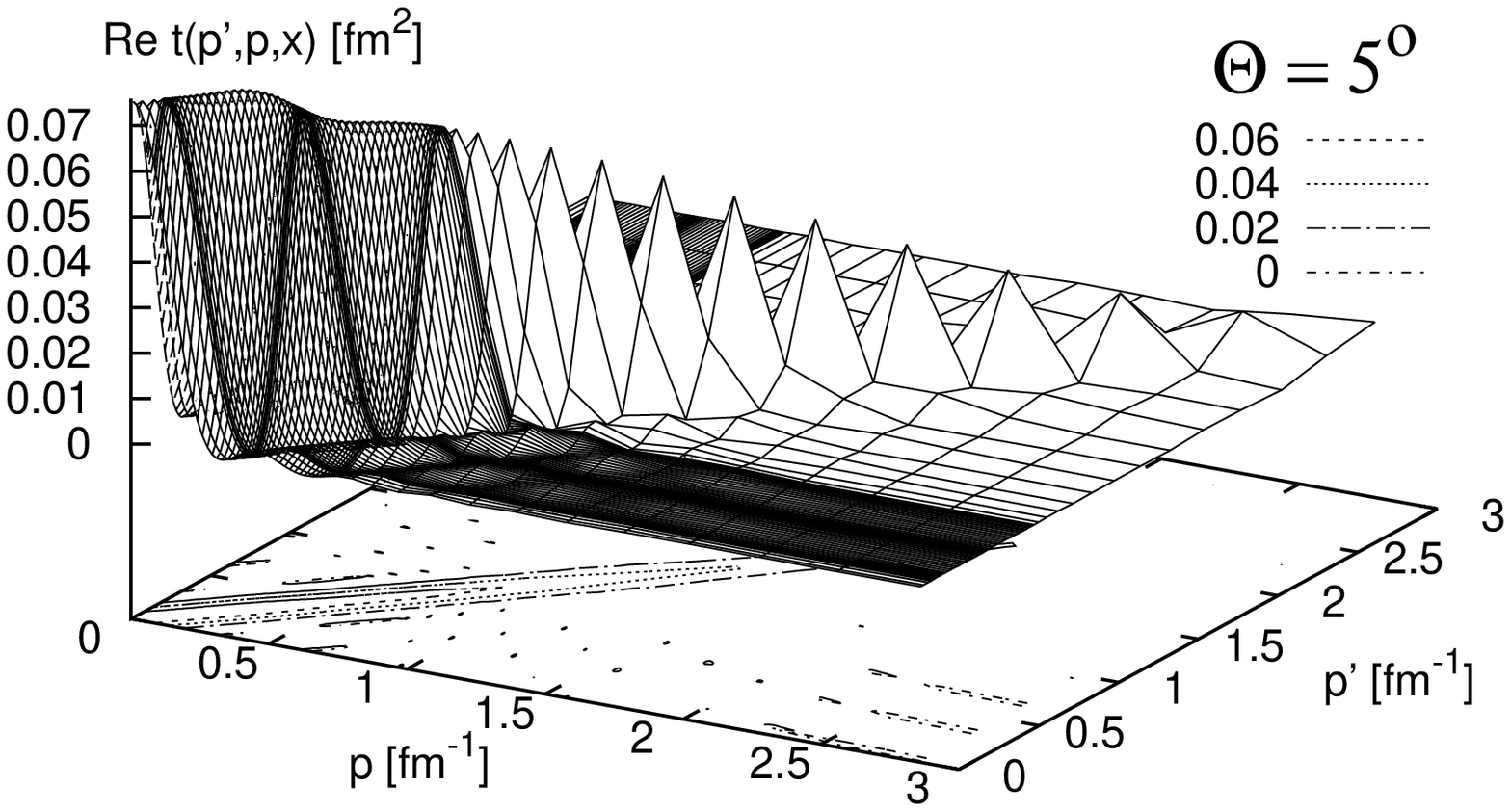}
\includegraphics[scale=0.4,clip=true,angle=-90]{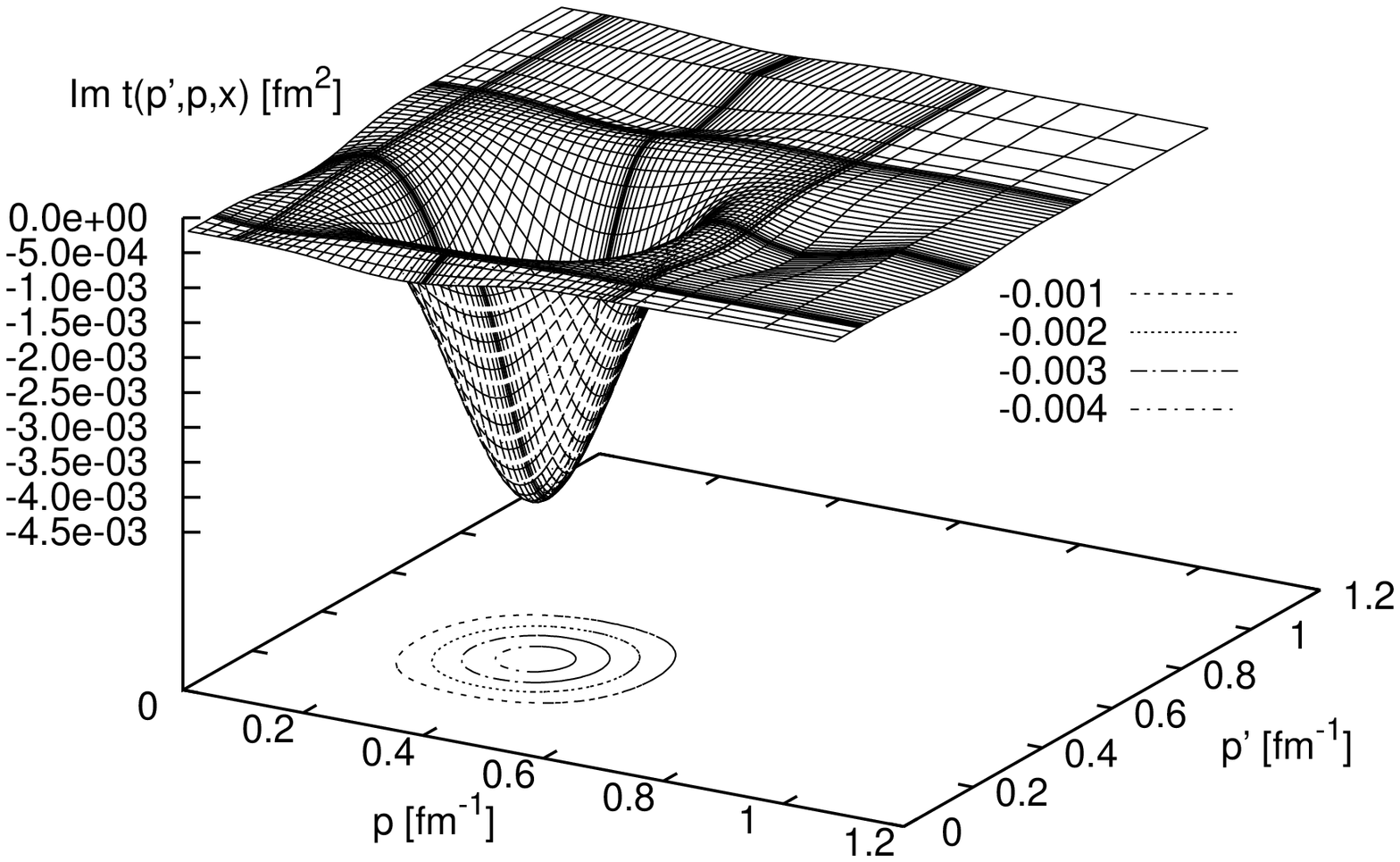}
\caption{The same as in Fig.\ref{fig5} but for the sharp cut-off screening of Eq.(\ref{eq.3sharp}) with 
R=20 fm at E=13 MeV.}
\label{fig15}
\end{figure}

\begin{figure}
\includegraphics[scale=0.4,clip=true,angle=-90]{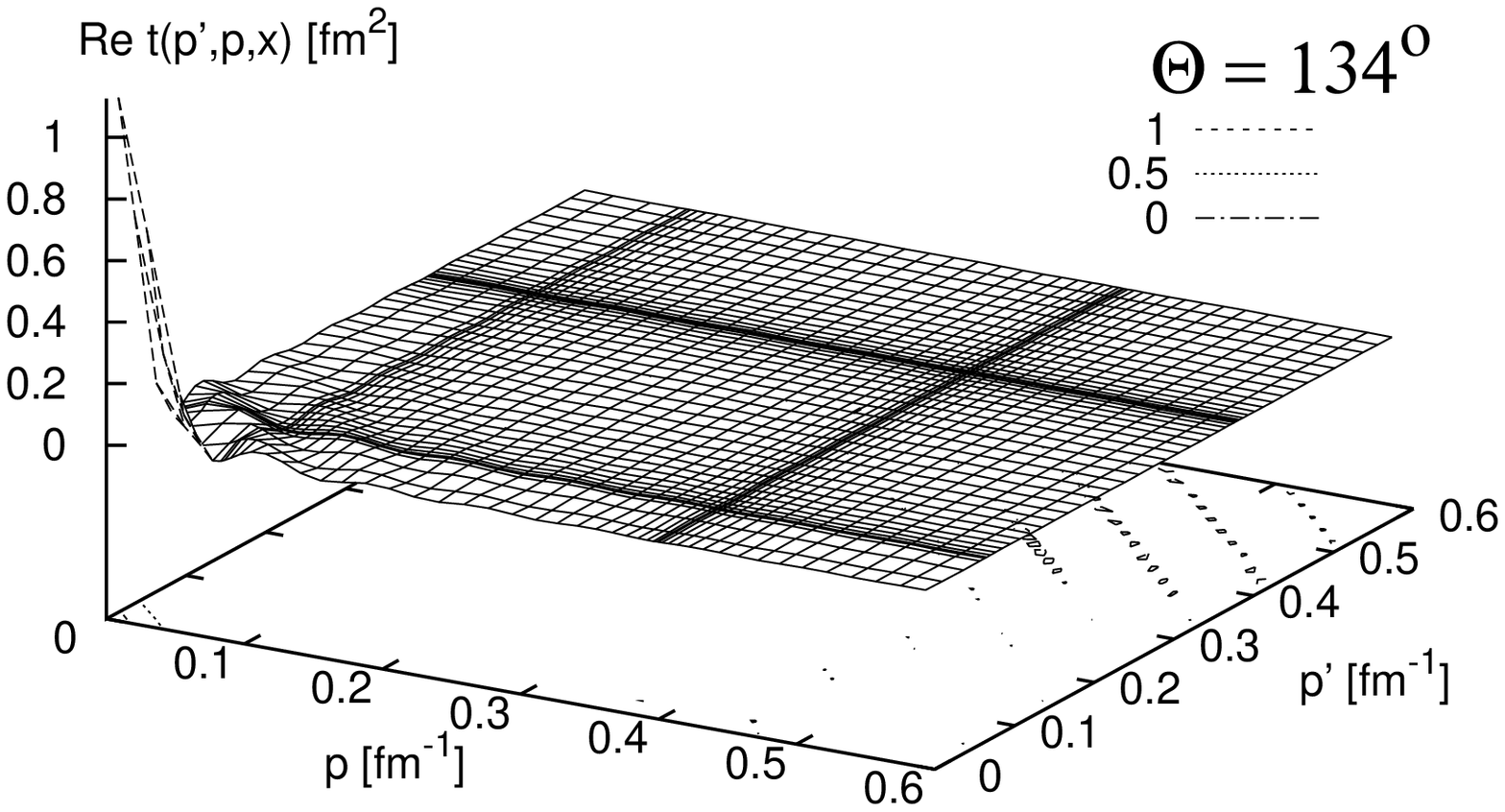}
\includegraphics[scale=0.4,clip=true,angle=-90]{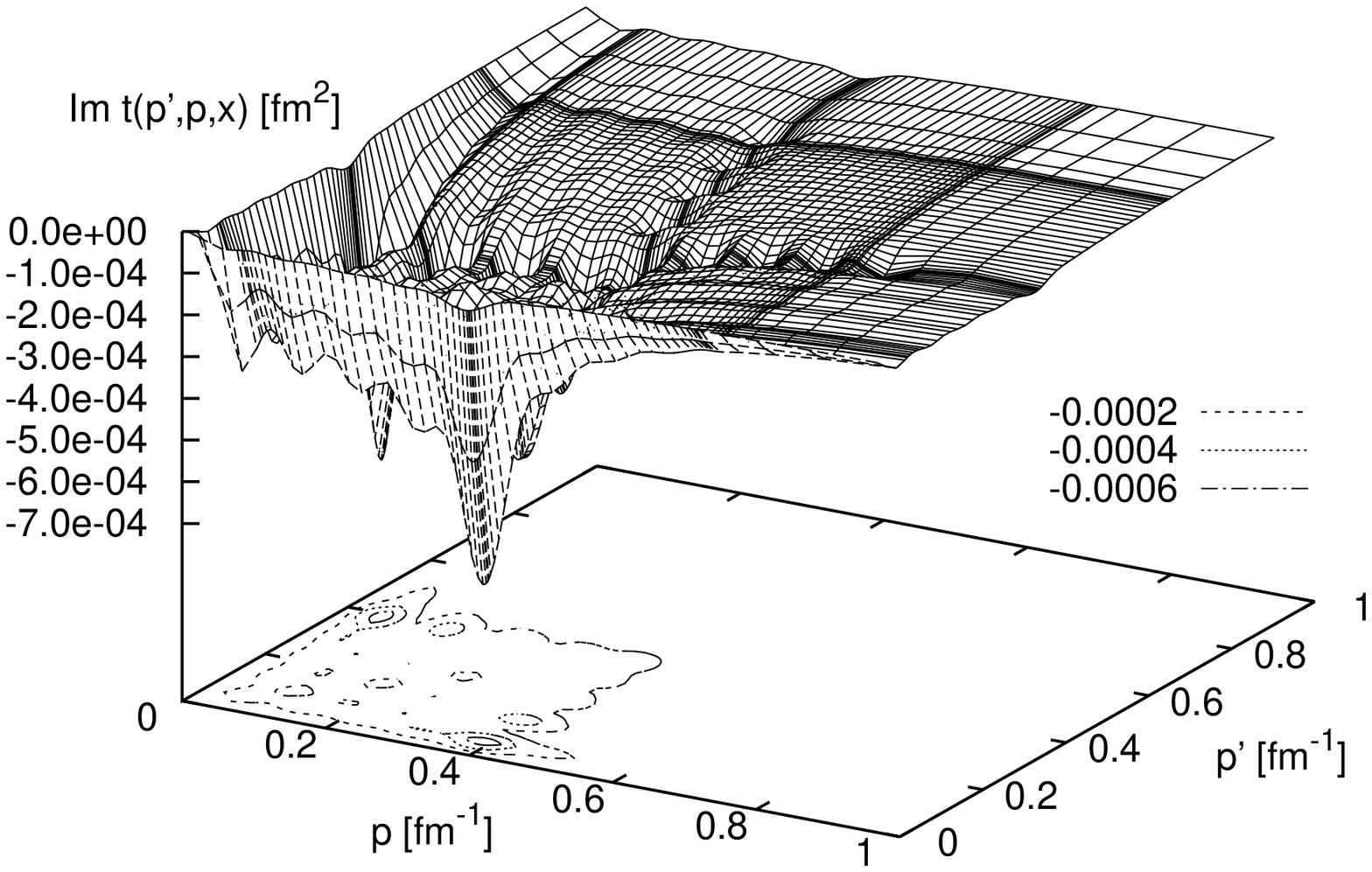}
\includegraphics[scale=0.4,clip=true,angle=-90]{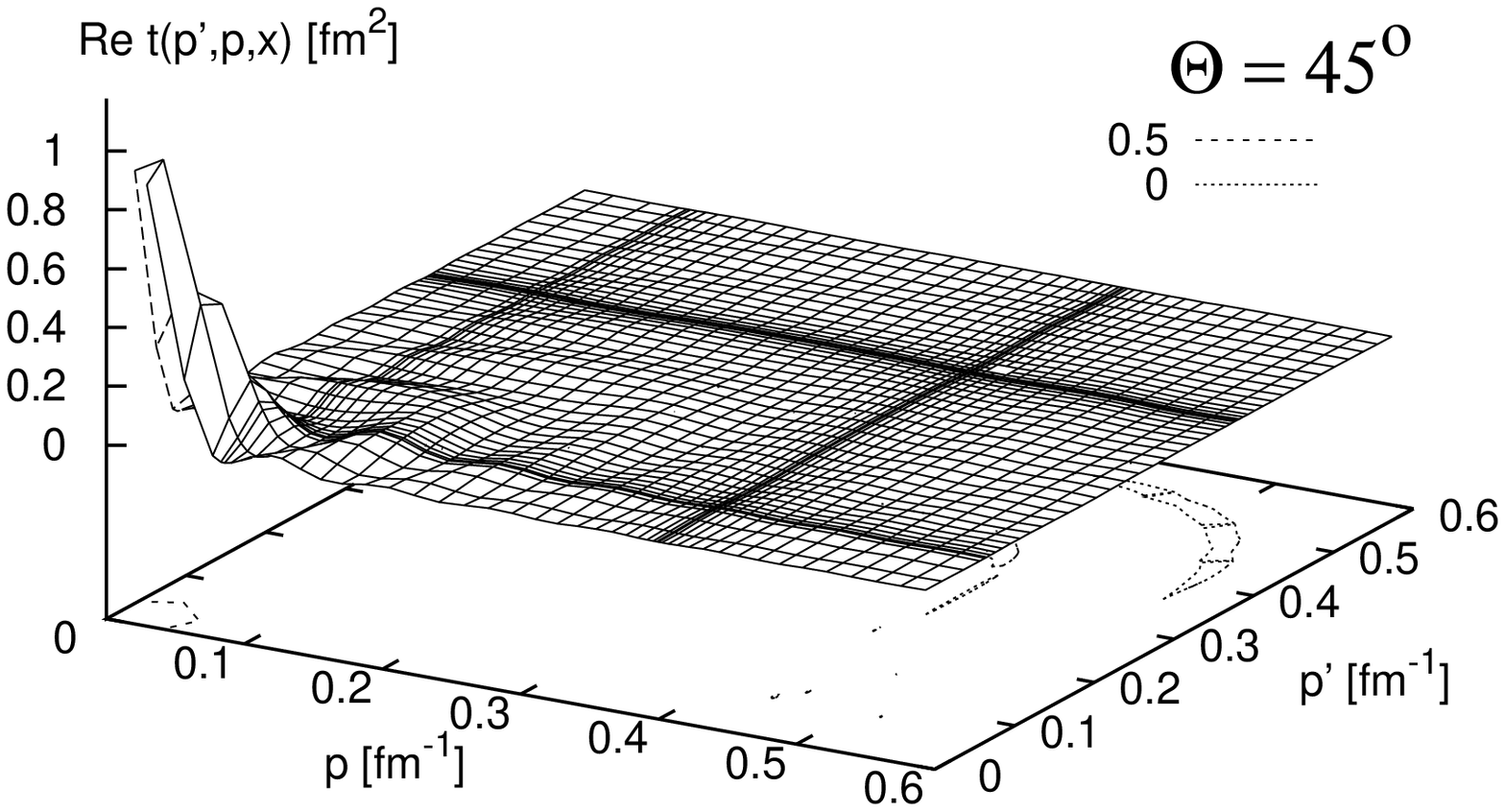}
\includegraphics[scale=0.4,clip=true,angle=-90]{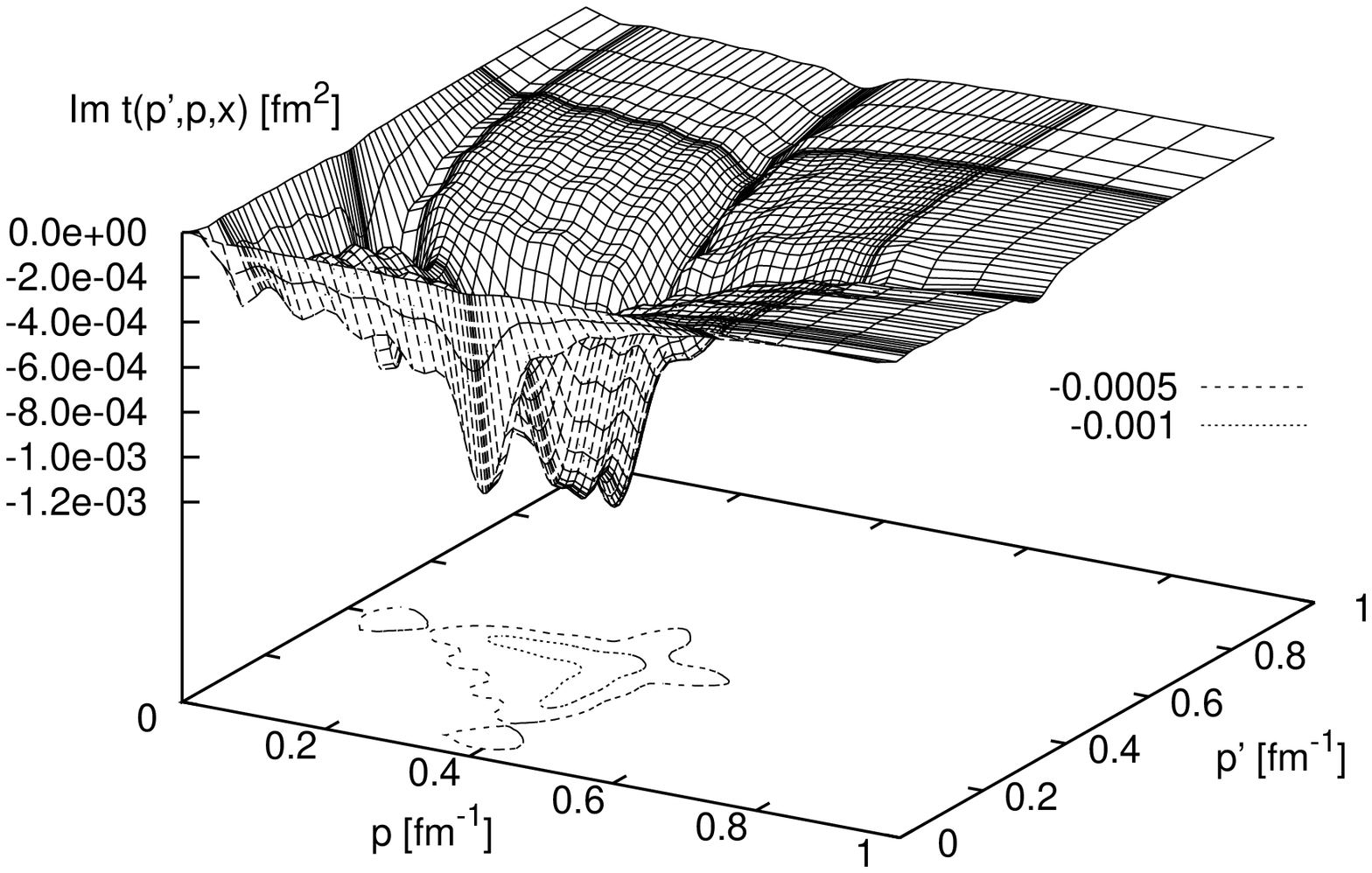}
\includegraphics[scale=0.4,clip=true,angle=-90]{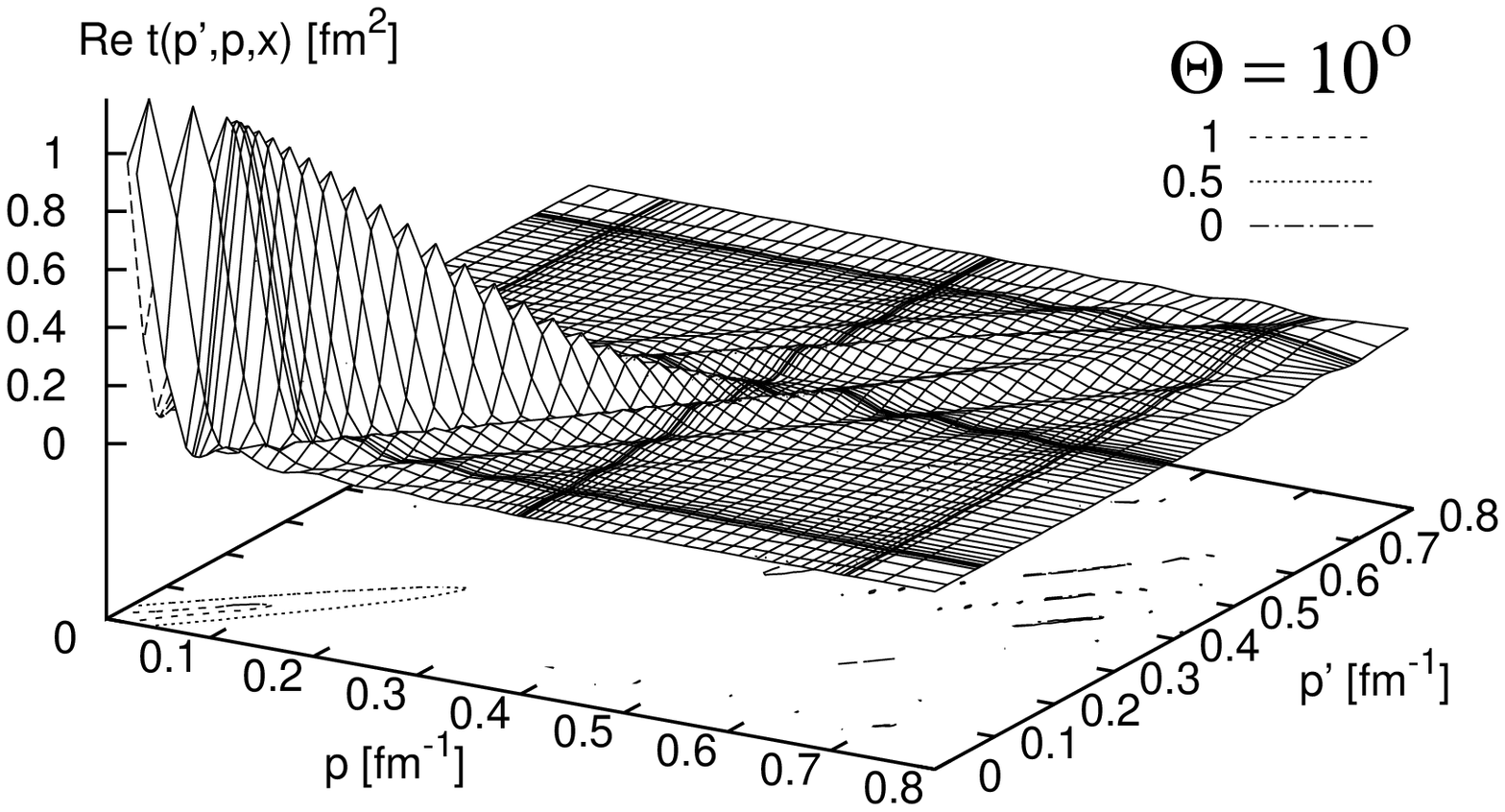}
\includegraphics[scale=0.4,clip=true,angle=-90]{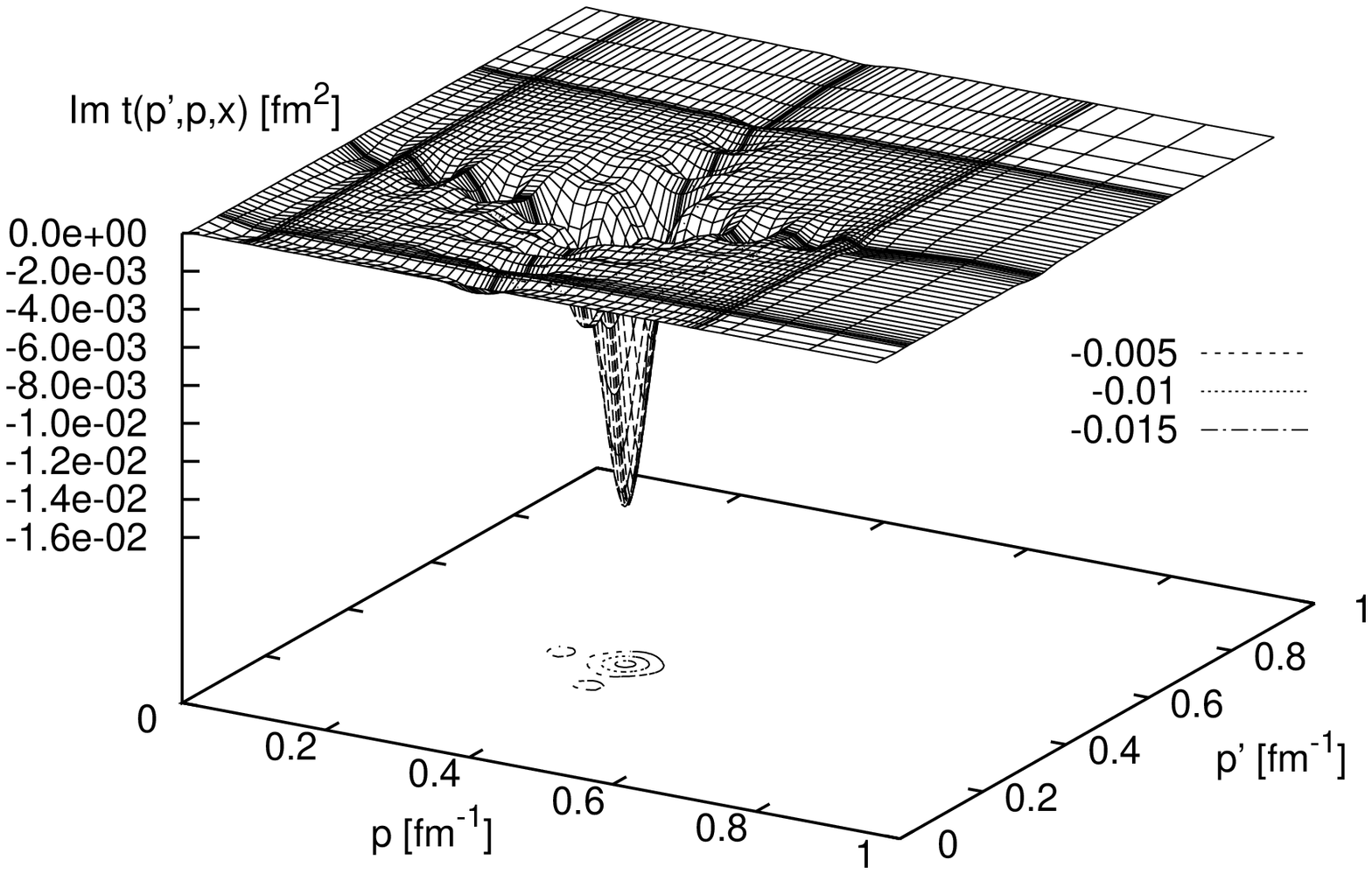}
\includegraphics[scale=0.4,clip=true,angle=-90]{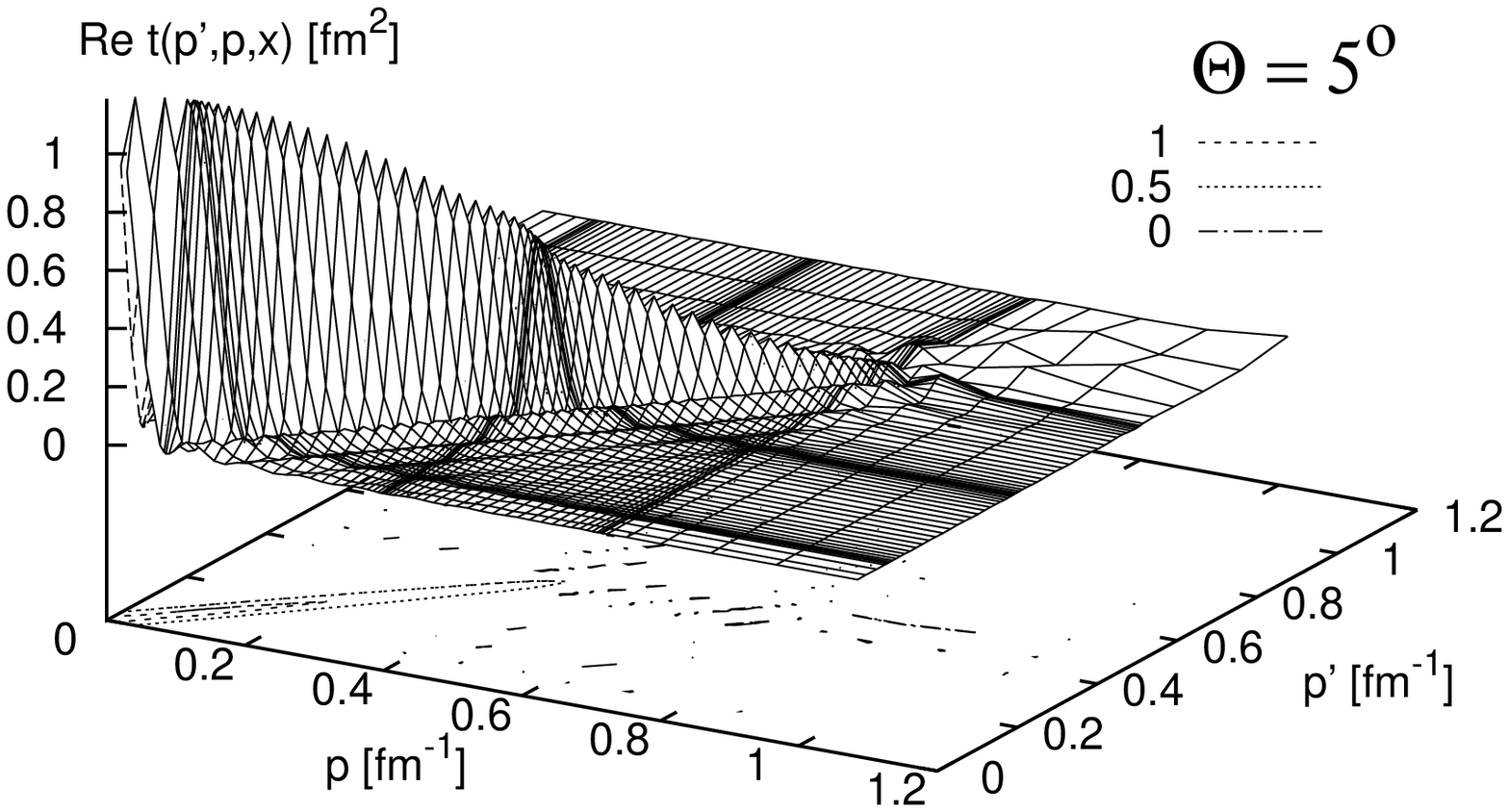}
\includegraphics[scale=0.4,clip=true,angle=-90]{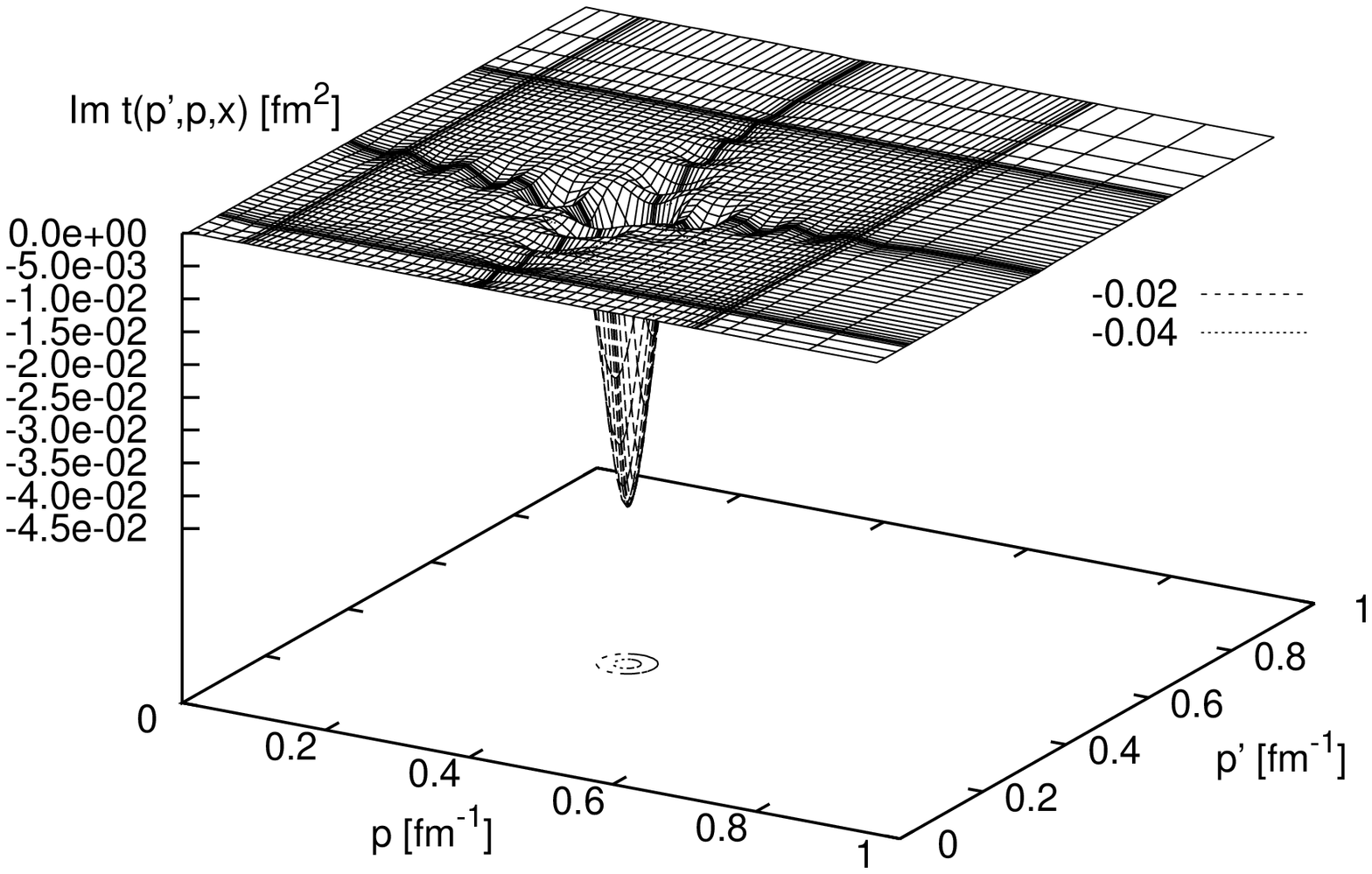}
\caption{The same as in Fig.\ref{fig5} but for the sharp cut-off screening of Eq.(\ref{eq.3sharp}) with
R=80 fm at E=13 MeV.}
\label{fig7}
\end{figure}

In Figs.~\ref{fig15} and~\ref{fig7} the t-matrix for the sharp cut-off screening at E=13 MeV is shown
for two values of cut-off parameter: R=20 fm and R=80 fm, respectively.
The general picture is similar
to that for the exponential screening, however, some oscillations are visible
for both real and imaginary parts of the t-matrix.
For the real part they are clearly visible at small momenta. For the 
imaginary part at bigger scattering angles they are relatively large, when compared to 
the absolute minimum of $Im(t)$ and decrease with decreasing angles.

\begin{figure}
\includegraphics[scale=0.4,clip=true,angle=-90]{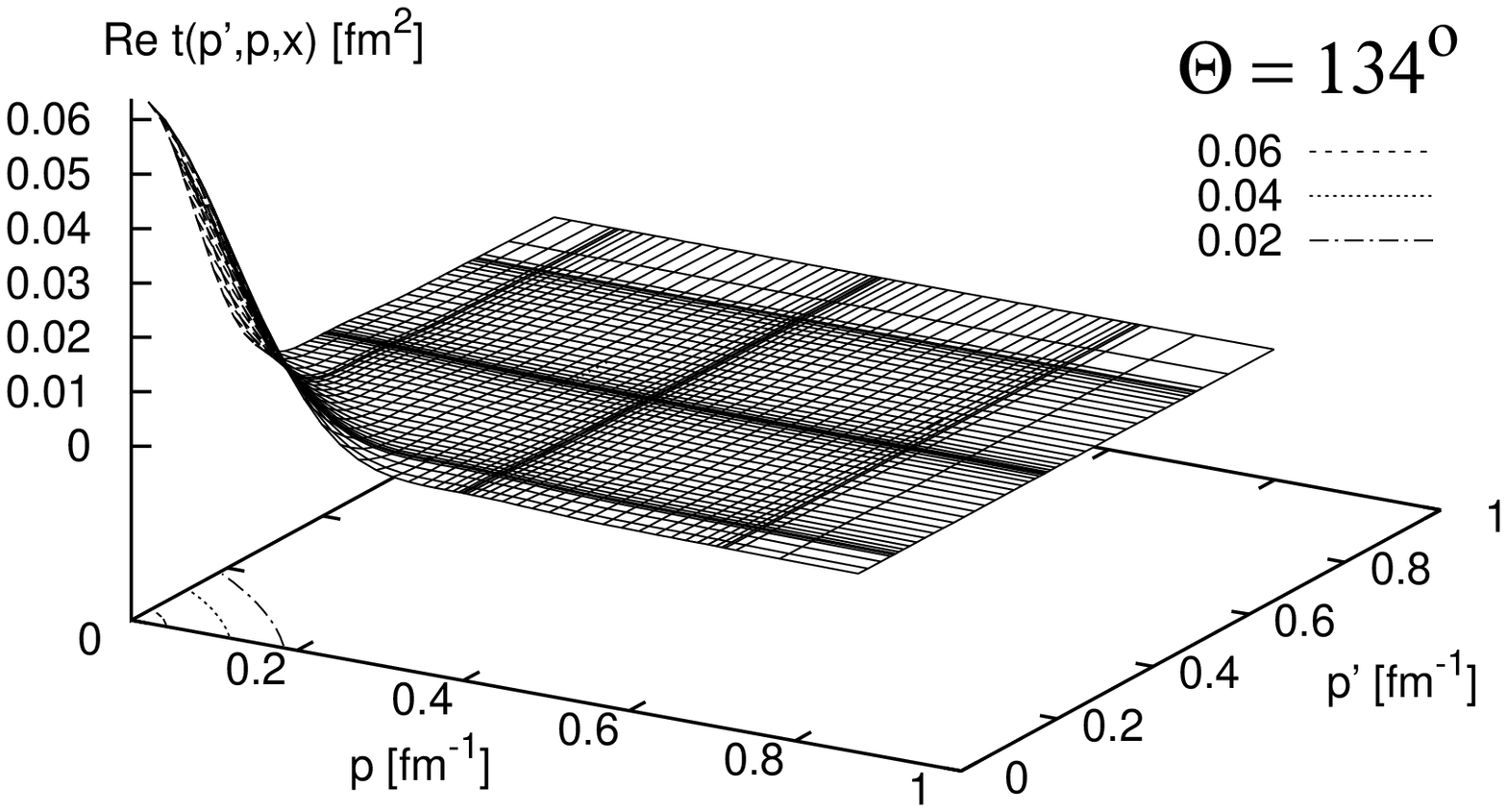}
\includegraphics[scale=0.4,clip=true,angle=-90]{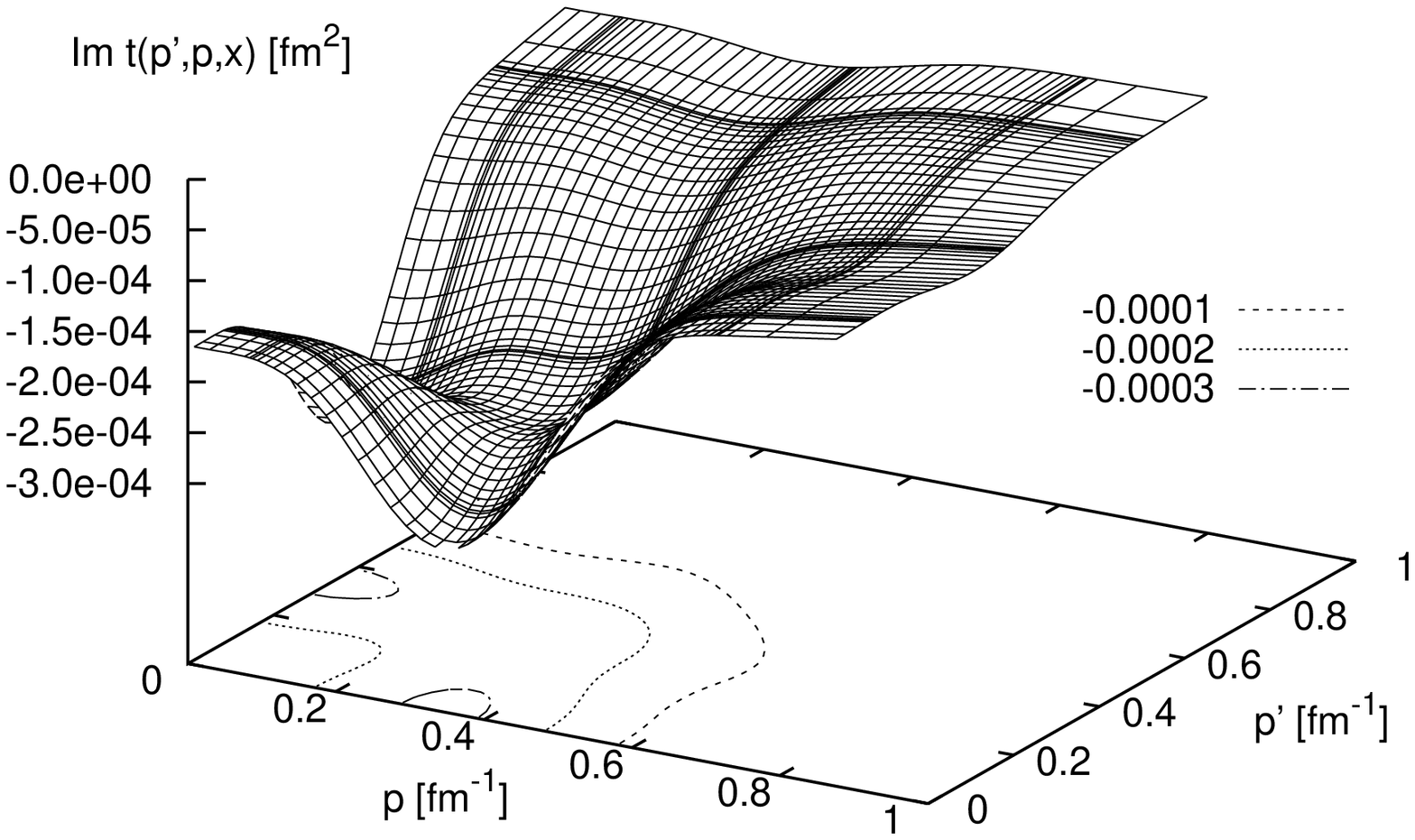}
\includegraphics[scale=0.4,clip=true,angle=-90]{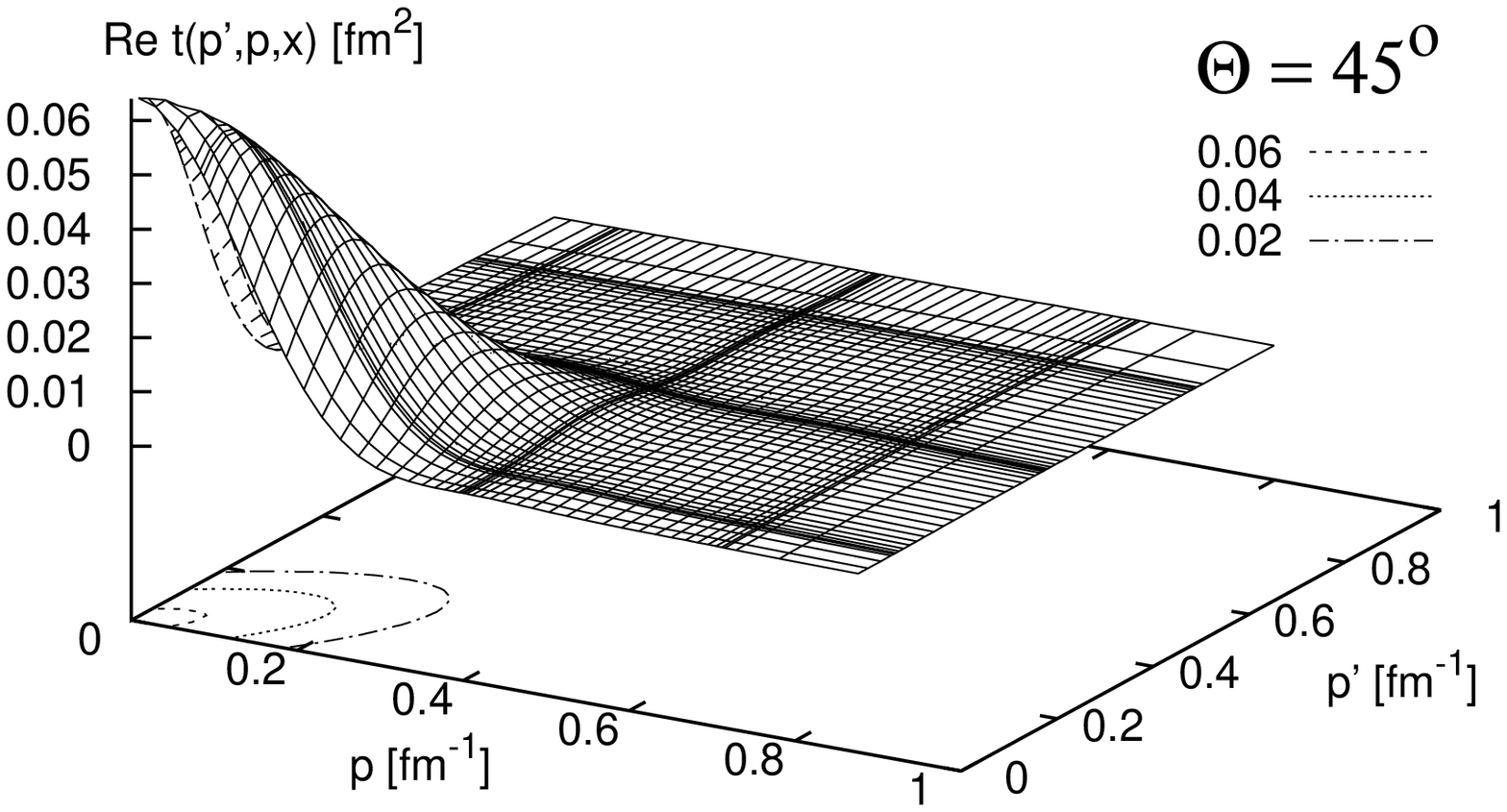}
\includegraphics[scale=0.4,clip=true,angle=-90]{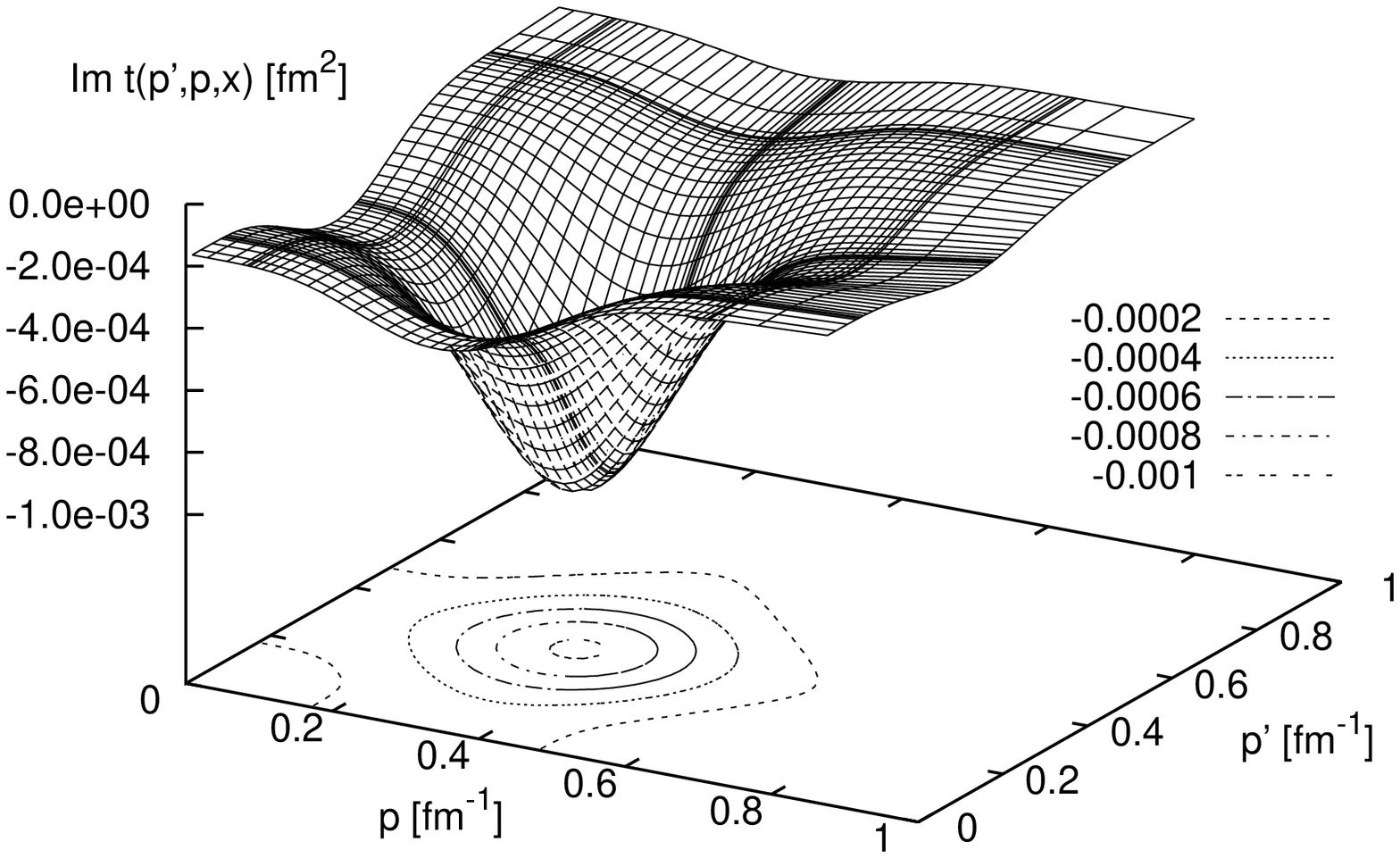}
\includegraphics[scale=0.4,clip=true,angle=-90]{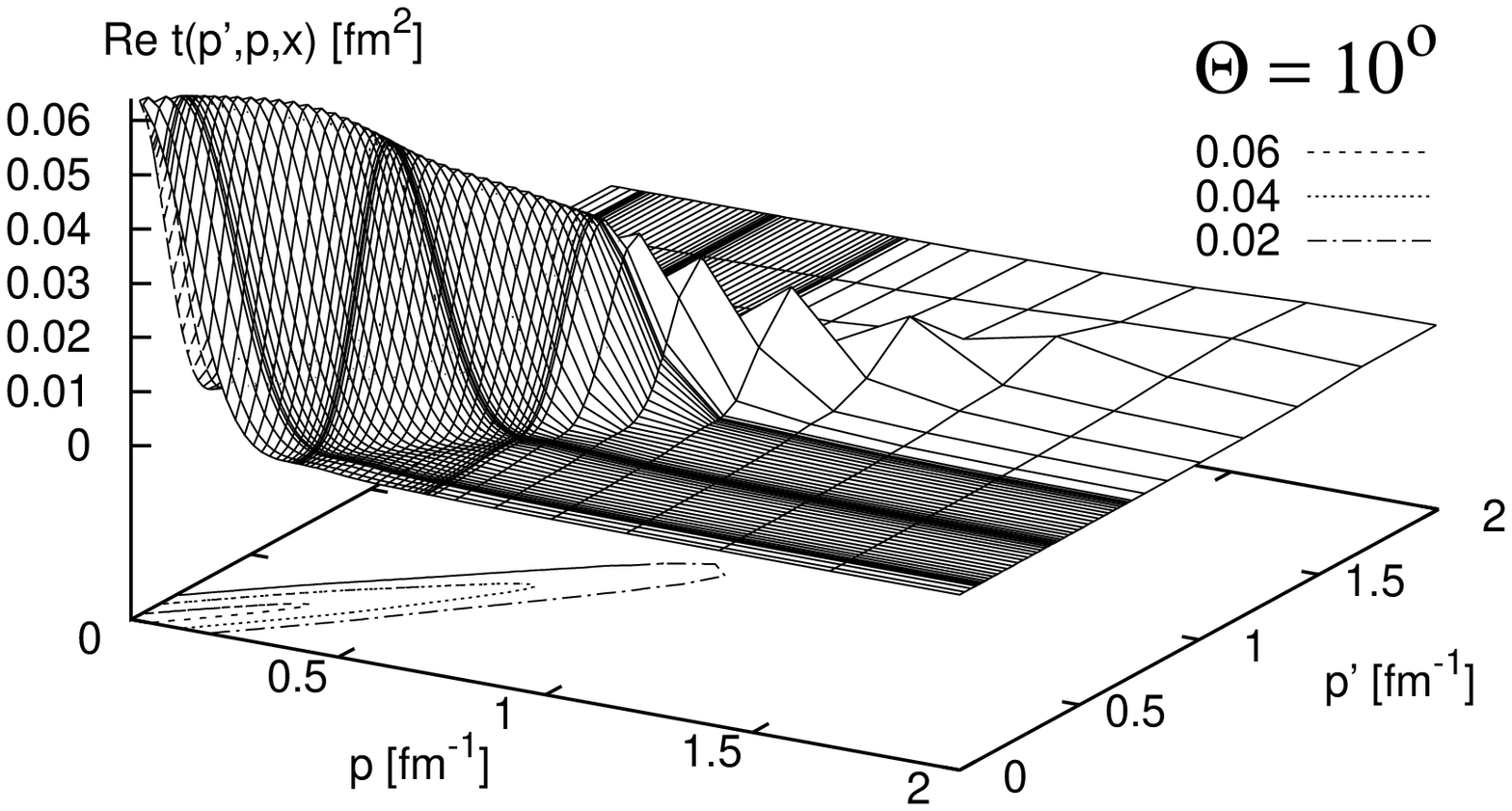}
\includegraphics[scale=0.4,clip=true,angle=-90]{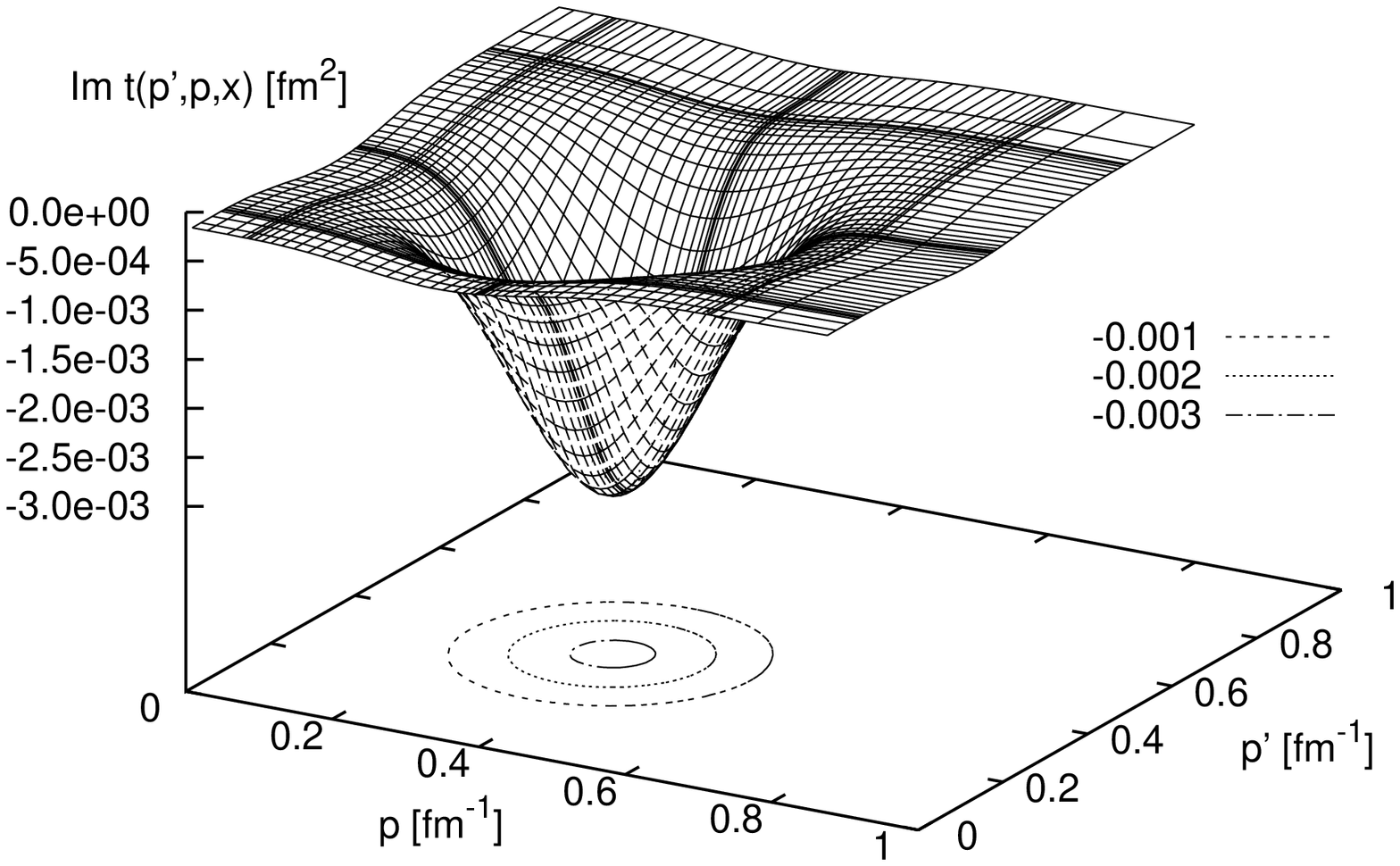}
\includegraphics[scale=0.4,clip=true,angle=-90]{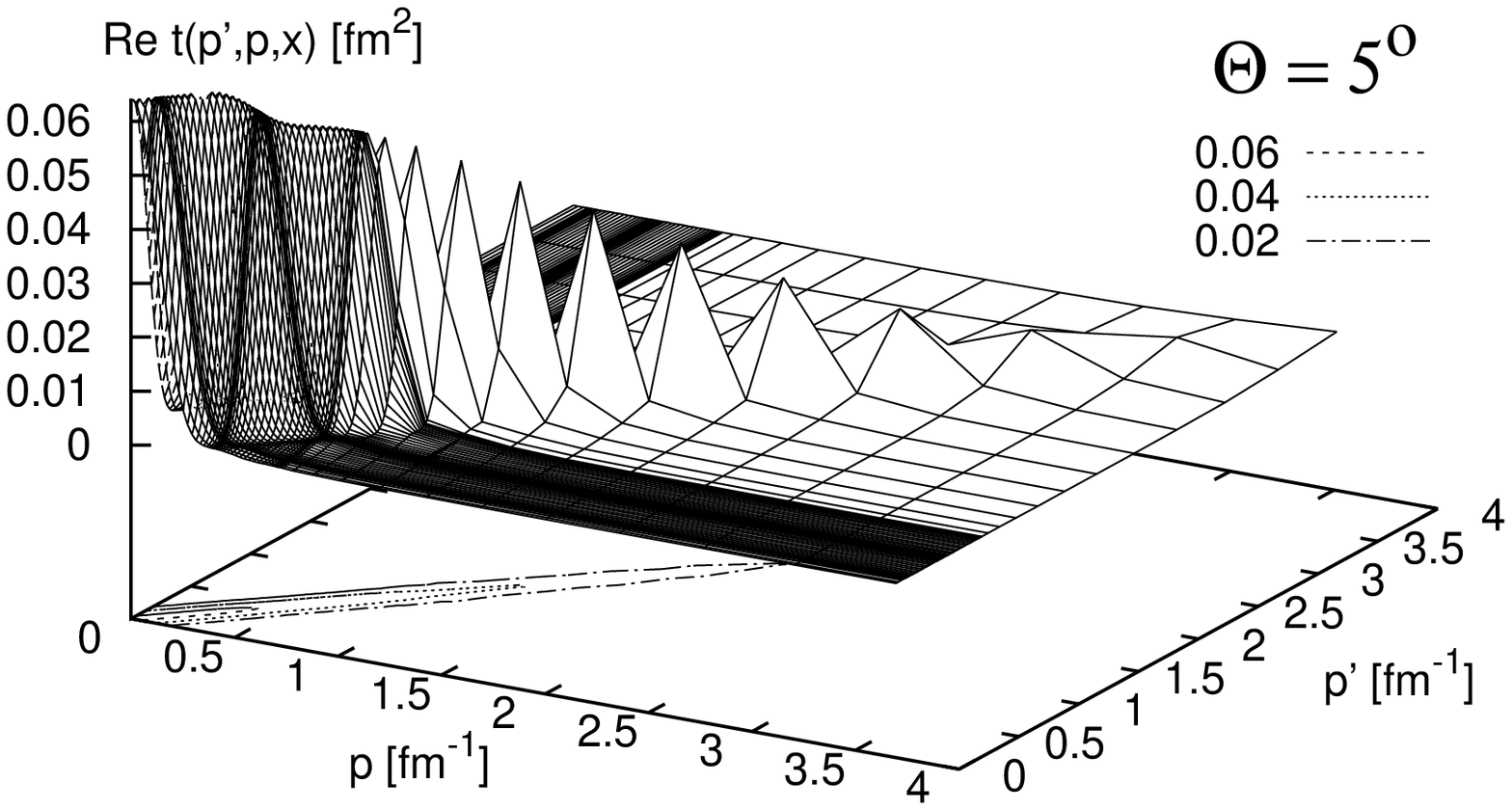}
\includegraphics[scale=0.4,clip=true,angle=-90]{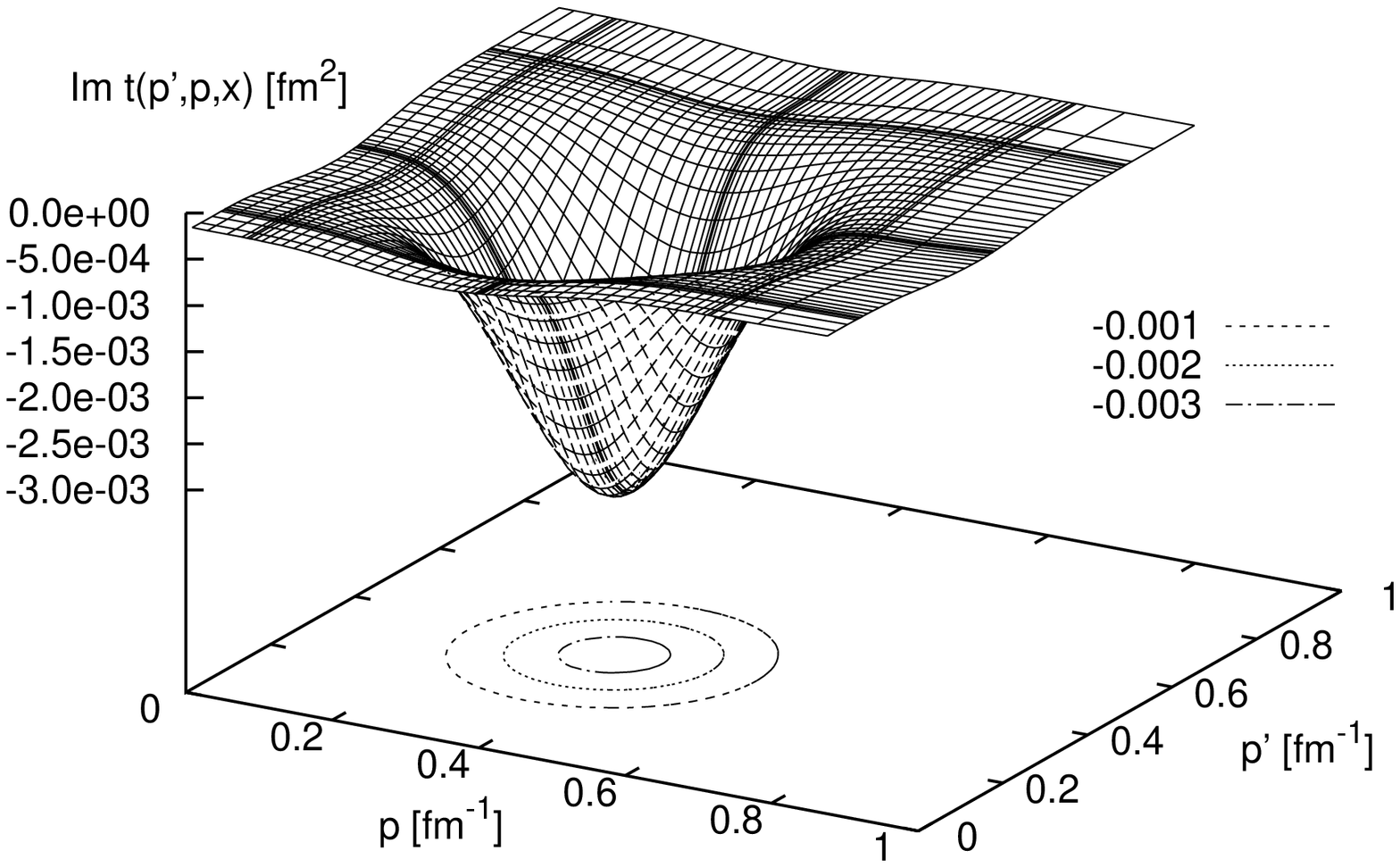}
\caption{The same as in Fig.\ref{fig5} but for the localized screening of Eq.(\ref{localized})
with R=9 fm.}
\label{fig13}
\end{figure}

\begin{figure}
\includegraphics[scale=0.4,clip=true,angle=-90]{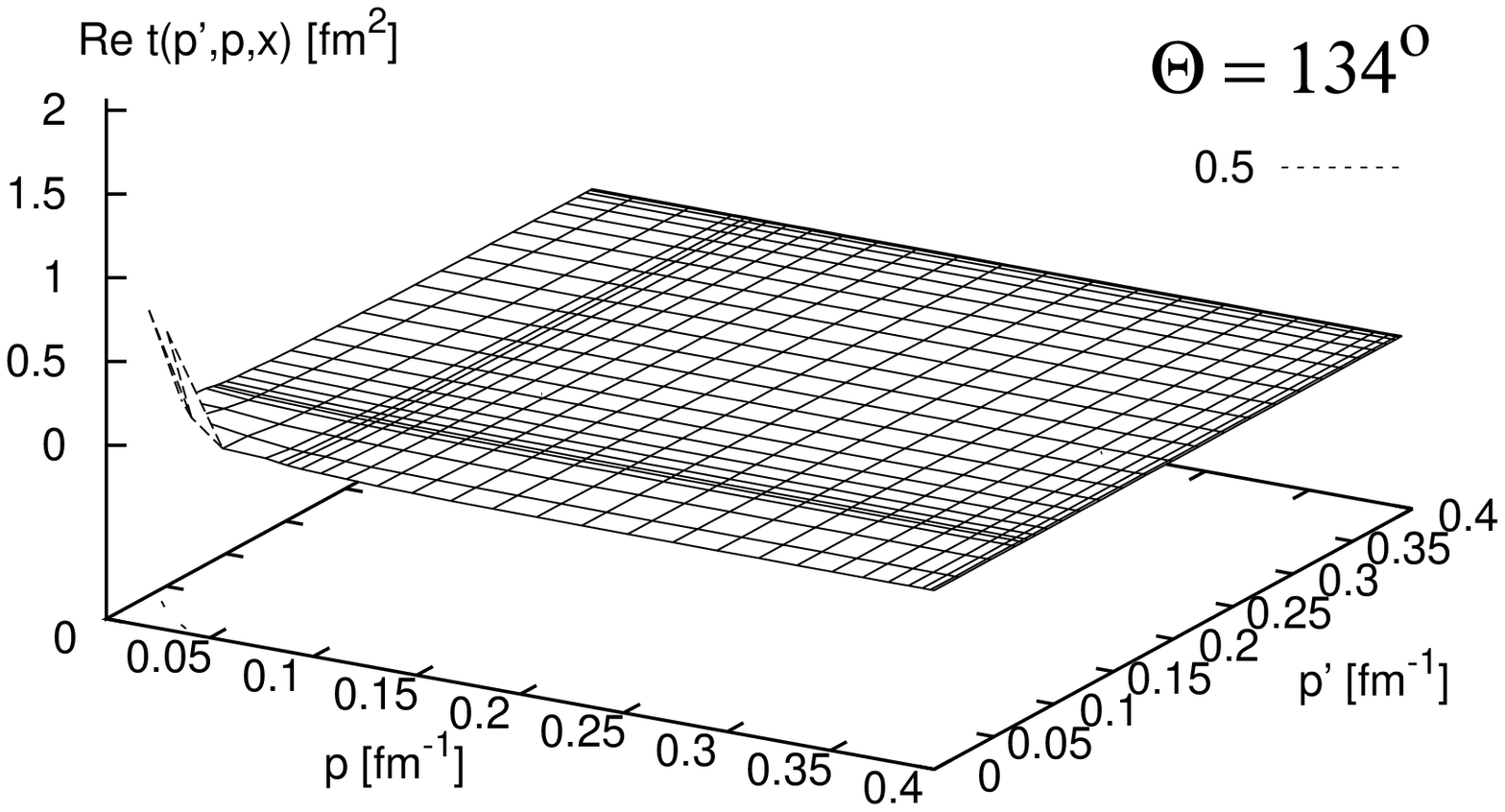}
\includegraphics[scale=0.4,clip=true,angle=-90]{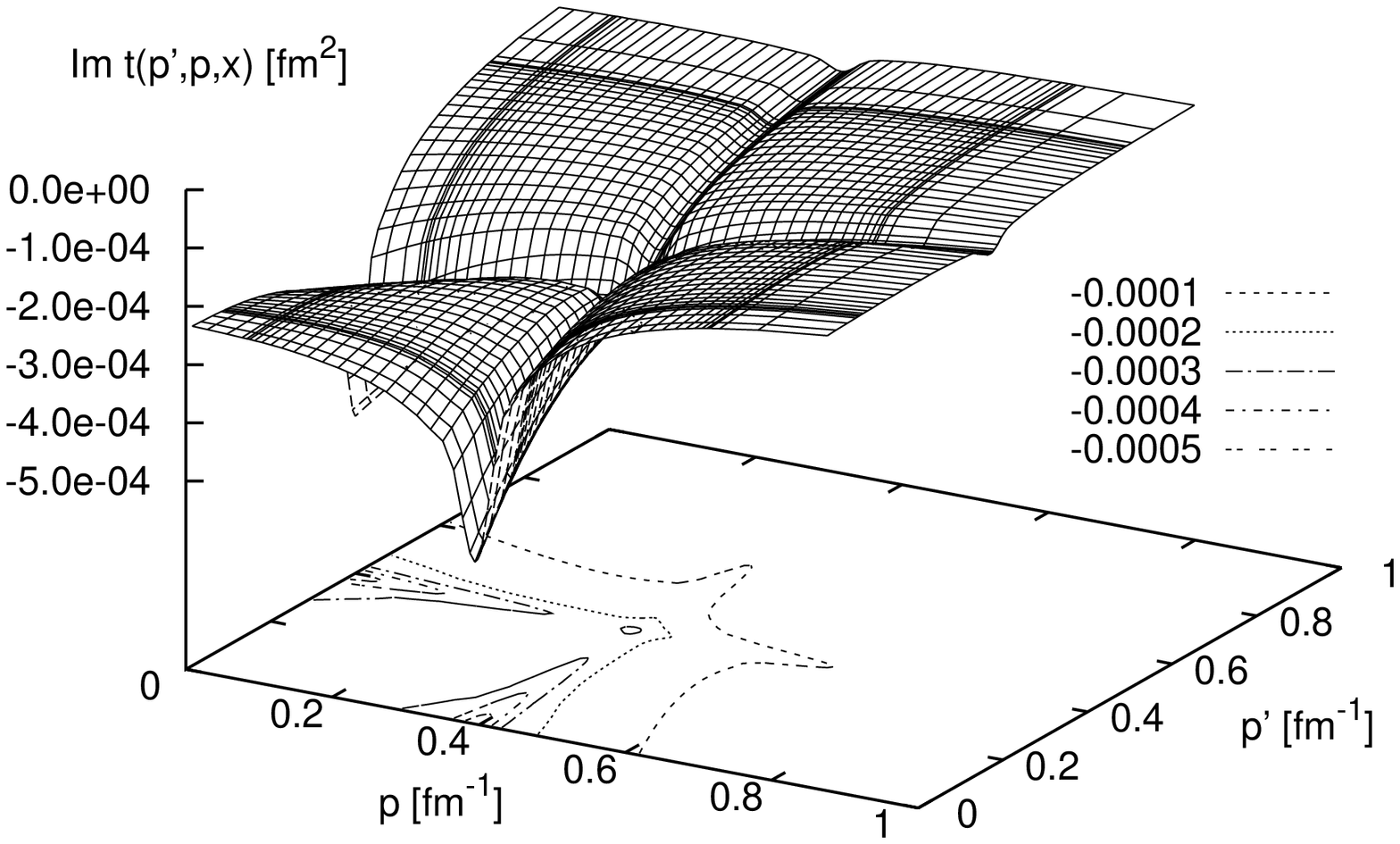}
\includegraphics[scale=0.4,clip=true,angle=-90]{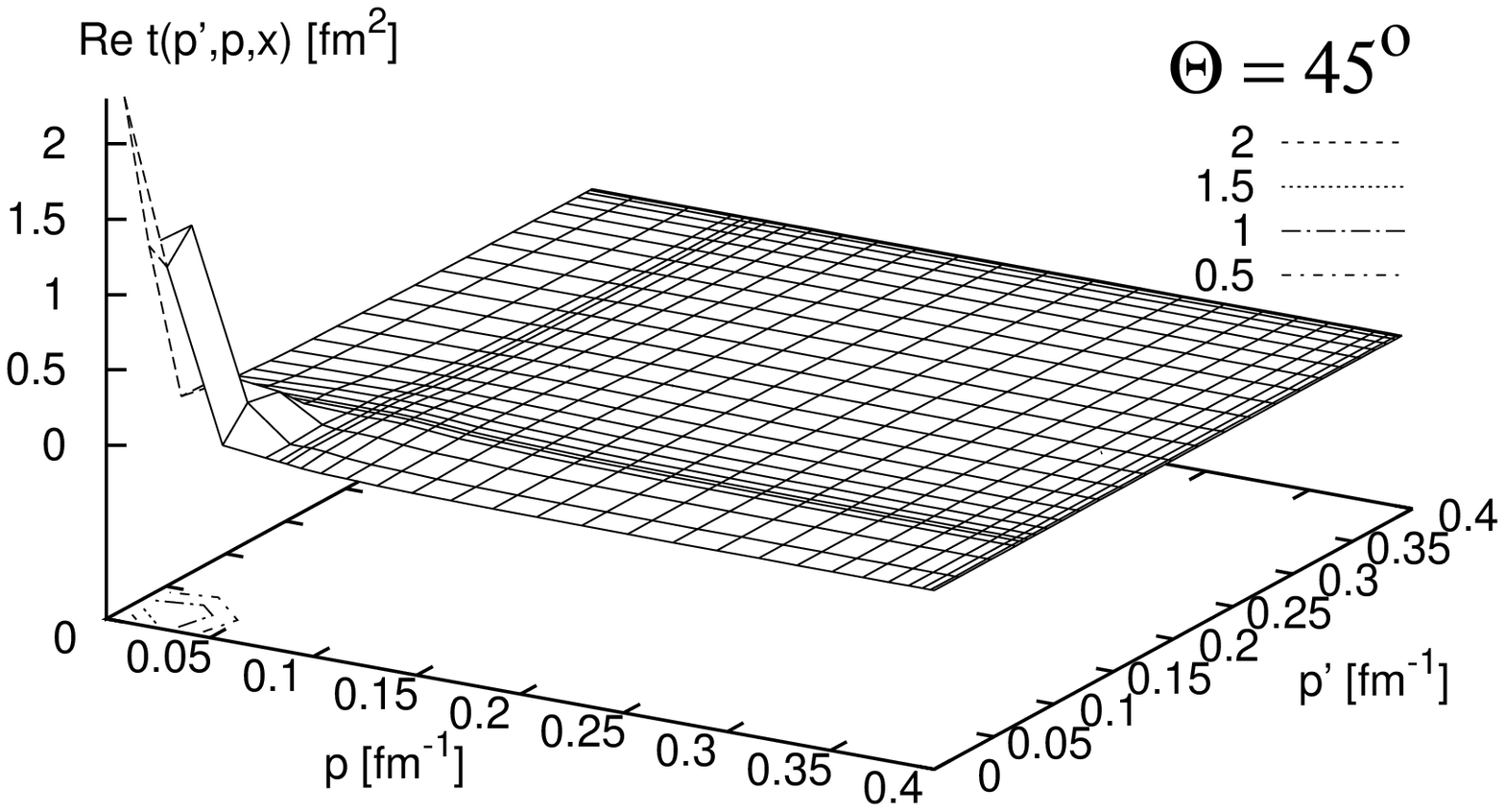}
\includegraphics[scale=0.4,clip=true,angle=-90]{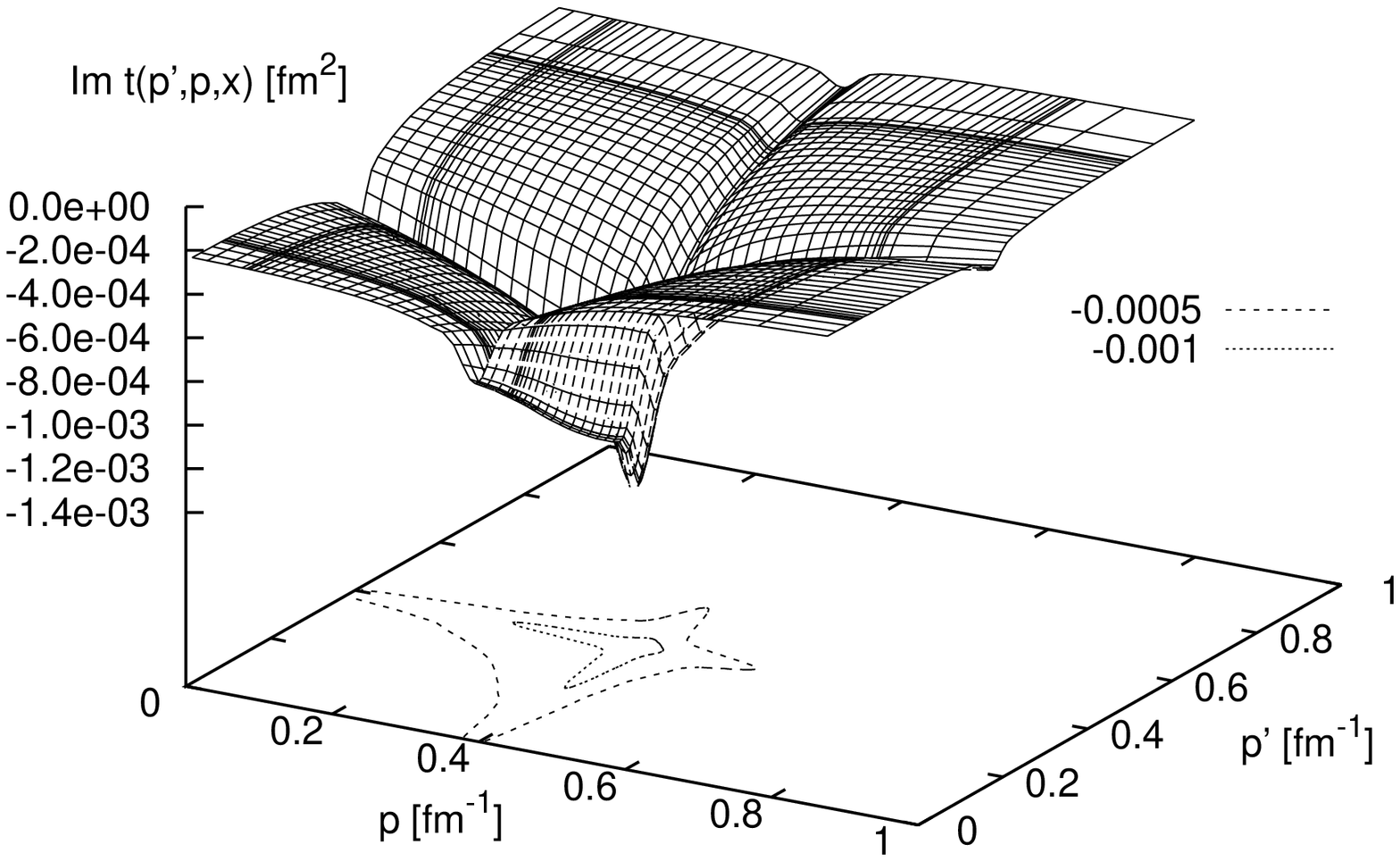}
\includegraphics[scale=0.4,clip=true,angle=-90]{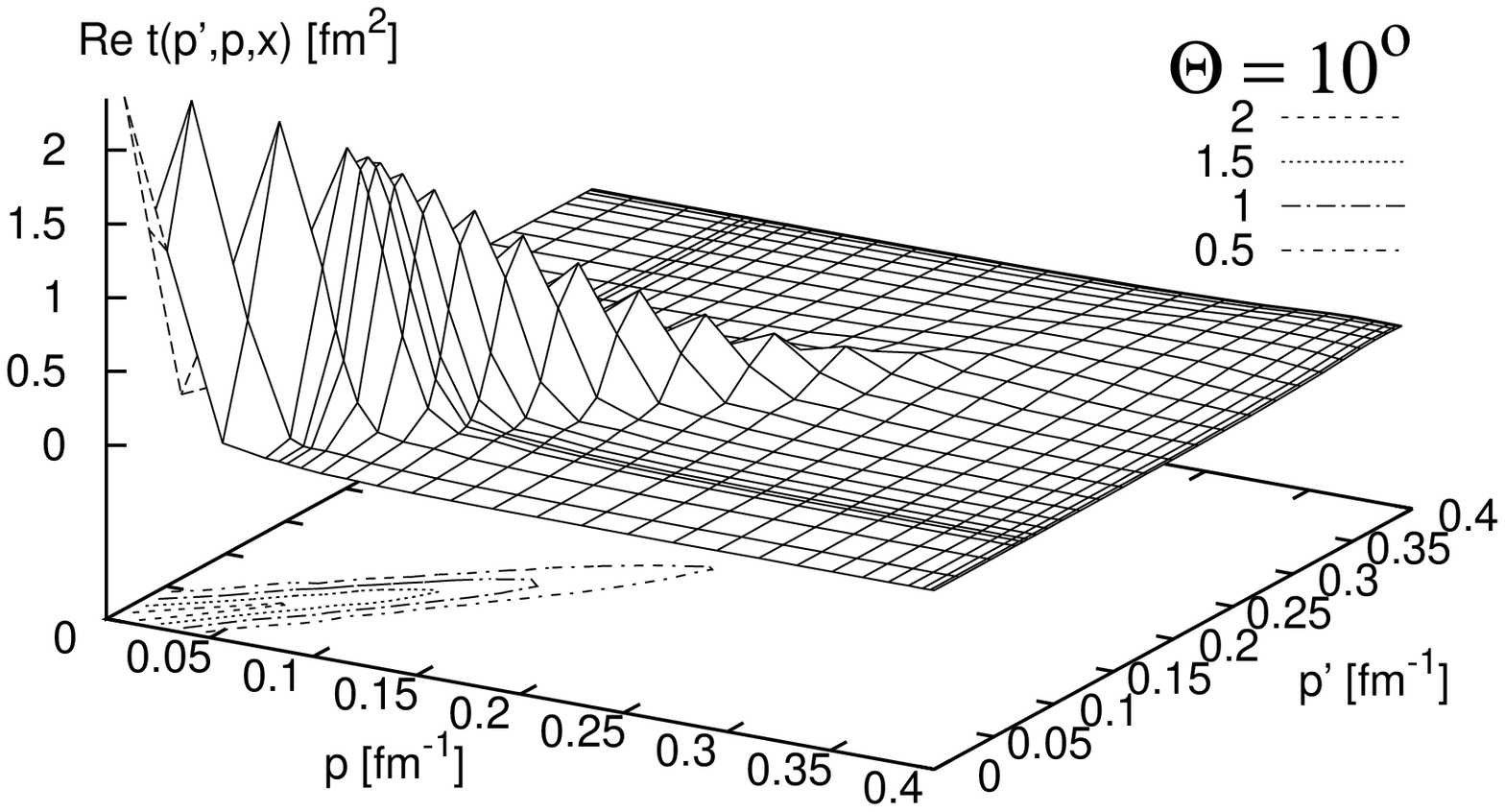}
\includegraphics[scale=0.4,clip=true,angle=-90]{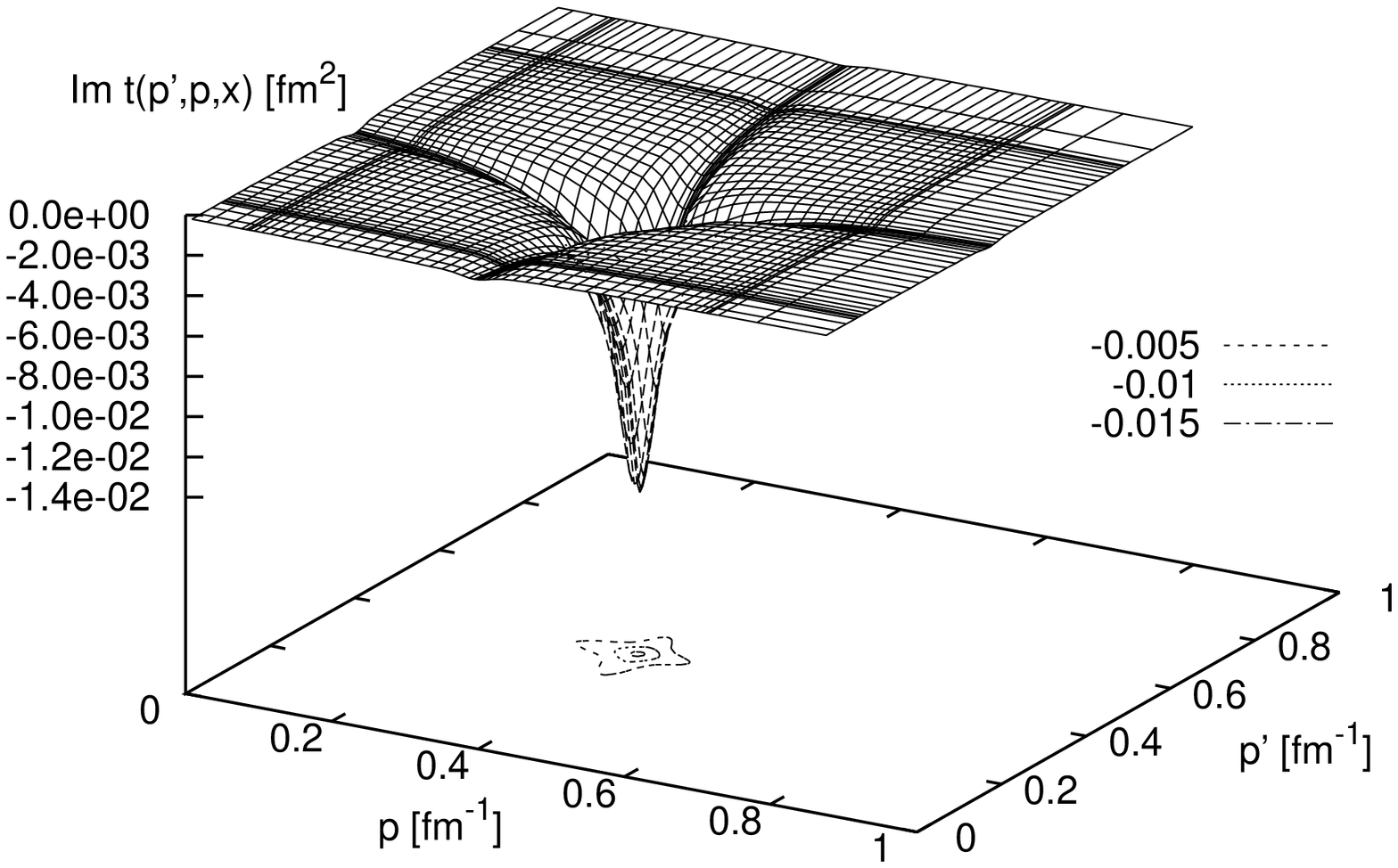}
\includegraphics[scale=0.4,clip=true,angle=-90]{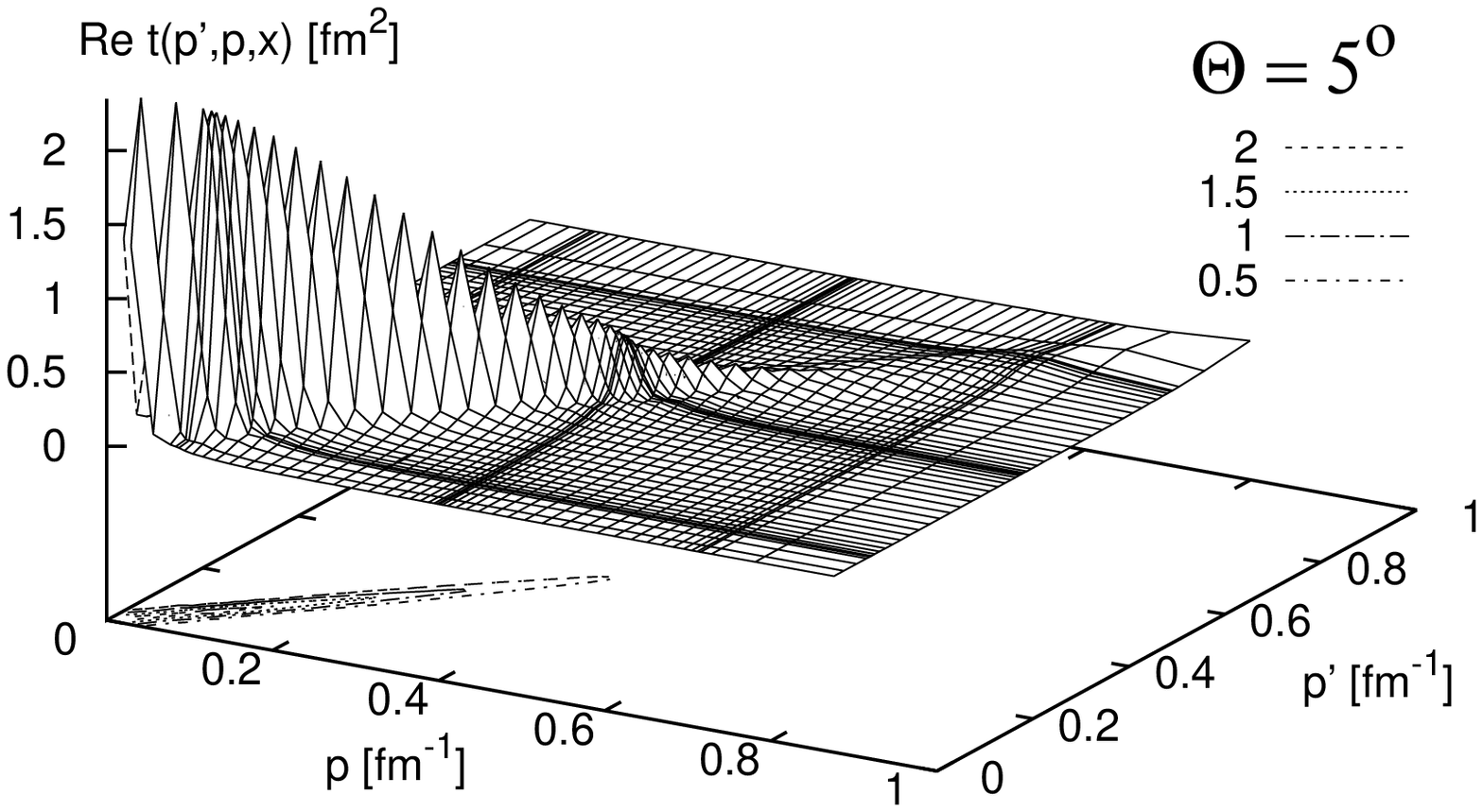}
\includegraphics[scale=0.4,clip=true,angle=-90]{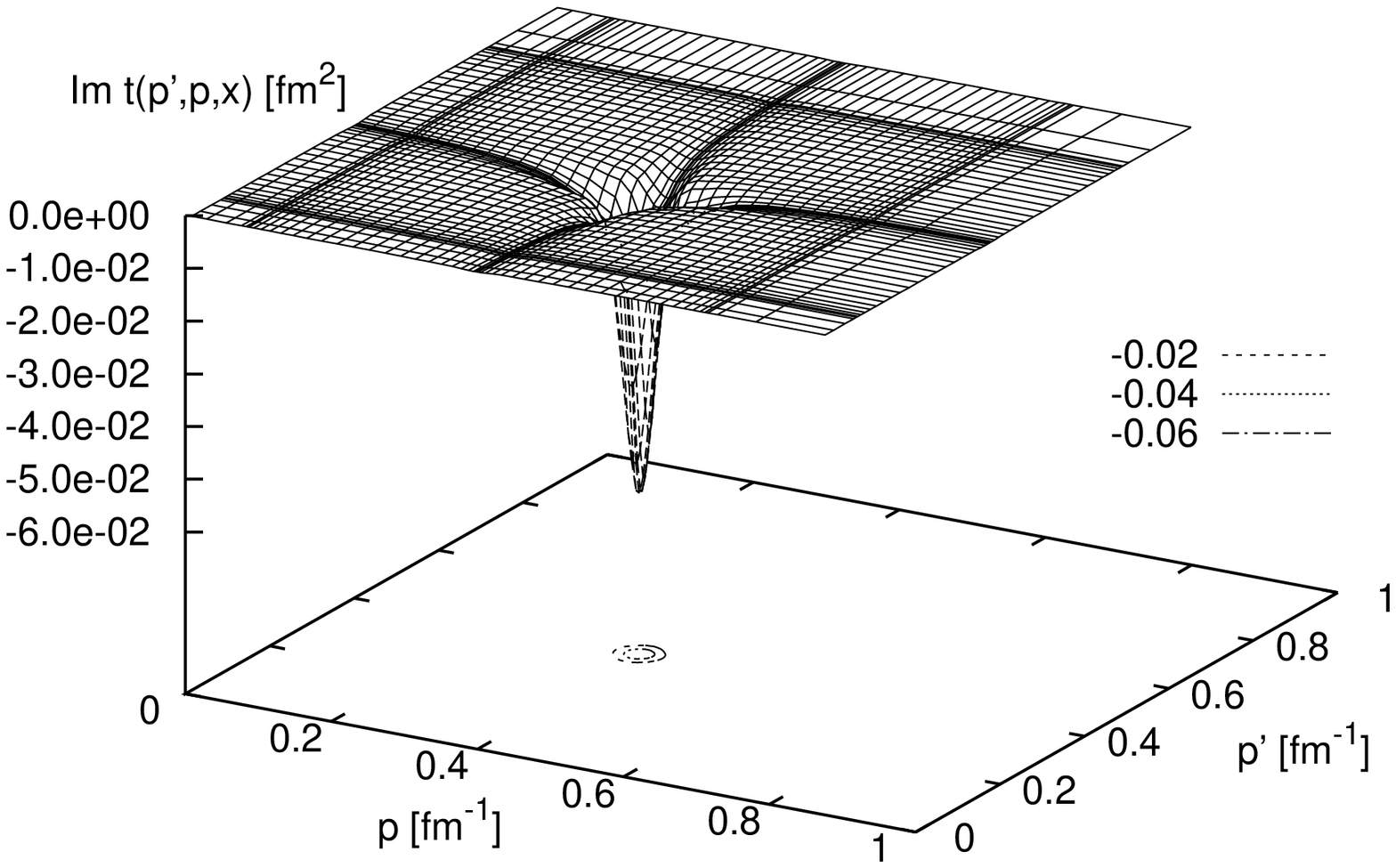}
\caption{The same as in Fig.\ref{fig5} but for the localized screening of Eq.(\ref{localized})
with R=55 fm.}
\label{fig14}
\end{figure}

The t-matrices obtained using the localized screening of Eq.~(\ref{localized})
are shown in Fig.~\ref{fig13} ($R=9$ fm) and Fig.~\ref{fig14} ($R=55$ fm). 
These values of the localized screening range parameter $R=9$~$(55)$ fm correspond roughly to the values of the screening 
radius $R=20$~$(120)$ fm for the exponential screening. 
The resulting t-matrices are very similar to those obtained with the
exponential screening at the same angles. 

\begin{figure}
\includegraphics[scale=0.4,clip=true,angle=-90]{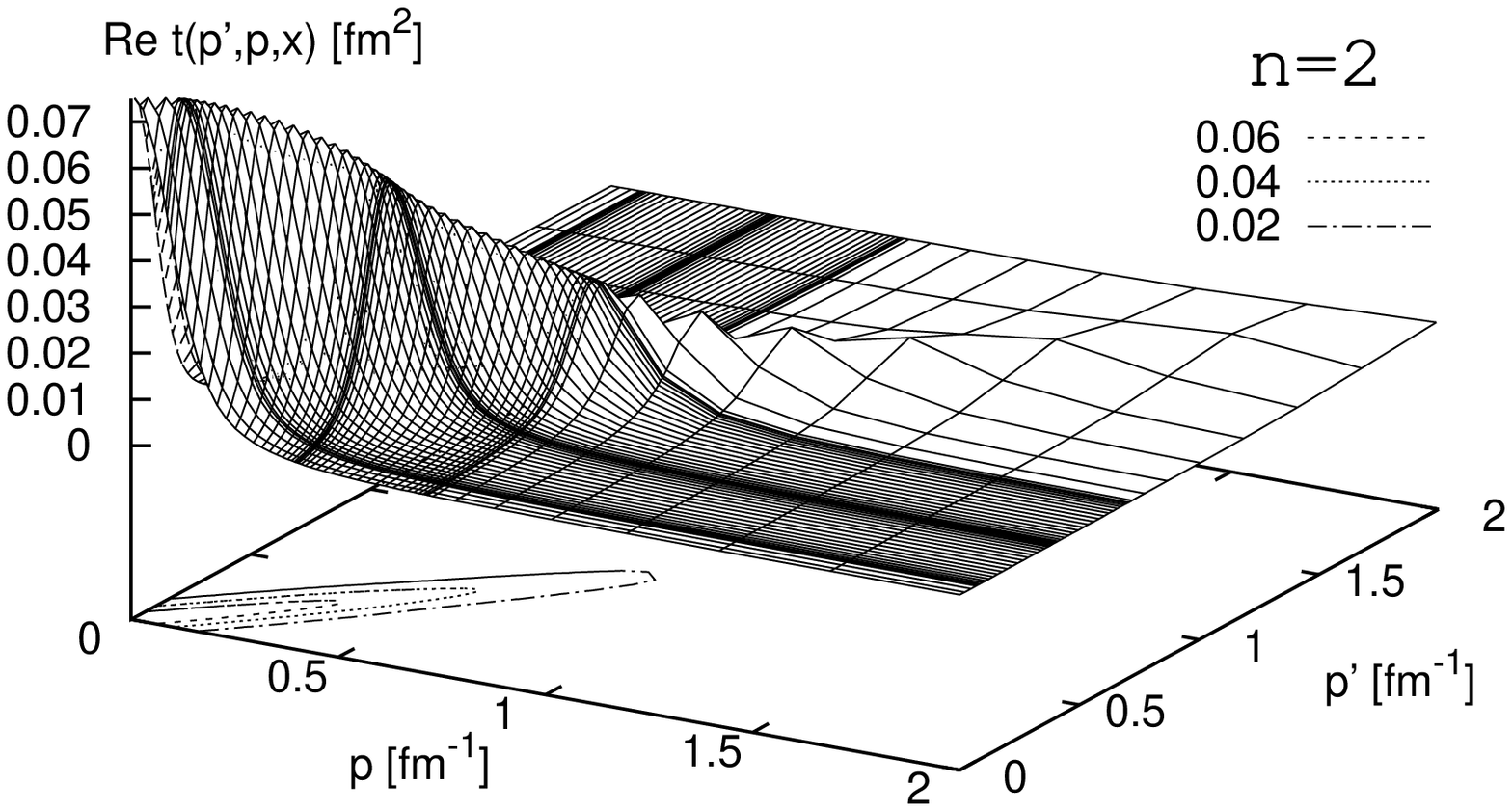}
\includegraphics[scale=0.4,clip=true,angle=-90]{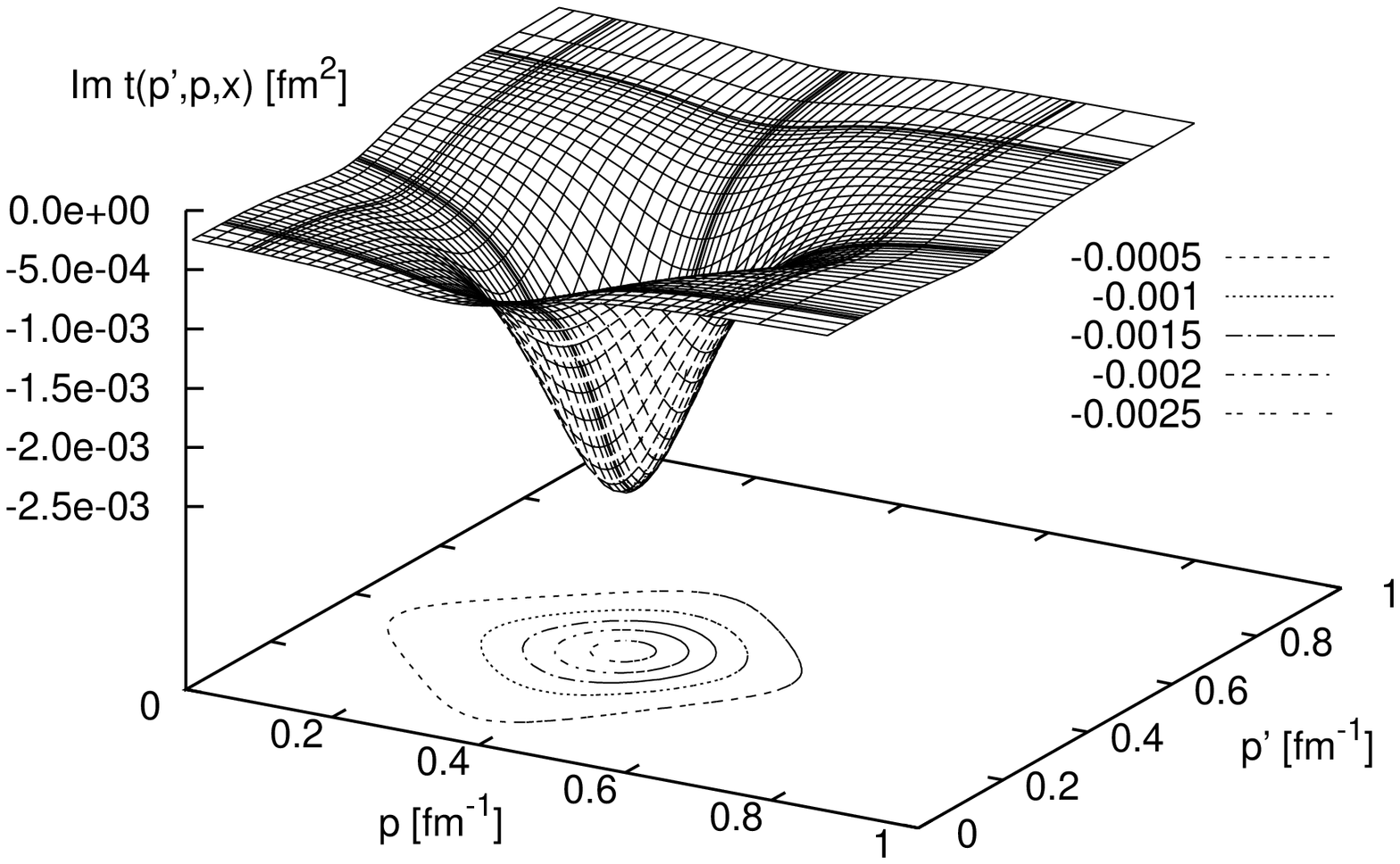}
\includegraphics[scale=0.4,clip=true,angle=-90]{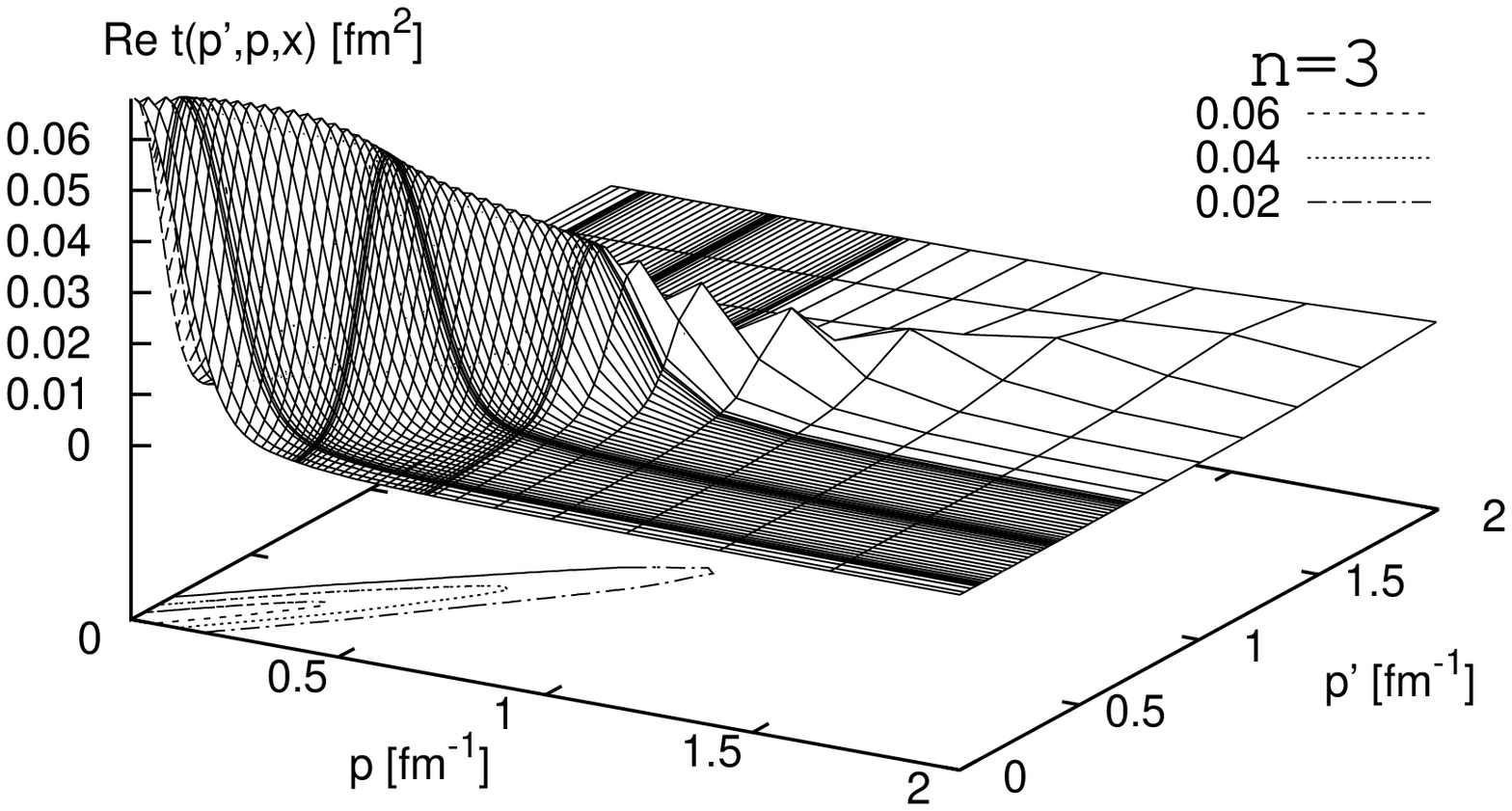}
\includegraphics[scale=0.4,clip=true,angle=-90]{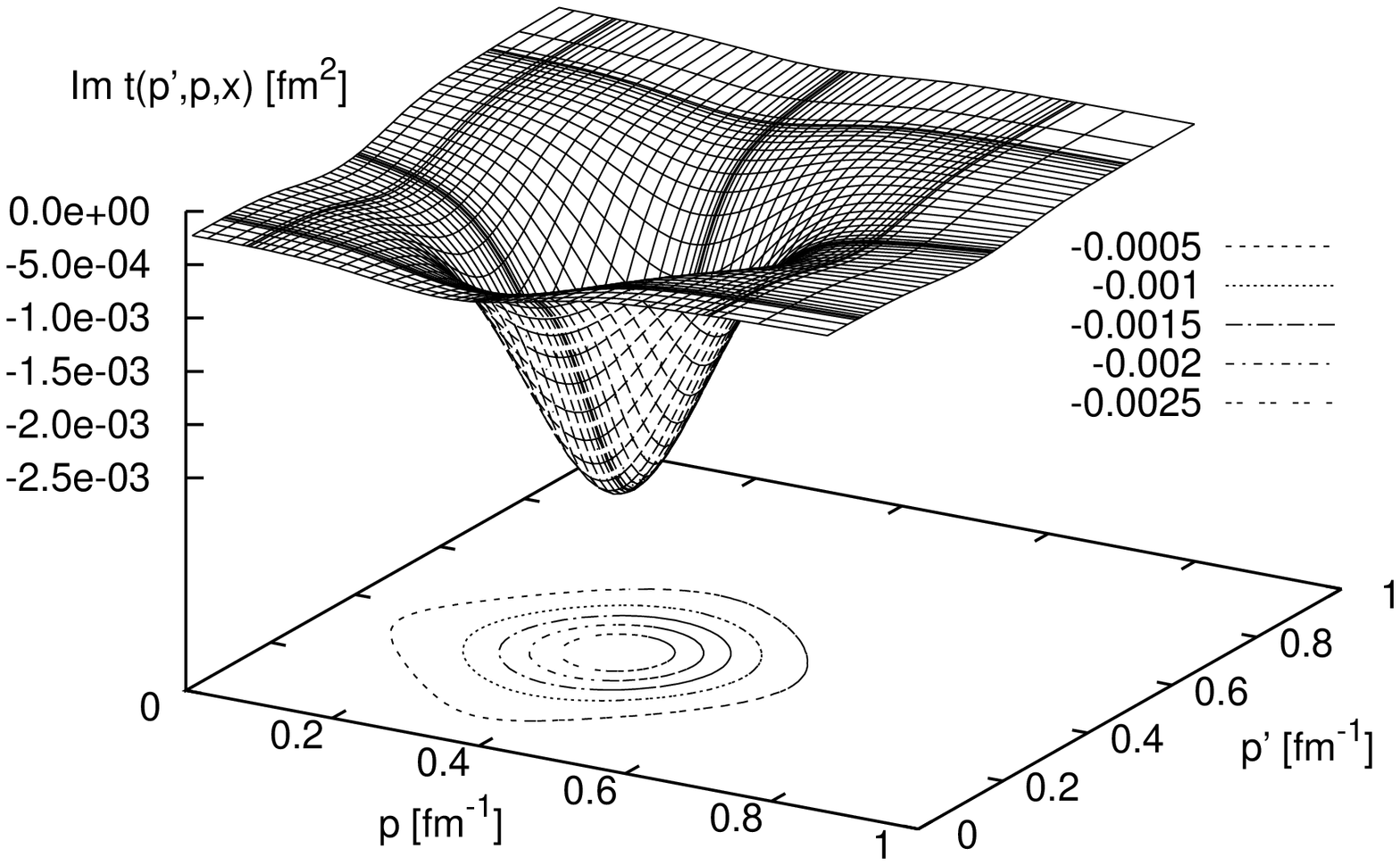}
\caption{The real and imaginary parts of the exponentially screened Coulomb t-matrix at E=13 MeV 
for the scattering angle $\theta$=$10^{\circ}$, R=20 fm 
and for two values of the screening parameter $n=2$~(up) and $n=3$~(down).}
\label{fig9}
\end{figure}

For the exponential screening we investigated the dependence of the 
screened Coulomb t-matrix on the value of the screening parameter $n$. 
We found only weak dependence, as exemplified in Fig.~\ref{fig9} where
t-matrices at $\theta=10^{\circ}$ are shown
for $n=2$ and $n=3$. This picture can be further supplemented by parts of Fig.~\ref{fig5} ($n=4$) and
Fig.~\ref{fig15} (sharp cut-off, what corresponds to infinite value of $n$) at the same angle. 
It is seen that there is only small decreasing of the height of the diagonal ridge
for the real part of the t-matrix at the smallest momenta. The minimum of the imaginary
part becomes narrower and deeper with increasing $n$.

\begin{figure}
\includegraphics[scale=0.4,clip=true,angle=-90]{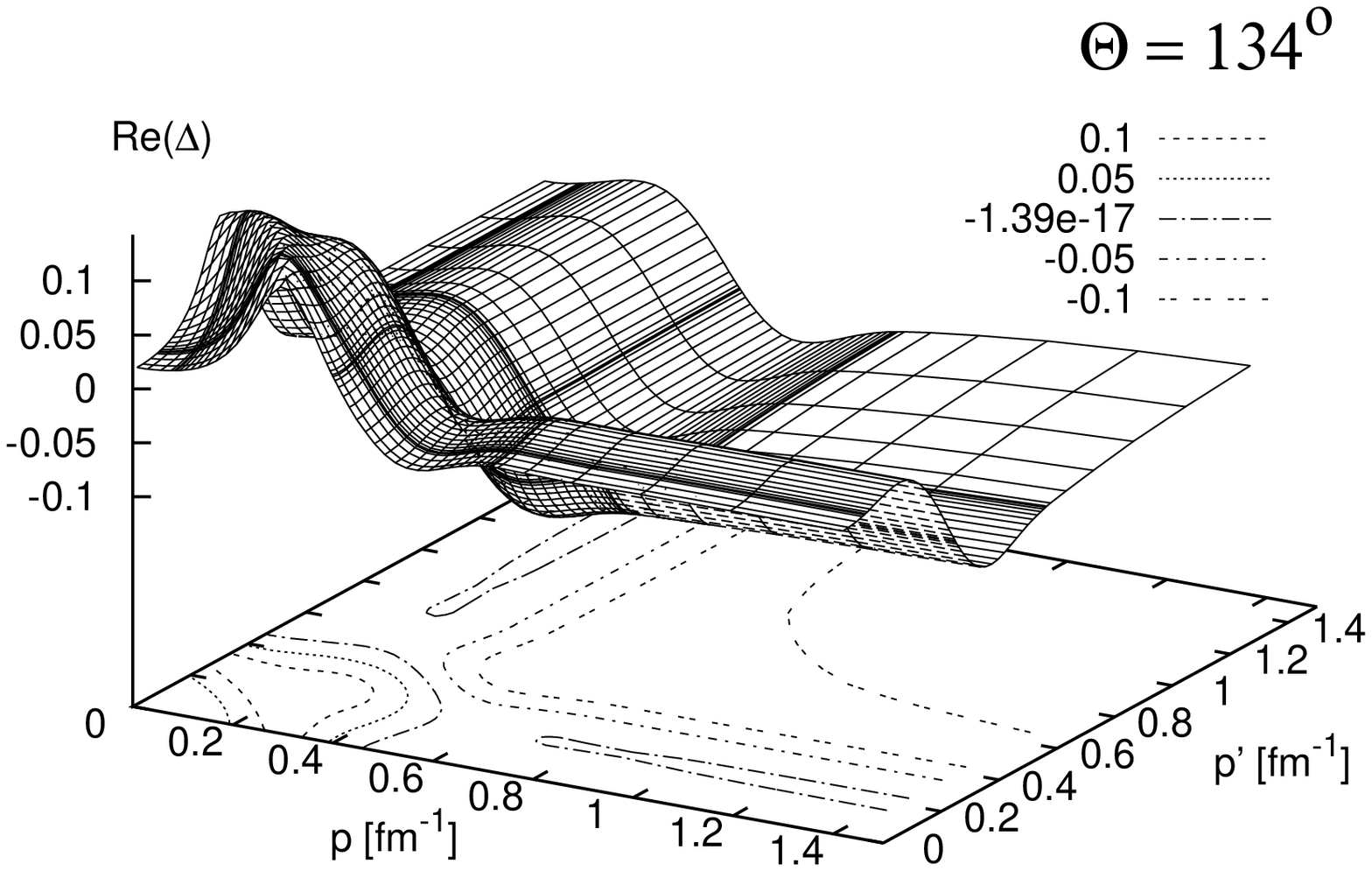}
\includegraphics[scale=0.4,clip=true,angle=-90]{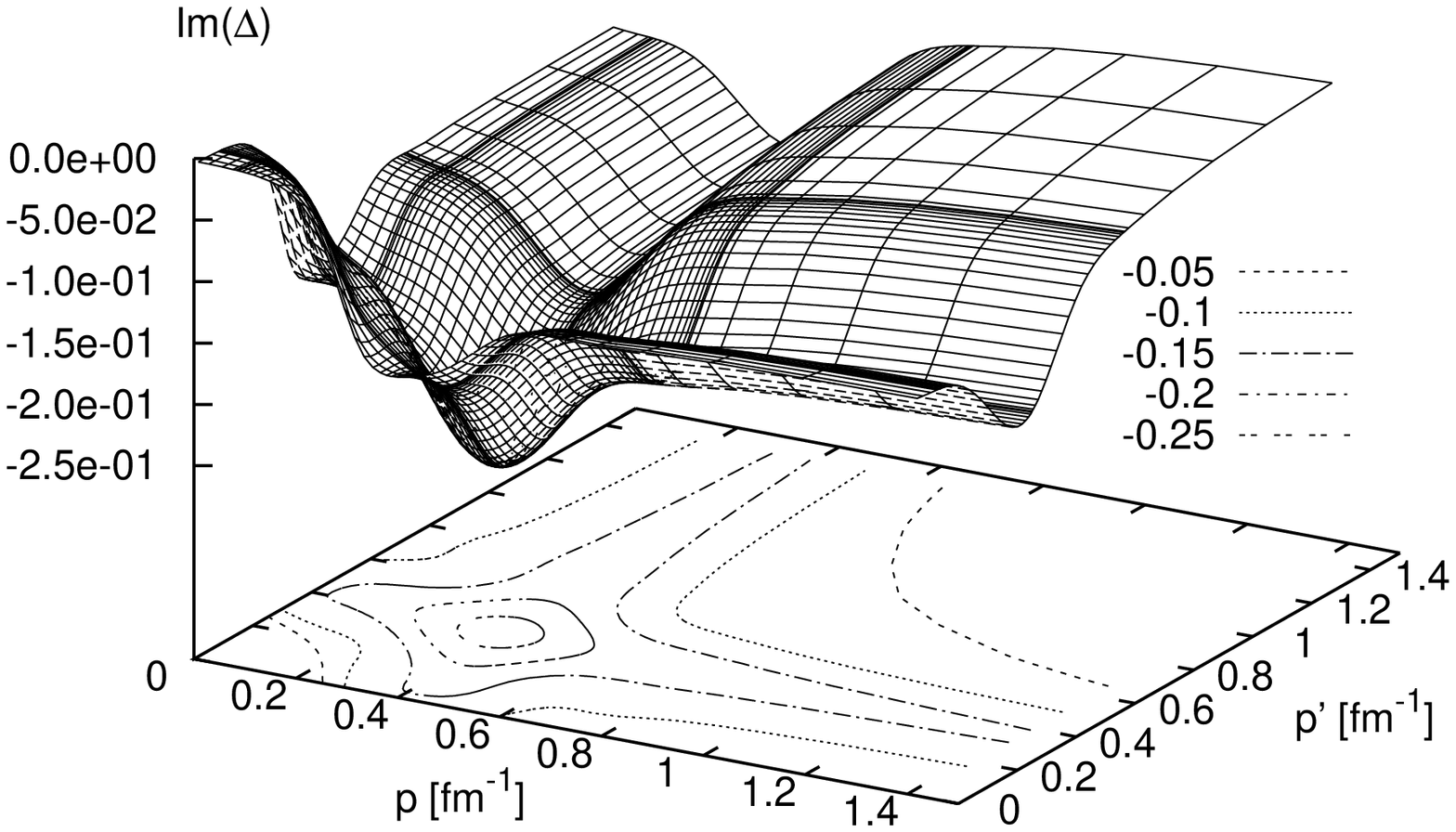}
\includegraphics[scale=0.4,clip=true,angle=-90]{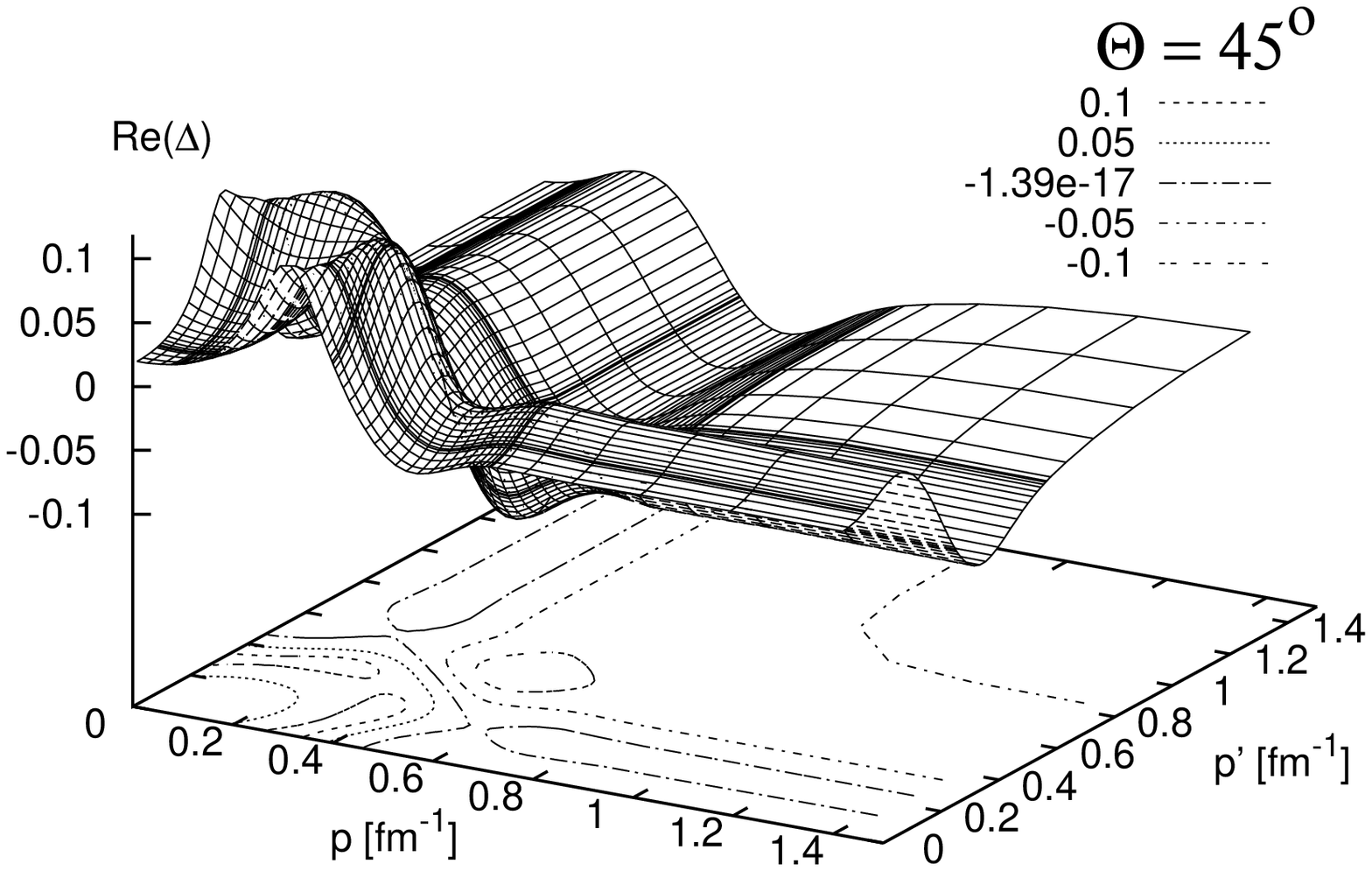}
\includegraphics[scale=0.4,clip=true,angle=-90]{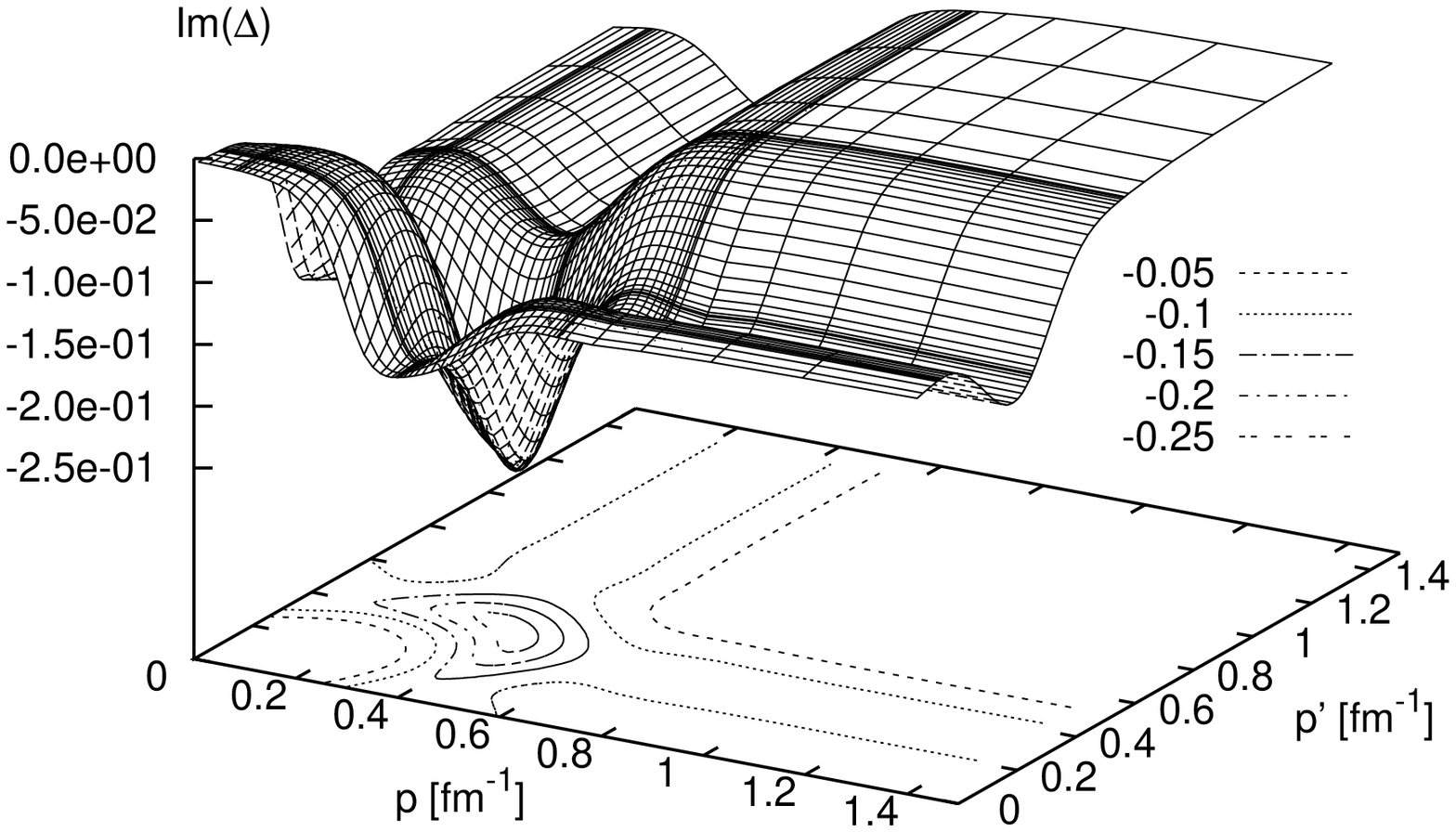}
\includegraphics[scale=0.4,clip=true,angle=-90]{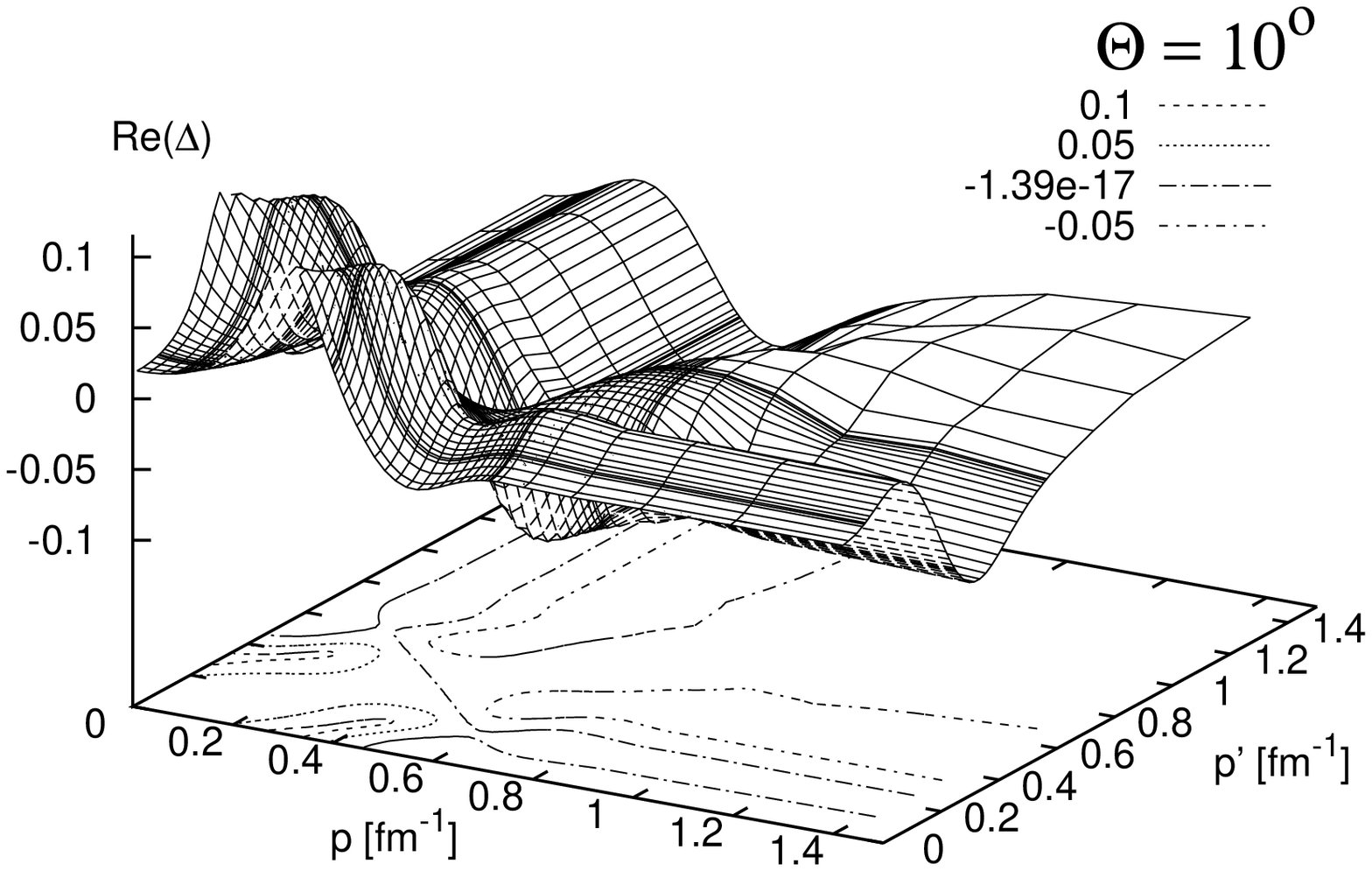}
\includegraphics[scale=0.4,clip=true,angle=-90]{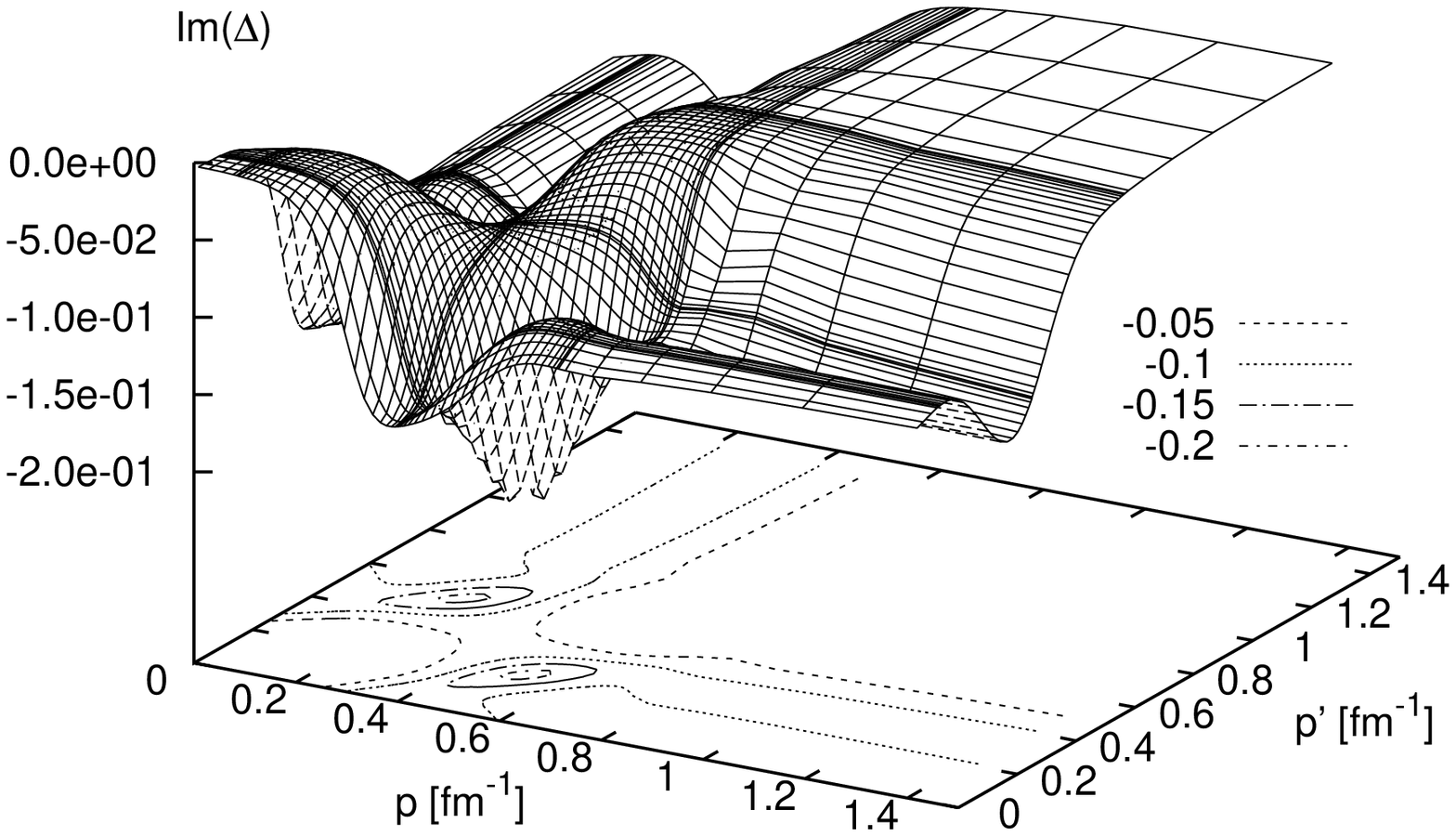}
\includegraphics[scale=0.4,clip=true,angle=-90]{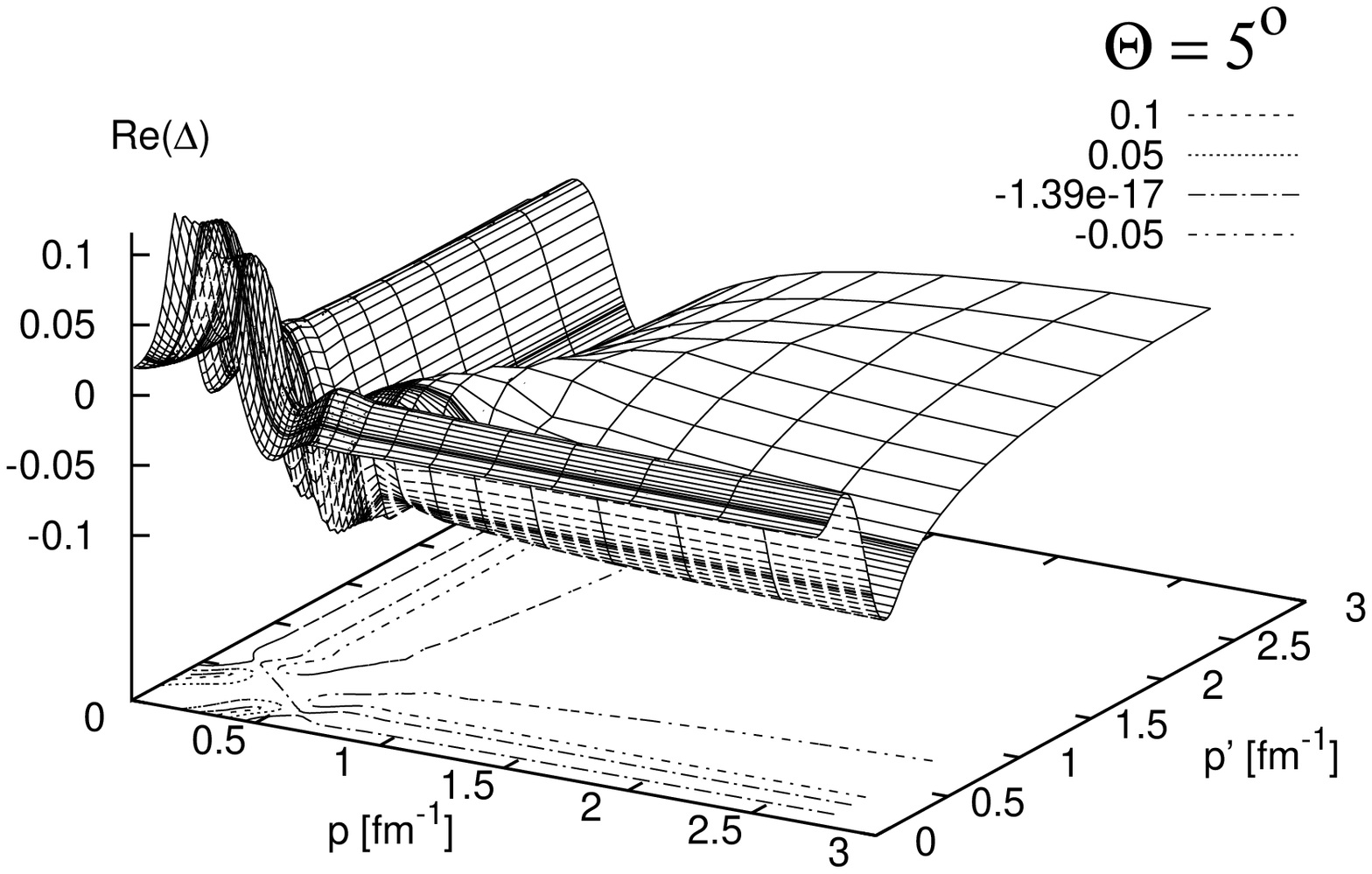}
\includegraphics[scale=0.4,clip=true,angle=-90]{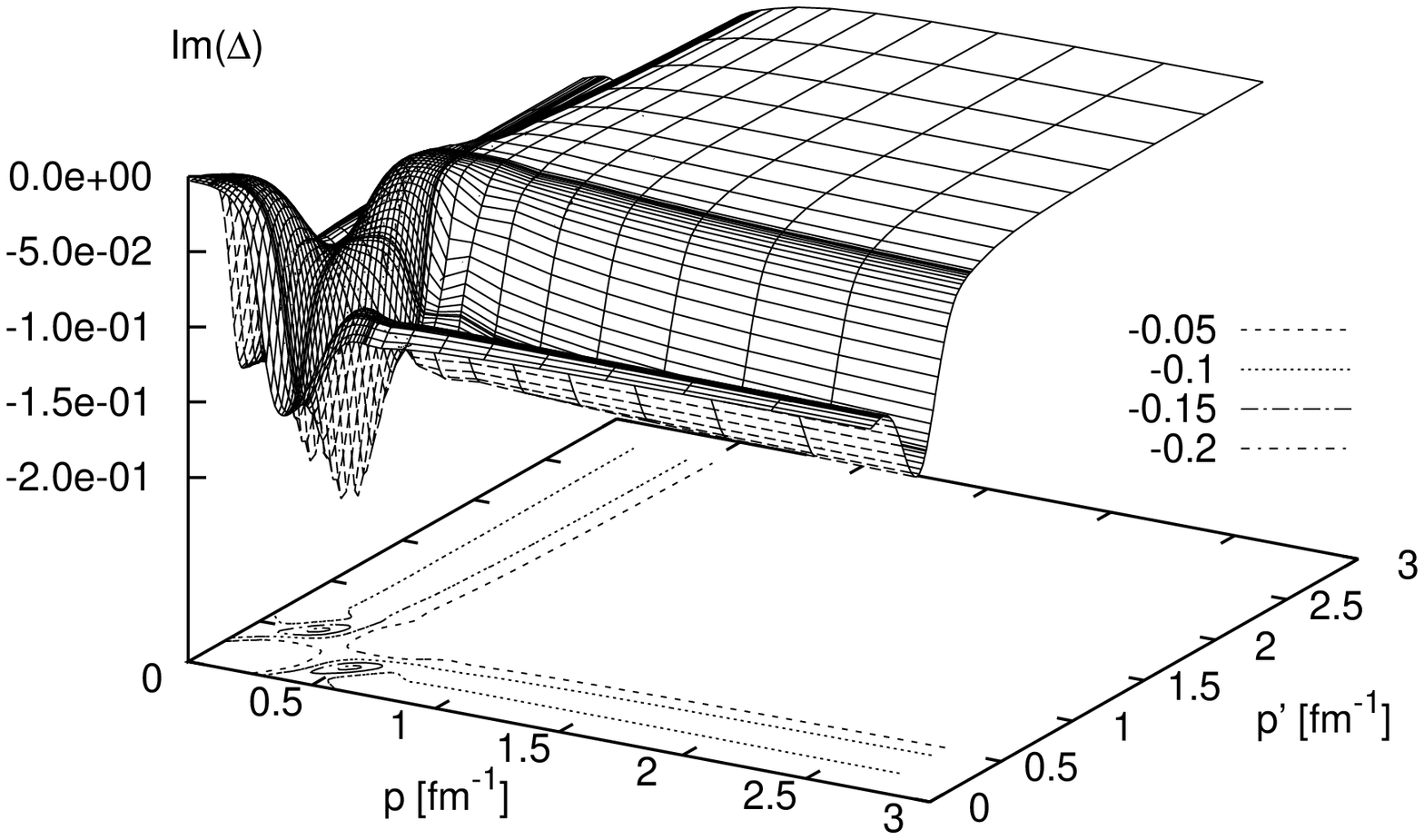}
\caption{The real (left) and imaginary (right) parts of the ratio
$\Delta=(t-v)/v$
(see text) for different 
scattering angles $\theta$:
$134^{\circ}$(1-st row), $45^{\circ}$(2-nd row),
$10^{\circ}$(3-rd row) and $5^{\circ}$(4-th row)
for the exponential screening with R=20 fm and n=4 at E=13 MeV.}
\label{fig10}
\end{figure}

Finally, we checked, how good is the approximation of the three-dimensional screened 
Coulomb t-matrix by the screened Coulomb potential alone.
To that aim we looked at the ratio \[\Delta \equiv \frac{t_c^R(p',p,x)-V_c^R(p',p,x)}{V_c^R(p',p,x)}  \]
In Fig.~\ref{fig10} we show $\Delta$ at different scattering angles,
obtained for the exponential screening with $R$=120 fm and $n$=4 at $E=13$ MeV. 
At all angles the real part of $\Delta$ does not exceed 12\%. The imaginary part of 
$\Delta$ is below 3\% for all $p$ and $p'$ values 
what emphasizes the smallness of the imaginary part and would indicate on validity of
approximation $t_c^R$ by pure screened Coulomb potential $V_c^R$.
The differences between the values of $t_c^R$ and $V_c^R$
are biggest for $p,p'<p_0$ and around $p=p_0$ or $p'=p_0$.
It means that in those regions of momenta the
approximation of the screened Coulomb t-matrix by the corresponding
screened Coulomb potential is rather poor.

\begin{figure}
\includegraphics[scale=0.9]{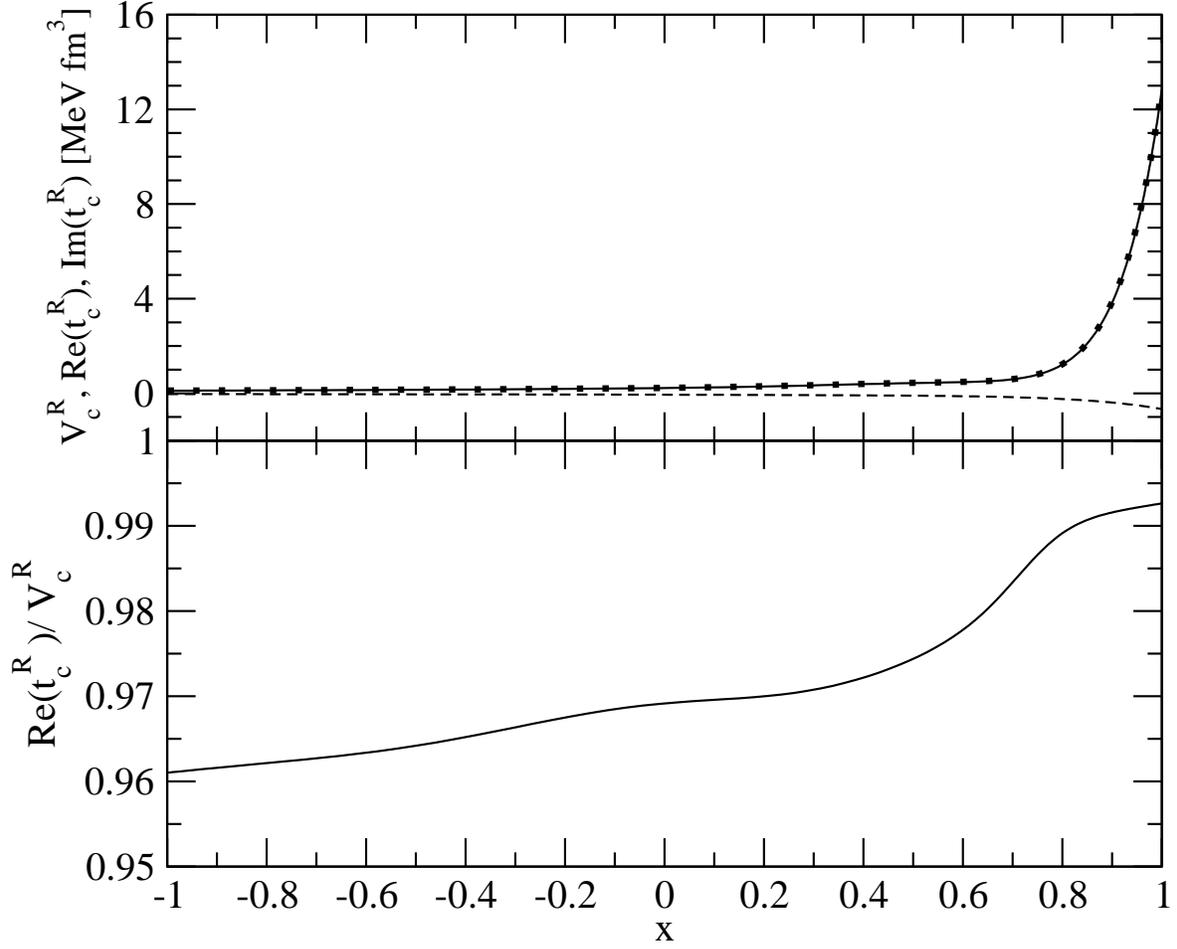}
\caption{
The real (solid line) and imaginary (dashed line) parts of the on-shell
exponentially screened Coulomb t-matrix $t_c^R(p_0,p_0,x)$ at E=13 MeV
as a function of the cosine of scattering angle $\theta$~($x=\cos(\theta)$)
(upper part). The dotted line is the corresponding screened Coulomb potential $V_c^R(p_0,p_0,x)$.
In the lower part of figure the ratio $Re(t_c^R)/V_c^R$ is shown. The parameters of the screening are 
R=20 fm and n=4.}
\label{fig2}
\end{figure}

\begin{figure}
\includegraphics[scale=0.9]{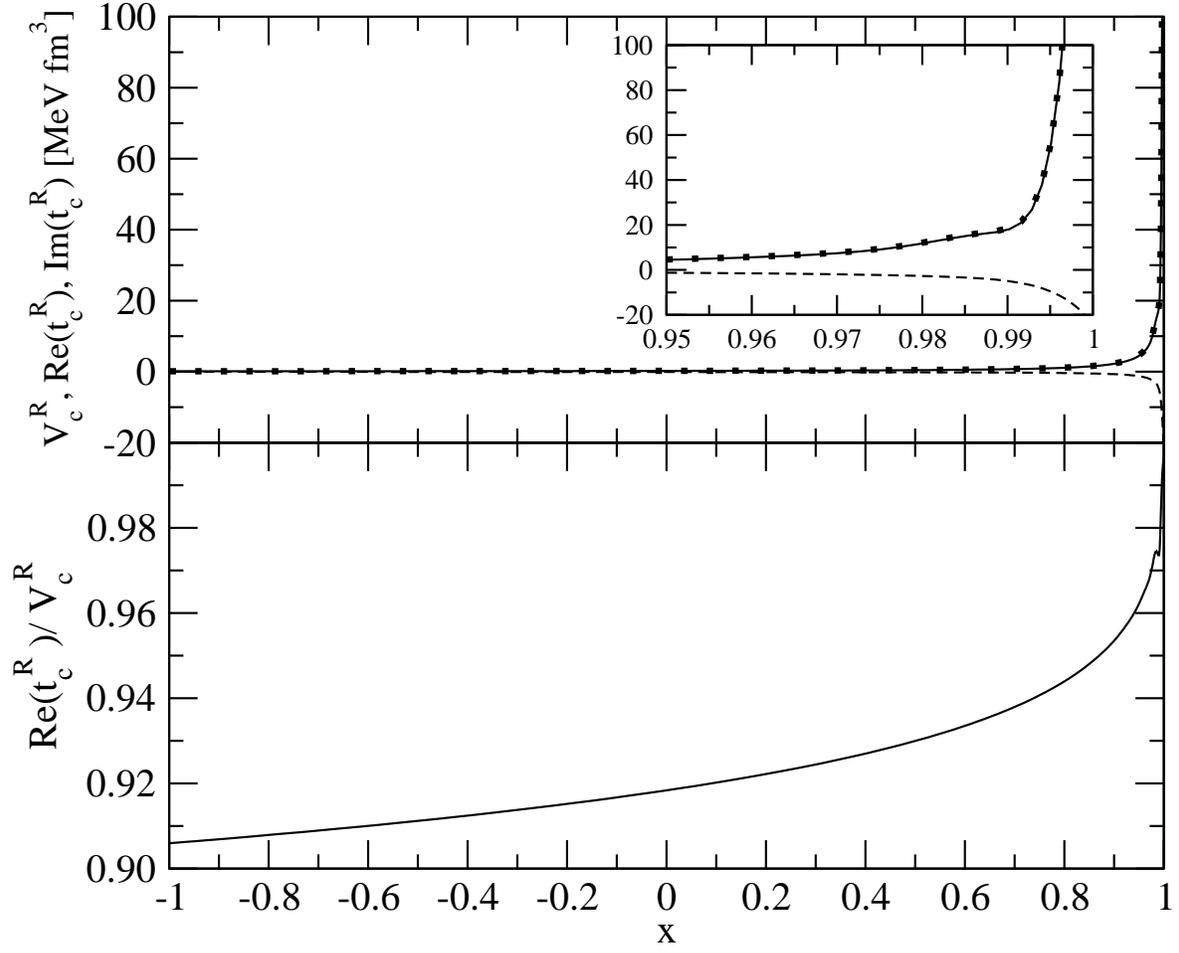}
\caption{The same as in Fig.\ref{fig2} but for R=120 fm.}
\label{fig3}
\end{figure}

For the on-shell screened Coulomb t-matrix elements $<p_0 \hat{p'} \vert t_c^R(\frac{p_0^2}{m}) \vert p_0 \hat p >$
obtained with exponential screening we show quality of the Born approximation
in Figs.~\ref{fig2} and~\ref{fig3}. There 
the screened Coulomb potential $V_c^R(p_0,p_0,x)$ together with t-matrix elements
are shown as a function of cosine of scattering angle $\theta$.
The screening parameters are $n=4$ and $R=20$ fm for Fig.~\ref{fig2} and
$n=4$ and $R=120$ fm for Fig.~\ref{fig3}

The real part of the screened Coulomb t-matrix (solid line) and the screened Coulomb potential
(dotted line) are close to each other, and their ratio does not
exceed 4\% of $V_c^R$ for the screening range $R=20$ fm and increases up to
about 10\% when the screening radius reaches $R=120$ fm.
In both cases the imaginary part of the t-matrix is
much smaller than its real part 
and becomes important only at very small scattering angles.
The real part of the t-matrix  is in all cases somewhat smaller
than the corresponding potential matrix element.

\section{The three-dimensional t-matrix from the partial wave decomposition}
\label{sec4}
In this section we would like to compare the 
on-shell elements of the three-dimensional screened Coulomb t-matrix obtained directly 
from Eq.(\ref{eq.1}) with its value 
derived from solutions of the partial wave decomposed Lippmann-Schwinger equation
$< p, l \vert t \vert p', l >$.
The latter is given as a sum of contributions from different angular momenta $l$
\begin{equation}
t^{pw}(\vec p, \vec{p'},x) 
\equiv \sum_{l=0}^{l_{max}} \frac{2l+1}{4\pi} P_l(x) 
< p, l \vert t \vert p', l > ,
\end{equation}
where $P_l(x)$ is the Legendre polynomial of the cosine of the scattering angle,
$x=\hat p \cdot \hat{p'}$, and $l_{max}$ should be high enough to give convergent result at given energy and screening. 
The $<p, l \vert t \vert p', l >$ partial wave t-matrix element
is a solution of the one-dimensional Lippmann-Schwinger equation
with the screened Coulomb potential driven by its partial wave element:
\begin{equation}
< p, l \vert V_c^R \vert p', l > \equiv \frac{2\alpha}{\pi} \int_0^{\infty} dr\; r\; j_l(pr) e^{-(\frac{r}{R})^n} j_l(p'r) \;.
\end{equation}
 
\begin{figure}
\includegraphics[scale=0.8]{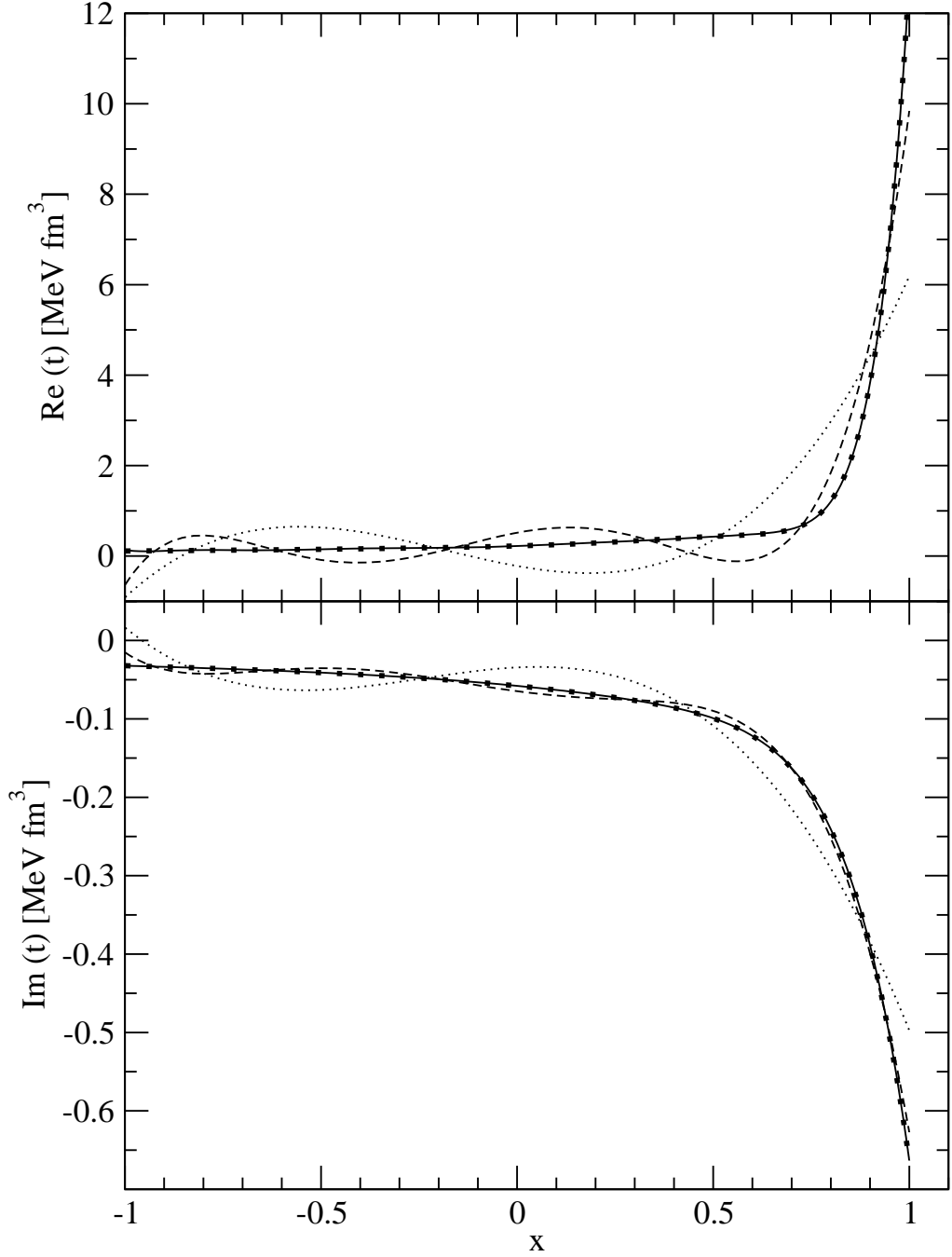}
\caption{
The convergence in the partial waves of the exponentially screened Coulomb t-matrix at E=13 MeV.
The real and imaginary parts of the on-shell elements of 
the three-dimensional screened Coulomb t-matrix are given by 
thick dotted line. The other lines represent 
partial wave generated results obtained by summation of partial waves up 
to 
$l_{max}=3$ (dotted),
$l_{max}=5$ (dashed) and 
$l_{max}=10$ (solid).
The screening parameters are $n=4$ and $R=20$ fm. }
\label{fig11}
\end{figure}

\begin{figure}
\includegraphics[scale=0.8]{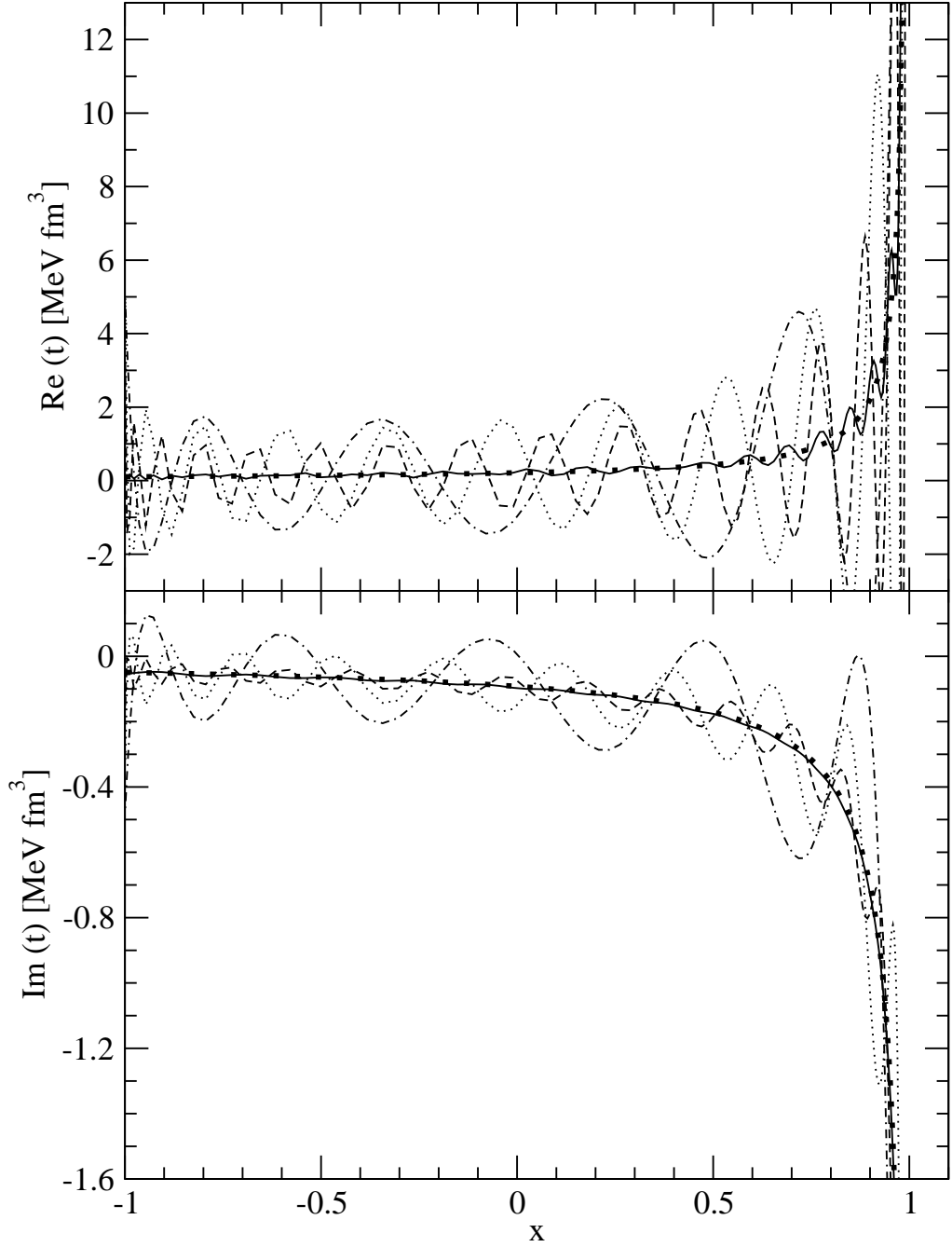}
\caption{The same as in Fig.~\ref{fig11} but for R=120 fm.
The partial waves generated results are obtained by summing up to
$l_{max}=10$ (dash-dotted),
$l_{max}=20$ (dotted),
$l_{max}=30$ (dashed) and
$l_{max}=50$ (solid). 
The thick dotted line represents as before the three-dimensional screened Coulomb t-matrix.}
\label{fig12}
\end{figure}

In Fig.\ref{fig11} we show convergence of the result with respect 
to $l_{max}$ for $l_{max}=3,5$ and $10$.
The partial wave generated screened Coulomb t-matrix is 
compared to the exponentially 
screened Coulomb t-matrix obtained directly from Eq.~(\ref{eq.1}) with $n=4$ and $R=20$ fm at E=13 MeV.
For such relatively  small screening radius $R$ it is sufficient 
to restrict to partial waves up to  $l_{max}=10$ only to
reproduce the full 3-dimensional t-matrix (thick dotted line).
However, 
the number of partial waves needed to reproduce the full three-dimensional t-matrix
increases rapidly with increasing screening radius.
This is exemplified in Fig.\ref{fig12} for $R=120$ fm, where
partial wave generated results with $l_{max}=10,20,30$ and $50$ 
are shown together with the full solution.
It is clearly seen that even taking $l_{max}=50$ is still not enough to describe 
with sufficient precision the full three-dimensional screened Coulomb 
t-matrix. This shows the importance of the direct three-dimensional solution of 
Eq.(\ref{eq.1}), 
which 
avoids problems caused by very slow convergence of the partial wave expansion
for big screening radii.

\section{Summary}
\label{summary}

In this paper we investigated numerically behaviour 
of the three-dimensional screened Coulomb t-matrix
using different types of screening.
Such t-matrix 
is solution of the three-dimensional Lippmann-Schwinger 
equation and is an important component of 
a recently developed novel approach to include the pp Coulomb 
force into the three-nucleon Faddeev calculations~\cite{witala,witala2}.
The direct application of the three-dimensional screened Coulomb t-matrix decreases 
substantially number of partial waves needed in 3N calculations
reducing them to the number of partial waves needed to get converged results in case when only 
nuclear part of the potential is acting.
The presented numerical procedure enables to get precise 
values of the three-dimensional screened Coulomb t-matrix
for any form of the screening.

We have also presented behaviour of the screened Coulomb t-matrix for 
different types of screening and for different
values of screening parameters.
The resulting t-matrices are similar.
Only the
t-matrix based on the sharply cut off Coulomb potential
reveals oscillations, which are absent for other 
types of screening. 
Those oscillations cause, that numerical requirements on  
computer resources are bigger in that case.
The behaviour of the real and imaginary parts of 
the three-dimensional screened Coulomb t-matrix with varying scattering angle is clearly seen in given examples.
The range of relative momenta, where the real part of $t$ 
is important rapidly grows with decreasing scattering angle for all types 
of screening. Contrary, the imaginary part of $t$ becomes more localized
with decreasing scattering angle but the depth of its minimum
grows.

The presented results exemplify, that the three-dimensional screened Coulomb
t-matrix 
can be obtained numerically in a reliable way for any type of screening
making it a valuable input 
in three-body calculations.

\section*{Acknowledgments}
This work was supported by the 2008-2011 Polish science funds as the
 research project No. N N202 077435. It was also partially supported by the
Helmholtz
Association through funds provided to the virtual institute ``Spin
and strong QCD''(VH-VI-231) and by
  the European Community-Research Infrastructure
Integrating Activity
``Study of Strongly Interacting Matter'' (acronym HadronPhysics2, Grant
Agreement n. 227431)
under the Seventh Framework Programme of EU.
 The numerical calculations were
performed on the supercomputer cluster of the JSC, J\"ulich,
Germany.

\section{Appendix}
\label{app1}
In the following tables we give values of parameters $a_k$ for the polynomial representations of the $I_{n,R}^{ratio}(q)=\sum_k a_k q^k$  
in case of exponential screening and two sets of $(n,R)$ parameters: $(n=2, R=20)$ in Tab.~\ref{tab1} and
$(n=4, R=120)$ in Tab.~\ref{tab2}. The $k$ (first column) gives a power of q which goes with the corresponding 
$a_k$ factor given in next columns for different ranges of q [fm$^{-1}$].  

\begin{table}[hp]
\begin{scriptsize}
\begin{center}
\begin{tabular}{|c|c|c|c|c|c|c|}
\hline
k & q$<$0.4   & 0.4$<$q$<$0.4825 & 0.4825$<$q$<$0.98 & 0.98$<$q$<$1.725 & 1.725$<$q$<$10.5 & 10.5$<$q$<$100 \\ 
\hline
0 & 0.5000125677171 & 2.400709537574  & 3.921200500830 & 1.100526033577 &  1.015402877247 & 1.000334483409 \\ 
\hline
1 & -0.006479230796918        & -11.71854218076  & -37.98859669416 & -0.2600475901233 & -0.1503793799690E-01  & -0.4682291329186E-04 \\ 
\hline
2 & 167.3502341712    & 42.56921618797  & 242.8850249646& 0.2987640252234 & 0.6468810517976E-02  & 0.2854148534117E-05 \\ 
\hline
3 & -20.31247086770           & -80.71301847419  & -974.8378775605 & -0.1809241210266 & -0.1511703363356E-02  & -0.9406333859035E-07 \\ 
\hline
4 & -12352.87853580 &78.54728957584 &2694.796660079 &0.5644212648115E-01 &0.2040113688732E-03  &0.1787524617018E-08 \\ 
\hline
5 &36456.32220409 &-31.11252021404 &-5362.934658952 &-0.7169472025225E-02 &-0.1584593947154E-04 &-0.1954522705295E-10 \\ 
\hline
6 &-524160.9711593 & - &7842.047876700 &- &0.6567867383073E-06 &0.1140554415044E-12 \\ 
\hline
7 &15708513.64668 &- &-8464.239235675 &- &-0.1124054147047E-07 &-0.2748652734470E-15 \\ 
\hline
8 &-165490836.2729 &- &6679.063924791 &- &- &- \\ 
\hline
9 &931536159.6442 &- &-3752.307134393 &- &- &- \\ 
\hline
10 &-3219232757.359 &- &1423.137799525 &- &- &- \\ 
\hline
11 &7104191720.650 &- &-326.9144001002 &- &- &- \\ 
\hline
12 &-9827857786.235 &- &34.37700645114 &- &-&- \\ 
\hline
13 &7804143302.566 &- &- &- &- &- \\ 
\hline
14 &-2723712664.926 &- &- &- &- &- \\ 
\hline
\end{tabular}
\caption{Parameters for the polynomial representation of $I_{n,R}^{ratio}(q)$ for exponential screening with $n=2$ and $R=20$.\label{tab1}}
\end{center}
\end{scriptsize}
\end{table}

\begin{table}[hp]
\begin{scriptsize}
\begin{center}
\begin{tabular}{|c|c|c|c|c|c|c|}
\hline
k& Q$<$0.0735   & 0.0735$<$q$<$0.11 & 0.11$<$q$<$0.148 & 0.148$<$q$<$0.4755 & 0.4755$<$q$<$1.49 & 1.49$<$q$<$100 \\ 
\hline
0 & 0.4430751038110 & -61.12120265116 & 17222.40652970 & 1.014501500770 & 1.002164118242 & 1.000039812743 \\ 
\hline
1 & 0.1624327164522 & 3122.555537517 & -1665157.950661 & -0.1499375990712 & -0.8401588857950E-02 & -0.1260392857265E-04  \\ 
\hline
2 & 5657.908456700 & -61815.86410617 & 72913086.23221 & 0.6806385287457 & 0.1422552507692E-01 & 0.1308082855220E-05 \\ 
\hline
3 & 37890.24852747 & 603243.9054240 & -1912425443.395 & -1.605992271344 & -0.1246326266740E-01 & -0.6209139096908E-07 \\ 
\hline
4 & -14459376.10638 & -2904646.753554 & 33474324944.51 & 1.925075383945 & 0.5526732459266E-02 & 0.1534201149986E-08 \\ 
\hline
5 & 613454088.3275 & 5524183.167489 & -412042007272.1 & -0.9274592364796 & -0.9817848874961E-03 & -0.2041445354694E-10 \\ 
\hline
6 & -34028042154.54 & - & 3658148517742. & - & - & 0.1386086594813E-12\\ 
\hline
7 & 1709808009779. & - & -0.2360571796460E+14 & - & - & -0.3765737840762E-15\\ 
\hline
8 & -0.5319386618092E+14 & - & 0.1098886587674E+15 & - & - & - \\ 
\hline
9 & 0.1051045675524E+16 & - & -0.3598714293204E+15 & - & - & - \\ 
\hline
10 & -0.1399718714098E+17 & - & 0.7868023497509E+15 & -  & -& -\\ 
\hline
11 & 0.1294364258655E+18 & - & -0.1030692803221E+16 & -  & -&- \\ 
\hline
12 & -0.8173922266757E+18 & - & 0.6113583130648E+15 & -  & -&- \\ 
\hline
13 & 0.3207993078724E+19 & -  & - & - & -&- \\ 
\hline
14 & -0.5891645069211E+19 & -  & - & - & -&- \\ 
\hline
\end{tabular}
\caption{The same as in Tab~\ref{tab1} but for $n=4$ and $R=120$. \label{tab2}}
\end{center}
\end{scriptsize}
\end{table}

\clearpage

\clearpage
\newpage
 
\clearpage
\newpage

\end{document}